\documentclass[12pt,a4paper]{article}

\usepackage{epsfig}
\usepackage{amsmath,amsfonts,amssymb}
\usepackage{t1enc}
\usepackage{cite}

\providecommand{\openone}{\leavevmode\hbox{\small1\kern-3.8pt\normalsize1}}

\newcommand{\Dpp}{\Delta^{++}}
\newcommand{\Dmm}{\Delta^{--}}
\newcommand{\Dppmm}{\Delta^{\pm \pm}}
\newcommand{\Dp}{\Delta^+}
\newcommand{\Dm}{\Delta^-}
\newcommand{\Dpm}{\Delta^\pm}
\newcommand{\Dmp}{\Delta^\mp}
\newcommand{\vt}{v_\Delta}
\newcommand{\remu}{r_{e\mu}}
\newcommand{\ptmiss}{p_T\!\!\!\!\!\!\!\!\not\,\,\,\,\,\,\,}
\newcommand{\tria}{{\tt Triada}}

\begin{document}

\begin{center}
\begin{Large}
{\bf Distinguishing seesaw models at LHC \\[2mm]
with multi-lepton signals}
\end{Large}

\vspace{0.5cm}
F. del Aguila, J. A. Aguilar--Saavedra  \\[0.2cm] 
{\it Departamento de F\'{\i}sica Te\'orica y del Cosmos and CAFPE, \\
Universidad de Granada, E-18071 Granada, Spain} \\[0.1cm]
\end{center}

\begin{abstract}
We investigate the LHC discovery potential for electroweak scale heavy neutrino singlets (seesaw I), scalar triplets (seesaw II) and fermion triplets (seesaw III). For seesaw I we consider a heavy Majorana neutrino coupling to the electron or muon. For seesaw II we concentrate on the likely scenario where the new scalars decay to two leptons. For seesaw III we restrict ourselves to heavy Majorana fermion triplets decaying to light leptons plus gauge or Higgs bosons, which are dominant except for unnaturally small mixings.
The possible signals are classified in terms of the charged lepton multiplicity, studying nine different final states ranging from one to six charged leptons. Using a fast detector simulation of signals and backgrounds, it is found that the trilepton channel
$\ell^\pm \ell^\pm \ell^\mp$ is by far the best one for scalar triplet discovery, and for fermion triplets it is as good as
the like-sign dilepton channel $\ell^\pm \ell^\pm$. For heavy neutrinos with a mass $O(100)$ GeV, this trilepton channel is also better than the usually studied like-sign dilepton mode.
In addition to evaluating the discovery potential, we make special emphasis on the discrimination among seesaw models if a positive signal is observed.
This could be accomplished not only by searching for signals in different final states, but also
by reconstructing the mass and determining the charge of the new resonances, which is possible in several cases. For high luminosities, further evidence is provided by the analysis of the production angular distributions in the cleanest channels with three or four leptons.
\end{abstract}

\section{Introduction}
\label{sec:1}

The near operation of the Large Hadron Collider (LHC) represents a remarkable opportunity to explore physics beyond the electroweak scale. In particular, 
physics at higher scales can be explored in the lepton sector~\cite{delAguila:2008iz}, where the only evidence of physics beyond the Standard Model (SM) has been found up to now, namely massive neutrinos. Many theories have been proposed to enlarge the SM incorporating tiny neutrino masses, as required by experimental data \cite{Amsler:2008zz}. Among them, seesaw models explain their smallness by introducing extra matter at a high scale. After integration of these heavy fields, the lepton number violating (LNV) dimension five operator \cite{Weinberg:1979sa}
\begin{equation}
(O_5)_{ij} = \overline{L_{iL}^c} \tilde \phi^* \tilde \phi^\dagger L_{jL}
\label{ec:O5}
\end{equation}
is generated, where
\begin{equation}
L_{iL} = \left( \!\begin{array}{c}\nu_{iL} \\ l_{iL}\end{array} \!\right) \,,\quad
i=1,2,3
\end{equation}
are the SM left-handed lepton doublets, $\phi$ the SM Higgs and $\tilde \phi=i\tau_2 \phi^*$, with $\tau_i$ the Pauli matrices. This operator yields Majorana masses for the neutrinos after spontaneous symmetry breaking. At higher energies neutrino masses can be generated from higher dimension operators involving extra fields (see for example
Ref.~\cite{Gogoladze:2008wz}) but when all heavy degrees of freedom are integrated the operator in Eq.~(\ref{ec:O5}) is recovered.

There are three types of tree-level seesaw mechanisms which originate the operator in
Eq.~(\ref{ec:O5}), which is the only five-dimensional one allowed by the
$\text{SU}(3) \times \text{SU}(2)_L \times \text{U}(1)_Y$ gauge symmetry.
The original seesaw \cite{Minkowski:1977sc,GellMann:1980vs,Yanagida:1979as,Mohapatra:1979ia}, also known as seesaw of type I, introduces right-handed neutrino singlets $N$ at a high scale. Type II seesaw \cite{Magg:1980ut,Cheng:1980qt,Gelmini:1980re,Lazarides:1980nt,Mohapatra:1980yp}
enlarges the SM with a complex scalar triplet $\Delta$ with hypercharge $Y=1$, and 
seesaw III \cite{Foot:1988aq,Ma:1998dn} introduces colourless fermionic triplets $\Sigma$ with $Y=0$.
Both seesaw I and II are present in left-right models \cite{Mohapatra:1974gc,Senjanovic:1975rk}. Combinations of seesaw I and III are predicted in some grand unified theories \cite{Bajc:2006ia,Dorsner:2006fx, Bajc:2007zf}, and can be implemented in left-right models as well \cite{Perez:2008sr}.

The three types of seesaw mechanism generate not only the dimension five operator which gives light neutrino masses, but also additional lepton number conserving (LNC) dimension six operators, which are different in each seesaw scenario \cite{Broncano:2002rw,Abada:2007ux} (see also Ref.~\cite{delAguila:2007ap}). Therefore,
seesaw models may in principle be discriminated, albeit indirectly, with precise low-energy measurements sensitive to these dimension six operators.
The seesaw messengers $N$, $\Delta$, $\Sigma$ can also be produced at LHC if their masses are not very large but of the order of the electroweak scale and, in the case of neutrino singlets, if their mixing with the SM leptons is of order 
$10^{-2}$ or larger. Therefore, LHC gives a unique chance to uncover the mechanism of neutrino mass generation if these heavy states are directly observed.

The production of heavy neutrinos~\cite{Datta:1993nm,Almeida:2000pz,Panella:2001wq,
Han:2006ip,delAguila:2007em,Bray:2007ru}, scalar triplets~\cite{Huitu:1996su,Gunion:1996pq,
Akeroyd:2005gt,Hektor:2007uu,Perez:2008zc,Perez:2008ha} and fermion triplets~\cite{Franceschini:2008pz}
and their possible signals have been extensively studied in the literature. In this paper we take a novel approach to their analysis.
Instead of classifying signals in terms of the particles produced in the hard process and studying one or more particular channels, we classify them by the signatures actually seen at the experiment, generating {\em all} the signal contributions. As the number of jets in the final state is not a good discriminant due to radiation and pile-up, signals are classified in terms of the charged lepton multiplicity, and the number of hard jets is considered only in few special cases as an extra information for the kinematical reconstruction. We believe that this is a more appropriate choice from the experimental point of view. An essential feature of a real experiment is that the observed final states receive in general contributions from several signal processes and, conversely, one given signal process contributes to several final states if, for example, one or more charged leptons are missed by the detector. Both effects are taken into account in our analysis, which includes the effects of radiation, pile-up and hadronisation, performed by a parton shower Monte Carlo, and uses a fast detector simulation.
Within a given seesaw scenario we simulate all signal processes and then
examine thoroughly the relevant final states,
nine in total, ranging from one to six charged leptons.

A second difference with respect to previous literature concerns the guiding principle of the analyses performed. Due to the profusion of new physics scenarios to be tested at the LHC it is expected that, at least in a first phase, searches will be inclusive to some extent and rather model-independent, in order to be sensitive to different types of new physics contributing to a given channel. For example, it is not likely that, in the absence of a strong physics case, a final state with two like-sign leptons, large missing energy, two jets with an invariant mass consistent with the $W$ mass and two additional jets will be examined right from the beginning of LHC operation. On the contrary, searches are expected, for example, for inclusive final states with two like-sign leptons, as it has already been done at Tevatron~\cite{Abulencia:2007rd}.
(With this philosophy, our classification of signals in terms of lepton multiplicity is perfectly suited.) Then, in our analyses we will not set fine-tuned kinematical cuts on many variables to enhance the signals, but our criteria for
variable selection and background suppression will be rather general, and in most cases valid for seesaw I, II and III signals. In this way, our results and procedures will be adequate for model-independent searches in the multi-lepton final states.
Of course, if some hint of new physics is found the analyses can be refined
and adapted to some particular scenario, in order to reconstruct the resonance masses and/or enhance the sensitivity.

Finally, we pay a special attention to the discrimination among models if a positive signal is found. This study is far more involved than the ``simple'' observation of a new physics signal from seesaw messengers in one or few particular channels, to which most previous literature has been devoted. On the contrary, model discrimination requires a systematic analysis of all possible final states and the reconstruction of the new resonances when possible. The three types of seesaw can give signals with one, two and three charged leptons, plus a variable number of jets (which, as stressed before, is not very indicative because of radiation and pile-up, added to the difficulty of the mass reconstruction in hadronic channels). 
Final states with four charged leptons $\ell^+ \ell^+ \ell^- \ell^-$ (with $\ell=e,\mu$, including all flavour combinations of the four leptons) only arise in seesaw II and III scenarios, and $\ell^\pm \ell^\pm \ell^\pm \ell^\mp$ signals only in seesaw III, as well as five and six lepton final states. Therefore,
a first straightforward discrimination of seesaw models results from considering the final states in which the signals can be seen and their statistical significance, which are not the same for seesaw I, II and III. Moreover, if a positive signal is found in a given channel, the mass reconstruction of the heavy resonance can be often performed and its charge measured, giving clear direct evidence for the production of the new particle. For example, doubly charged scalars $\Dpp$ produced in seesaw II models often give from their decay a signal consisting of two like-sign leptons with
an invariant mass very close to $M_{\Dpp}$, whereas the spectrum of like-sign dilepton invariant masses in seesaw I and III does not exhibit any peak.
For high integrated luminosities, the opening angle distribution can also be tested in the cleanest channels, giving evidence for the scalar or fermionic nature of the heavy states produced.

We also have to point out that,
since the main interest in the arrival of the LHC era is in early discoveries (and a luminosity of 300 fb$^{-1}$ will not be available until several years of LHC operation, in the optimistic scenarios), we have concentrated our study on seesaw messengers with masses close to the electroweak scale, which in the case of seesaw II and III would be quickly discovered after quality data are available. Thus, 
for heavy neutrino singlets we have assumed a mass $m_N = 100$ GeV, while for scalar and triplet fermions the mass has been taken as 300 GeV. 
We have found that the best channel for discovery, balancing large signal branching ratios and small backgrounds, is the trilepton one $\ell^\pm \ell^\pm \ell^\mp$. For scalar triplets of this mass, $5\sigma$ discovery would be possible with only 3.6 fb$^{-1}$ of luminosity for normal neutrino mass hierarchy (NH) and 0.9 fb$^{-1}$ for inverted hierarchy (IH). For fermionic triplets the luminosity needed is 2.5 fb$^{-1}$, while for heavy neutrino singlets it is much larger, around 180 fb$^{-1}$. A clean model discrimination would be possible within the first LHC year with 10 fb$^{-1}$.

In this paper we work out the minimal seesaw I--III scenarios without new interactions. It is worth mentioning that the latter can mediate new production mechanisms of seesaw messengers, for example, heavy neutrinos in models with left-right symmetry and an additional $W_R$ boson~\cite{Keung:1983uu,Datta:1992qw,
Ferrari:2000sp,Gninenko:2006br}, models with an additional $Z'$~\cite{delAguila:2007ua,Huitu:2008gf} or with new scalar
doublets~\cite{Ma:2000cc,BarShalom:2008gt}.
The same type of signals can also be produced by fourth generation neutrinos\cite{CuhadarDonszelmann:2008jp}.
New interactions can also lead to predictive indirect seesaw signals, as for example in type II seesaw in the context of supersymmetric models~\cite{Joaquim:2006uz,Joaquim:2006mn,Hirsch:2008gh}.
Alternatives to the seesaw mechanism are also possible, for example in $R$-parity violating supersymmetric models,
leading to interesting connections between neutrino physics
and collider observables~\cite{Hirsch:2000ef,Hirsch:2003fe,Hirsch:2008ur}
(for reviews see Refs.~\cite{Romao:1999bu,Romao:2007ny} and references there in).

The rest of this paper is organised as follows. In section~\ref{sec:2} we write down the relevant Lagrangian terms for the three seesaw types. We concentrate on the interactions 
mediating the production and decay processes considered in this work using a notation that, while being general, is intended to be standard and of easy use for readers not familiar with the models introduced.
In section~\ref{sec:3} we describe in detail the common features of the analyses, which are performed in sections \ref{sec:4}, \ref{sec:5} and \ref{sec:6} for heavy neutrinos, scalar triplets and fermion triplets, respectively. We begin these sections with
a short introduction, and conclude them with a summary highlighting the main results, so that the reader may skip the details.
Our conclusions are drawn in section~\ref{sec:7}. In the Appendix we give the Feynman rules used in the Monte Carlo generators.

\section{General framework}
\label{sec:2}

In this section we introduce the different seesaw models, set our conventions and write the Lagrangian terms relevant to the production and decay processes studied here. We keep the notation as simple as possible and similar for the three types of seesaw mechanism. Constraints on the mixing of the new fermions and scalars are briefly reviewed at the end of each subsection. The Feynman rules are collected in the Appendix.

\subsection{Seesaw I}
\label{sec:2.1}

Type-I seesaw is usually implemented by adding three right-handed current eigenstates
$N'_{iR}$, $i=1,2,3$, transforming as singlets under the SM gauge group. This allows to write a Yukawa interaction for neutrinos analogous to the one for charged leptons,
\begin{equation}
\mathcal{L}_{\text{Y}} = -Y_{ij} \, \overline{L'_{iL}} N'_{jR} \, \tilde \phi +\text{H.c.}\,, 
\end{equation}
where $Y$ is a $3 \times 3$ matrix of couplings and $L'_{iL}$ the SM lepton doublets (in the weak eigenstate basis). This interaction generates a mass term upon spontaneous symmetry 
breaking
\begin{equation}
\phi = \left( \!\begin{array}{c} \phi^+ \\ \phi^0 \end{array} \!\right)
\to \frac{1}{\sqrt 2} \left( \!\begin{array}{c} 0 \\ v \end{array} \!\right) \,,\quad
\tilde \phi \equiv i\tau_2 \phi^*
\to \frac{1}{\sqrt 2} \left( \!\begin{array}{c} v \\ 0 \end{array} \!\right) \,,
\end{equation}
with $v=246$ GeV. Since $N'_{iR}$ are SM singlets, gauge symmetry allows a Majorana mass term
\begin{equation}
\mathcal{L}_\text{M} = -\frac{1}{2} M_{ij} \overline{N'_{iL}} N'_{jR} + \text{H.c.} \,,
\end{equation}
with $M$ a $3\times 3$ symmetric matrix and $N'_{iL} \equiv N_{iR}^{'c}$.\footnote{We avoid writing parentheses in charge conjugate fields to simplify the notation, and write $\psi_L^c \equiv (\psi_L)^c$, 
$\psi_R^c \equiv (\psi_R)^c$.}
Defining $\nu'_{iR} \equiv \nu_{iL}^{'c}$, where $\nu'_{iL}$ are the SM neutrino eigenstates, the full neutrino mass term reads
\begin{eqnarray}
\mathcal{L}_\mathrm{mass} & = & - \frac{1}{2} \,
\left(\bar \nu'_L \; \bar N'_L \right)
\left( \! \begin{array}{cc}
0 & \frac{v}{\sqrt 2} Y \\ \frac{v}{\sqrt 2} Y^T & M
\end{array} \! \right) \,
\left( \!\! \begin{array}{c} \nu'_R \\ N'_R \end{array} \!\! \right)
\; + \mathrm{H.c.} \,.
\label{ec:massI}
\end{eqnarray}
The neutrino gauge interactions are the same as in the SM,
\begin{eqnarray}
\mathcal{L}_W & = & - \frac{g}{\sqrt 2} \left(
 \bar l'_{iL} \gamma^\mu \nu'_{iL} \, W_\mu^- 
+ \bar \nu'_{iL} \gamma^\mu l'_{iL} \, W_\mu^+  \right) \,, \nonumber \\
\mathcal{L}_Z & = & - \frac{g}{2 c_W} \, \bar \nu'_{iL} \gamma^\mu \nu'_{iL} \, Z_\mu \,,
\label{ec:Ngauge1}
\end{eqnarray}
with $l'_{iL}$ the charged lepton weak eigenstates and $c_W$ the cosine of the weak mixing angle. The interaction with the Higgs boson $H$, with the usual normalisation
$\phi^0 = (v+H+i\chi)/\sqrt 2$, is
\begin{eqnarray}
\mathcal{L}_H & = & - \frac {1}{\sqrt 2} \left( \bar \nu'_{iL} Y_{ij} N'_{jR}
+ \bar N'_{jR} Y^\dagger_{ji} \nu'_{iL} \right) \, H \,.
\label{ec:Nhiggs1}
\end{eqnarray}
Then, the relevant interaction terms for the heavy neutrino mass eigenstates
$N_i \simeq N'_{iR}$ can be obtained by diagonalising the mass matrix in Eq.~(\ref{ec:massI}) and rewriting the interactions in the mass eigenstate basis (for details see for example Refs.~\cite{delAguila:2005pf,delAguila:2006dx}). For a heavy Majorana neutrino $N$ (dropping the subindex) and $l=e,\mu,\tau$ we have
\begin{eqnarray}
\mathcal{L}_W & = & - \frac{g}{\sqrt 2}  \left( V_{l N} \, \bar l \gamma^\mu P_L N \; W_\mu^- 
  + V_{l N}^*  \, \bar N \gamma^\mu  P_L l \; W_\mu^+ \right) \,, \nonumber \\
\mathcal{L}_Z & = & - \frac{g}{2 c_W} \left( V_{l N} \, \bar \nu_l \gamma^\mu  P_L N
   + V_{l N}^* \, \bar N \gamma^\mu P_L \nu_l \right) Z_\mu \,, \nonumber \\[1mm]
\mathcal{L}_H & = & - \frac{g \, m_N}{2 M_W} \left( V_{l N} \, \bar \nu_l  P_R N
+ V_{l N}^* \, \bar N  P_L \nu_l \right) H \,,
\label{ec:Nint}
\end{eqnarray}
where $m_N$ is the heavy neutrino mass and
\begin{equation}
V_{lN} \simeq \frac{Y_{lN} v}{\sqrt 2 m_N}
\end{equation}
is the mixing between the charged lepton $l$ and the heavy neutrino $N$. Due to the Majorana character of $N$ and $\nu_l$, the last terms in the $Z$, $H$ Lagrangians can be rewritten,
\begin{eqnarray}
\mathcal{L}_Z & = & - \frac{g}{2 c_W} \, \bar \nu_l \gamma^\mu \left(  
V_{l N} P_L - V_{l N}^* P_R \right) N \; Z_\mu \,, \nonumber \\[1mm]
\mathcal{L}_H & = & - \frac{g \, m_N}{2 M_W} \, \bar \nu_l \left( V_{l N} P_R
+ V_{l N}^* P_L \right) N \; H \,.
\label{ec:Nint2}
\end{eqnarray}
In the absence of any particular symmetry in the Yukawa couplings, light neutrino masses $m_\nu$ are of the order $Y^2 v^2  / 2 m_N$, and the heavy neutrino mixings are $V_{lN} \sim \sqrt{m_\nu / m_N}$. Hence, for a heavy neutrino within LHC reach, say with a mass $m_N \sim 100$ GeV,
its seesaw-type contribution to light neutrino masses is of the order of $300\, Y^2$ GeV,
requiring very small Yukawas $Y\sim 10^{-6}$ to reproduce light neutrino masses $m_\nu \sim 0.1$ eV. Moreover, the natural order of magnitude of the mixings is $O(10^{-6})$, too small to give observable signals. In models with approximate flavour symmetries
the total $Y^2 v^2  / 2 m_N$ contribution to light neutrino masses can be suppressed
\cite{Buchmuller:1991tu,Ingelman:1993ve,Gluza:2002vs,Pilaftsis:2005rv} and the mixings decoupled from the light/heavy mass ratio \cite{Tommasini:1995ii}. In the specific realisations considered in the literature, the symmetry advocated to naturally reproduce light neutrino masses with electroweak scale heavy neutrinos is lepton number~\cite{Kersten:2007vk}, in which case the heavy neutrinos are quasi-Dirac instead of Majorana particles and the phenomenology is different, in particular their decays. In this work we will not address how to build a realistic model in which heavy Majorana neutrinos appear with non-negligible mixings, but we will
simply take the Lagrangian in Eqs.~(\ref{ec:Nint}) as a phenomenological one, in order to investigate the LHC potential to discover heavy Majorana neutrinos.
The prospects for Dirac neutrinos at the electroweak scale, which appear more naturally in seesaw models, are discussed elsewhere \cite{corto}. 

Even if we put aside the connection between heavy neutrino mixing and light neutrino masses, the former must be small due to present experimental constraints. Electroweak precision data set limits on mixings involving a single charged lepton~\cite{Langacker:1988up,Nardi:1994iv,
Bergmann:1998rg,Bekman:2002zk,delAguila:2008pw}. Using the latest experimental data, the constraints at 90\% confidence level (CL) are~\cite{delAguila:2008pw}
\begin{equation}
\sum_{i=1}^3 |V_{e N_i}|^2 \leq 0.0030 \,, \quad
\sum_{i=1}^3 |V_{\mu N_i}|^2 \leq 0.0032 \,, \quad
\sum_{i=1}^3 |V_{\tau N_i}|^2 \leq 0.0062 \,,
\label{ec:eps1}
\end{equation}
which in particular imply constraints on the individual mixings $V_{lN}$ of a heavy neutrino $N$. These constraints are particularly important since they determine the heavy neutrino production cross sections at LHC (see next section). From lepton flavour-violating (LFV) processes \cite{Korner:1992an,Ilakovac:1994kj,
Illana:2000ic,Tommasini:1995ii} one has constraints involving two charged leptons,
\begin{equation}
\left| \sum_{i=1}^3 V_{e N_i} V_{\mu N_i}^* \right| \leq 0.0001 \,, \quad
\left| \sum_{i=1}^3 V_{e N_i} V_{\tau N_i}^* \right| \leq 0.01 \,, \quad
\left| \sum_{i=1}^3 V_{\mu N_i} V_{\tau N_i}^* \right| \leq 0.01 \,.
\label{ec:eps2}
\end{equation}
In the absence of some cancellation among heavy neutrino contributions (which could be possible~\cite{delAguila:2005mf}) this bound implies that a heavy neutrino cannot have sizeable mixings with the electron and muon simultaneously. Notice that the constraints in Eqs.~(\ref{ec:eps2}) can be easily evaded, for example, if each heavy neutrino mixes with a different charged lepton.

For Majorana neutrinos coupling to the electron the experimental
bound on neutrinoless double beta decay also constrains their direct exchange
\cite{Aalseth:2004hb}
\begin{equation}
\left| \, \sum_{i=1}^3 \, V_{e N_i}^2 \, \frac{1}{m_{N_i}} \, \right|
\, <  5 \times 10^{-8} \; {\rm GeV}^{-1} \,.
\label{ec:beta}
\end{equation}
If $V_{e N_j}$ saturate the bound in Eq. (\ref{ec:eps1}),
this limit can be fulfilled either demanding that $m_{N_j}$ are at the TeV scale \cite{Benes:2005hn} and then beyond LHC reach, or that there is a
cancellation among the different terms in Eq.~(\ref{ec:beta}), as it
may happen in definite models \cite{Ingelman:1993ve}, in particular for (quasi)Dirac
neutrinos.


\subsection{Seesaw II}
\label{sec:2.2}

In type II seesaw light neutrinos acquire masses from a gauge-invariant Yukawa interaction of the left-handed lepton doublets with a scalar triplet $\Delta$ of hypercharge $Y=1$ (with $Q=T_3+Y$). 
Writing the triplet in Cartesian components $\vec \Delta = (\Delta^1 ,\, \Delta^2 ,\, \Delta^3)$, the Yukawa interaction reads
\begin{equation}
\mathcal{L}_{\text{Y}} = \frac{1}{\sqrt 2} Y_{ij} \, \overline{\tilde L_{iL}}  \,
(\vec \tau \cdot \vec \Delta) \, L_{jL} +\mathrm{H.c.} \,,
\label{ec:LLyuk}
\end{equation}
with
\begin{equation}
{\tilde L}_{jL} = i \tau_2 \left( \!\begin{array}{c}\nu_{jL}^c \\ l_{jL}^c
\end{array} \!\right)
\end{equation}
and $Y$ a symmetric matrix of Yukawa couplings. We assume without loss of generality that the charged lepton mass matrix is diagonal, and drop primes on the neutrino fields, which are taken in the flavour basis $\nu_e$, $\nu_\mu$, $\nu_\tau$.
The triplet charge eigenstates are related to the Cartesian components by
\begin{equation}
\Dpp = \frac{1}{\sqrt 2} (\Delta^1-i \Delta^2) \,, \quad
\Dp = \Delta^3 \,, \quad
\Delta^0 = \frac{1}{\sqrt 2} (\Delta^1+i \Delta^2) \,.
\end{equation}
When the neutral triplet component acquires a vacuum expectation value (vev) $\langle \Delta^0 \rangle = \vt$, the Yukawa interaction in Eq.~(\ref{ec:LLyuk}) induces a neutrino mass term
\begin{eqnarray}
\mathcal{L}_\mathrm{mass} & = & -Y_{ij}^* \vt \, \bar \nu_{iL} \, \nu_{jR}
+ \mathrm{H.c.} \nonumber \\
& \equiv & -\frac{1}{2} M_{ij} \, \bar \nu_{iL} \, \nu_{jR} + \mathrm{H.c.} \,,
\label{ec:massII}
\end{eqnarray}
where we have again introduced the notation $\nu_{iR} \equiv \nu_{iL}^{c}$, and
\begin{equation}
M_{ij} = 2 Y_{ij}^* \vt
\label{ec:Dnumass}
\end{equation}
are the matrix elements of the light neutrino Majorana mass matrix. These equations imply the relation
\begin{equation}
Y_{ij} = \frac{1}{2\vt} (V_\text{MNS}^* M^\text{diag} V_\text{MNS}^\dagger)_{ij}
\label{ec:YMNS}
\end{equation}
which allows to determine the triplet Yukawa couplings from the diagonal neutrino mass matrix $M^\text{diag}$, the triplet vev and the Maki-Nakagawa-Sakata mixing matrix \cite{Pontecorvo:1957cp,Maki:1962mu} $V_\text{MNS}$.

The triplet Yukawa interaction in Eq.~(\ref{ec:LLyuk}) also generates triplet couplings to the charged leptons. The relevant terms for our analysis are
\begin{eqnarray}
\mathcal{L}_{\Delta ll} & = &  Y_{ij} \, \overline{l_{iL}^c} \, l_{jL} \, \Dpp
+ Y_{ij}^* \, \overline{l_{jL}} \, l_{iL}^c \, \Dmm \,, \notag \\[1mm]
\mathcal{L}_{\Delta l\nu} & = & \sqrt 2 Y_{ij} \, \overline{\nu_{iR}} \, l_{jL} \, \Dp
+ \sqrt 2 Y_{ij} \, \overline{l_{jL}} \, \nu_{iR} \, \Dm \,,
\end{eqnarray}
where $\Dmm = (\Dpp)^\dagger$ and $\Dm = (\Dp)^\dagger$, as usual.

The gauge interactions of the triplet components are obtained from the kinetic term
\begin{equation}
\mathcal{L}_{\text{K}} = (D^\mu \vec \Delta)^\dagger \cdot (D_\mu \vec \Delta) \,,
\label{ec:Dkin}
\end{equation}
where the covariant derivative is
\begin{equation}
D_\mu = \partial_\mu + ig \vec T \cdot \vec W_\mu + ig' Y B_\mu
\end{equation}
with
\begin{equation}
T_1 = \left( \! \begin{array}{ccc} 0 & 0 & 0 \\ 0 & 0 & -i \\ 0 & i & 0 \end{array} \! \right) \,, \quad
T_2 = \left( \! \begin{array}{ccc} 0 & 0 & i \\ 0 & 0 & 0 \\ -i & 0 & 0 \end{array} \! \right) \,, \quad
T_3 = \left( \! \begin{array}{ccc} 0 & -i & 0 \\ i & 0 & 0 \\ 0 & 0 & 0 \end{array} \! \right) \,,
\label{ec:T123}
\end{equation}
and $\vec W_\mu = (W_\mu^1 ,\, W_\mu^2 ,\, W_\mu^3)$, the $\text{SU}(2)_L$ gauge fields, with
\begin{equation}
W^+_\mu = \frac{1}{\sqrt 2} (W^1_\mu -i W^2_\mu) \,, \quad
W^-_\mu = \frac{1}{\sqrt 2} (W^1_\mu +i W^2_\mu) \,.
\end{equation}
The gauge interactions mediating scalar triplet pair production processes in our analysis are
\begin{eqnarray}
\mathcal{L}_W & = & -ig \left[ (\partial^\mu \Dmm) \Dp - \Dmm (\partial^\mu \Dp)
\right]  W_\mu^+ \,, \nonumber \\[1mm]
& & -ig  \left[ (\partial^\mu \Dm) \Dpp - \Dm (\partial^\mu \Dpp) \right] W_\mu^- \,,\nonumber \\[1mm]
\mathcal{L}_Z & = & \frac{ig}{c_W} (1-2s_W^2) \left[ (\partial^\mu \Dmm) \Dpp - \Dmm (\partial^\mu \Dpp)
\right] Z_\mu  \nonumber \\
& & -\frac{ig}{c_W} s_W^2 \left[ (\partial^\mu \Dm) \Dp - \Dm (\partial^\mu \Dp) \right] Z_\mu \,,
\nonumber \\[1mm]
\mathcal{L}_\gamma & = & i2e \left[ (\partial^\mu \Dmm) \Dpp - \Dmm (\partial^\mu \Dpp)
\right] A_\mu  \nonumber \\[1mm]
& & + ie \left[ (\partial^\mu \Dm) \Dp - \Dm (\partial^\mu \Dp) \right] A_\mu \,.
\end{eqnarray}

Constraints on the triplet parameters are much less important than for heavy neutrino singlets, because the new scalars can be produced at LHC by unsuppressed gauge interactions. Electroweak precision data set an upper limit on the triplet vev $\vt$. The most recent bound obtained from a global fit is~\cite{delAguila:2008ks}
\begin{equation}
\vt < 2~\text{GeV} \,,
\end{equation}
which is much less stringent than the one derived from neutrino masses, Eq.~(\ref{ec:Dnumass}), if $Y_{ij}$ are of the order of the charged lepton Yukawa couplings. The relative values of $\vt$ and $Y_{ij}$, whose product is fixed by Eq.~(\ref{ec:Dnumass}), determine the decays of the scalars (a detailed discussion can be found in section~\ref{sec:5}). Limits on $Y_{ij}$ arise from four-fermion processes like $\mu^- \to e^+ e^- e^-$, $\tau \to 3\ell$, as well as LFV processes as $\mu \to e \gamma$
\cite{Abada:2007ux,Akeroyd:2006bb}. Constraints on products of two $Y_{ij}$ are of the order $10^{-2}$ or larger \cite{Abada:2007ux}, much weaker than the expected size for these couplings, except
for the product
\begin{eqnarray}
|Y_{ee} Y_{e\mu}^* | & < & 2.4 \times 10^{-5} \,,
\end{eqnarray}
which is obtained from $\mu^- \to e^+ e^- e^-$. These constraints are automatically satisfied if $Y_{ij}$ are of order $10^{-3}$, which is consistent with our assumption that the dilepton modes dominate the charged scalar decays (see section~\ref{sec:5}).


\subsection{Seesaw III}
\label{sec:2.3}

In type III seesaw the SM is usually enlarged with three leptonic triplets $\Sigma_j$, each composed by three Weyl spinors of zero hypercharge. We choose the spinors to be
right-handed under Lorentz transformations, following the notation in Refs.~\cite{Abada:2007ux,Abada:2008ea} to some extent.
Writing the triplets in Cartesian components $\vec \Sigma_j = (\Sigma_j^1 ,\, \Sigma_j^2 ,\, \Sigma_j^3)$ and using standard four-component notation, the triplet Yukawa interaction with the lepton doublets takes the form
\begin{equation}
\mathcal{L}_{\text{Y}} = -Y_{ij} \, \bar L'_{iL} (\vec \Sigma_j \cdot \vec \tau)
\, \tilde \phi +\text{H.c.} \,,
\label{ec:Tyuk}
\end{equation}
with $Y$ a $3\times 3$ matrix of Yukawa couplings.
The triplet Majorana mass term mediating the seesaw is
\begin{equation}
\mathcal{L}_\text{M} = -\frac{1}{2} \,  M_{ij} \, \overline{\vec \Sigma_i^c} \cdot \vec \Sigma_j
+\text{H.c.} \,,
\label{ec:TMaj}
\end{equation}
with $M$ a $3\times 3$ symmetric matrix. Notice that all the members 
$\Sigma_j^1$, $\Sigma_j^2$, $\Sigma_j^3$ of the triplet $\Sigma_j$ have the same mass term. For each triplet $\Sigma_j$, the charge eigenstates are related to the Cartesian components by
\begin{equation}
\Sigma_j^+ = \frac{1}{\sqrt 2} (\Sigma_j^1-i \Sigma_j^2) \,, \quad
\Sigma_j^0 = \Sigma_j^3 \,, \quad
\Sigma_j^- = \frac{1}{\sqrt 2} (\Sigma_j^1+i \Sigma_j^2) \,.
\label{ec:T:chdef}
\end{equation}
The physical particles are charged Dirac fermions $E'_j$ and neutral Majorana fermions $N'_j$ (as before, we use primes for the weak interaction eigenstates),
\begin{equation}
E'_j = \Sigma_j^- + \Sigma_j^{+c} \,,\quad N'_j = \Sigma_j^0 + \Sigma_j^{0c} \,.
\label{ec:T:EN}
\end{equation}
Then, for our choice of right-handed chirality for the triplets we have
\begin{equation}
E'_{jL} = \Sigma_j^{+c} \,,\quad E'_{jR} = \Sigma_j^- \,,\quad
N'_{jL} = \Sigma_j^{0c} \,,\quad N'_{jR} = \Sigma_j^{0} \,.
\end{equation}
After spontaneous symmetry breaking the terms in Eqs.~(\ref{ec:Tyuk}), (\ref{ec:TMaj}) lead to the neutrino mass matrix
\begin{eqnarray}
\mathcal{L}_{\nu,\mathrm{mass}} & = & - \frac{1}{2} \,
\left(\bar \nu'_L \; \bar N'_L \right)
\left( \! \begin{array}{cc}
0 & \frac{v}{\sqrt 2} Y \\ \frac{v}{\sqrt 2} Y^T & M
\end{array} \! \right) \,
\left( \!\! \begin{array}{c} \nu'_R \\ N'_R \end{array} \!\! \right)
\; + \mathrm{H.c.} \,,
\label{massIII}
\end{eqnarray}
similar to the one for type-I seesaw in Eq.~(\ref{ec:massI}). The mass matrix for charged leptons, also including the $3\times 3$ SM Yukawa matrix $Y^l$, reads
\begin{eqnarray}
\mathcal{L}_{l,\mathrm{mass}} & = & -
\left(\bar l'_L \; \bar E'_L \right)
\left( \! \begin{array}{cc}
\frac{v}{\sqrt 2} Y^l & v Y \\ 0 & M
\end{array} \! \right) \,
\left( \!\! \begin{array}{c} l'_R \\ E'_R \end{array} \!\! \right)
\; + \mathrm{H.c.}
\label{massIIIl}
\end{eqnarray}
The gauge interactions of the new triplets can be obtained from the kinetic term
\begin{equation}
\mathcal{L}_{\text{K}} = i \, \overline{\vec \Sigma_j} \cdot \gamma^\mu D_\mu \, \vec \Sigma_j \,,
\label{ec:Tkin}
\end{equation}
where a sum over $j=1,2,3$ is understood. The covariant derivative is
\begin{equation}
D_\mu = \partial_\mu + ig \vec T \cdot \vec W_\mu \,,
\end{equation}
with the $T$ matrices defined in Eqs.~(\ref{ec:T123}).
The $B_\mu$ term is absent because the triplets have zero hypercharge.
With the definitions in Eqs.~(\ref{ec:T:chdef}) and (\ref{ec:T:EN}), the gauge interactions in the weak eigenstate basis are
\begin{eqnarray}
\mathcal{L}_W & = & - g \left( \bar E'_j \gamma^\mu N'_j \, W_\mu^-
+ \bar N'_j \gamma^\mu E'_j \, W_\mu^+ \right) \,, \notag \\
\mathcal{L}_Z & = & g c_W \, \bar E'_j \gamma^\mu E'_j \, Z_\mu \,, \notag \\
\mathcal{L}_\gamma & = & e \, \bar E'_j \gamma^\mu E'_j \, A_\mu \,,
\label{ec:Tgauge}
\end{eqnarray}
In the mass eigenstate basis, the interactions with the photon are obviously the same. For the $W$ and $Z$ bosons they can be obtained
after diagonalisation of the charged lepton and neutrino mass matrices.
Rewriting Eqs.~(\ref{ec:Tgauge}) and the SM interactions in terms of the mass eigenstates
$E$, $N$ (dropping the subindices) and SM leptons $l$, $\nu_l$, the relevant terms are
\begin{align}
\mathcal{L}_W & = - g \left( \bar E \gamma^\mu N \, W_\mu^-
   + \bar N \gamma^\mu E \, W_\mu^+ \right) \notag \\
&  - \frac{g}{\sqrt 2} \left( V_{l N} \, \bar l \gamma^\mu  P_L N \; W_\mu^- 
   + V_{l N}^* \, \bar N \gamma^\mu  P_L l \; W_\mu^+ \right) \notag \\
&  - g \left( V_{lN} \, \bar E \gamma^\mu  P_R \nu_l \; W_\mu^- 
    + V_{lN}^* \, \bar \nu_l \gamma^\mu  P_R E \; W_\mu^+ \right) \,, \notag
\displaybreak \\[1mm]
\mathcal{L}_Z & = g c_W \, \bar E \gamma^\mu E \, Z_\mu \notag \\
&  + \frac{g}{2c_W} \left( V_{lN} \, \bar \nu_l \gamma^\mu P_L N
+ V_{lN}^* \, \bar N \gamma^\mu P_L \nu_l \right) Z_\mu \notag \\
&  + \frac{g}{\sqrt 2 c_W} \left( V_{lN} \, \bar l \gamma^\mu P_L E
+ V_{lN}^* \, \bar E \gamma^\mu P_L l \right) Z_\mu \,, \notag \\[1mm]
\mathcal{L}_\gamma & = e \, \bar E \gamma^\mu E \, A_\mu \,,
\label{ec:Tgauge2}
\end{align}
at first order in the small mixing
\begin{equation}
V_{lN} \simeq - \frac{Y_{lN} v}{\sqrt 2 m_N} \,.
\end{equation}
Notice that the couplings of the heavy neutrino $N$ to the SM leptons are the same as in seesaw I except for a sign change in the neutral current term. For the heavy charged lepton they are similar but a factor $\sqrt 2$ larger, and the charged current coupling has opposite chirality. The interactions with the SM Higgs can be obtained from the Yukawa term in Eq.~(\ref{ec:Tyuk}),
\begin{eqnarray}
\mathcal{L}_H & = & - \frac {1}{\sqrt 2} \left( \bar \nu'_{iL} Y_{ij} N'_{jR}
+ \bar N'_{jR} Y^\dagger_{ji} \nu'_{iL} \right) \, H \nonumber \\
& & - \left( \bar l'_{iL} Y_{ij} E'_{jR}
+ \bar E'_{jR} Y^\dagger_{ji} l'_{iL} \right) \, H \,.
\label{ec:Thiggs1}
\end{eqnarray}
Rewriting the weak eigenstates as a function of the mass eigenstates $E$, $N$, $l$ and $\nu_l$ we find the interactions
\begin{eqnarray}
\mathcal{L}_H & = &  \frac{g \, m_\Sigma}{2 M_W} \left( V_{lN} \, \bar \nu_l  P_R N
+ V_{lN}^* \, \bar N  P_L \nu_l \right) H \nonumber \\
& & +\frac{g \, m_\Sigma}{\sqrt 2 M_W} \left( V_{lN} \, \bar l  P_R E
+ V_{lN}^* \, \bar E  P_L l \right) H \,,
\end{eqnarray}
where $m_\Sigma=m_E = m_N$ is the common triplet mass.
Finally, using the fact that $N$ and $\nu_l$ are Majorana fermions their neutral and scalar interactions can be rewritten,
\begin{eqnarray}
\mathcal{L}_{Z\nu N} & = & \frac{g}{2 c_W} \, \bar \nu_l \gamma^\mu \left(  
V_{l N} P_L - V_{l N}^* P_R \right) N \; Z_\mu \,, \nonumber \\[1mm]
\mathcal{L}_{H \nu N} & = & \frac{g \, m_\Sigma}{2 M_W} \, \bar \nu_l \left( V_{l N} P_R
+ V_{l N}^* P_L \right) N \; H \,.
\end{eqnarray}

As in the case of heavy neutrino singlets, limits on the mixing of new fermion triplets arise from electroweak precision data. The most recent constraints are~\cite{delAguila:2008pw}
\begin{equation}
\sum_{i=1}^3 |V_{e N_i}|^2 \leq 0.00036 \,, \quad
\sum_{i=1}^3 |V_{\mu N_i}|^2 \leq 0.00029 \,, \quad
\sum_{i=1}^3 |V_{\tau N_i}|^2 \leq 0.00073
\label{ec:Tlim1}
\end{equation}
at 90\% CL.
Additional limits result from the non-observation of LFV processes like $\mu \to e \gamma$, $\tau \to e \gamma$ and $\tau \to \mu \gamma$. A global fit
allows to obtain the 90\% CL bounds~\cite{Abada:2007ux,Abada:2008ea}
\begin{equation}
\left| \sum_{i=1}^3 V_{e N_i} V_{\mu N_i}^*    \right| \leq 1.1 \times 10^{-6} \,, \quad
\left| \sum_{i=1}^3 V_{e N_i} V_{\tau N_i}^*   \right| \leq 0.0012 \,, \quad
\left| \sum_{i=1}^3 V_{\mu N_i} V_{\tau N_i}^* \right| \leq 0.0012 \,,
\end{equation}
although the limits in Eqs.~(\ref{ec:Tlim1}) and the Schwarz inequality imply stronger constraints by a factor of two for the products of the mixing with the tau lepton and the electron or muon, 
\begin{equation}
\left| \sum_{i=1}^3 |V_{e N_i} V_{\tau N_i}^* \right|\leq 0.0005 \,, \quad
\left| \sum_{i=1}^3 |V_{\mu N_i} V_{\tau N_i}^* \right| \leq 0.0005  \,.
\end{equation}
These limits are not relevant for heavy triplet production at LHC, which takes place through gauge interactions of order unity. For the decay, the bounds on the mixing are six orders of magnitude above the critical values $V_{lN} \sim O(10^{-8})$ for which the gauge boson decay modes begin to be suppressed with respect to other decays.

\section{General features of the analyses}
\label{sec:3}

Our estimations of the discovery potential for seesaw messengers are performed by simulating their production at LHC, together with the SM processes which may constitute a background. The signals and backgrounds are calculated with matrix element generators, which produce event samples which are feeded into {\tt Pythia 6.4}~\cite{Sjostrand:2006za} to add initial and final state radiation (ISR, FSR) and pile-up, and perform hadronisation for each event. After this, a fast detector simulation program \cite{atlfast} is used, whose purpose is to simulate how the event would be seen in the ATLAS detector. The tagging of $b$ and $\tau$ jets is performed with the {\tt ATLFASTB} package, selecting efficiencies of 60\% and 50\%, respectively, and the corresponding mistag rates: for $b$ tagging, it is about
1\% for light jets and 15\% for $c$ jets, and for $\tau$ tagging it is about 1\%
for non-tau jets.
The analysis of the events is then performed in terms of charged leptons, jets and missing energy.

The heavy neutrino singlet signals, described further in section~\ref{sec:4}, are generated with the {\tt Alpgen} extension of Ref.~\cite{delAguila:2007em}. For scalar and fermion triplet production a new generator \tria\ has been developed. 
For seesaw II it calculates $\Dpp \Dmm$, $\Dppmm \Dmp$ and $\Dp \Dm$ production with subsequent leptonic decays $\Dppmm \to l_i^\pm l_j^\pm$, $\Dpm \to l_i^\pm \nu$ (see section~\ref{sec:5}), where $l_i=e,\mu,\tau$. (We will use $l$ when referring to electrons, muons and tau leptons, reserving $\ell$ for electrons and muons only.)
For seesaw III it calculates $E^+ E^-$ and $E^\pm N$ production with all decays to light leptons and gauge or Higgs bosons, as described in detail in section~\ref{sec:6}. This is a cumbersome task since it involves 289 different final states with 128 different matrix elements for $E^+ E^-$ and 748 final states with 72 matrix elements for $E^\pm N$, which have to be generated with their corresponding weights.
Matrix elements for the $2 \to n$ processes ($n=4$ for scalar triplet and $n=6$ for fermion triplet production) are calculated using {\tt HELAS} \cite{helas}, including all spin and finite width effects. Integration in phase space is performed with {\tt VEGAS} \cite{vegas}.
The output of the program, in the form of unweighted events, includes the colour structure necessary to interface it with {\tt Pythia}.

An equally important ingredient for a correct estimation of the discovery potential of a signal is the background evaluation. Recent parton-level calculations~\cite{Han:2006ip,Perez:2008zc,Perez:2008ha,Franceschini:2008pz} underestimate SM backgrounds because they cannot account for several physical effects present in real experiments. One such effect is the appearance of extra jets in the final state from ISR, FSR and especially pile-up. In this way,
processes with a small number of partons (and typically larger cross sections) give important contributions to final states with a larger number of jets, due to pile-up. An illustrative example will be found in section~\ref{sec:6.9}: $4/5$ of the total contribution of $Wnj$ production (where $nj$ stands for $n$ jets at the partonic level) to a final state with four hard jets is due to multiplicities $n=0,1,2$ at the partonic level. Another physical effect which cannot be included in parton-level calculations is the production of isolated charged leptons from $b$ quark decays. As we found in previous work~\cite{delAguila:2007em}, $t \bar t nj$ and $b \bar b nj$ are the main SM backgrounds giving two like-sign leptons $\ell^\pm \ell^\pm$ ($\ell=e,\mu$). In the former case, this happens when the $t \bar t$ pair decays semileptonically, and one $b$ quark gives the second lepton, while in the latter both charged leptons result from $b$ decays. Likewise, three leptons $\ell^\pm \ell^\pm \ell^\mp$ can be produced in the dilepton decay of a $t \bar t$ pair, the third lepton coming from a $b$ quark. Charged leptons from $b$ quark decays typically have small transverse momentum, but the cross section for $t \bar t nj$ production is three orders of magnitude larger than for the signals considered and even larger for $b \bar b nj$, $1.4~\mu\text{b}$. Besides, signals sometimes have leptons which are not very energetic, {\em e.g.} when they originate from $\tau$ decays. In any case, suppressing these backgrounds without eliminating important signal contributions is not easy. A third effect which makes a parton level calculation insufficient is that sometimes a charged lepton is missed by the detector, for example because it is located inside a hadronic jet and thus it is not isolated. This is important, for instance, for like-sign dilepton signals, where one of the main backgrounds is $WZnj$ production when one of the leptons from the $Z$ decay is missed by the detector.

In order to have predictions for SM backgrounds as accurate as possible we use {\tt Alpgen}~\cite{Mangano:2002ea} to generate hard events
\begin{table}[t]
\begin{center}
\begin{small}
\begin{tabular}{llcc}
Process & Decay & $L$ & Events \\
\hline
$t \bar t nj$, $n=0,\dots,6$   & semileptonic           & 300 fb$^{-1}$   & 60.8 M \\
$t \bar t nj$, $n=0,\dots,6$   & dileptonic             & 300 fb$^{-1}$   & 15.2 M \\
$b \bar b nj$, $n=0,\dots,3$   & all                    & 0.075 fb$^{-1}$ & 116 M  \\
$c \bar c nj$, $n=0,\dots,3$   & all                    & 0.075 fb$^{-1}$ & 145 M  \\
$tj$                          & $W \to l \nu$       & 300 fb$^{-1}$   & 9.5 M  \\
$t\bar b$                     & $W \to l \nu$       & 300 fb$^{-1}$   & 540 K  \\
$tW$                          & all                    & 300 fb$^{-1}$   & 16 M   \\
$t \bar t t \bar t$           & all                    & 300 fb$^{-1}$   & 1.6 K   \\
$t \bar t b \bar b$           & all                    & 300 fb$^{-1}$   & 340 K   \\
$Wnj$, $n=0,1,2$              & $W \to l \nu$       & 10 fb$^{-1}$    & 557.4 M \\
$Wnj$, $n=3,\dots,6$           & $W \to l \nu$       & 30 fb$^{-1}$    & 10 M \\
$W b \bar b nj$, $n=0,\dots,4$ & $W \to l \nu$       & 300 fb$^{-1}$   & 5.2 M \\
$W c \bar c nj$, $n=0,\dots,4$ & $W \to l \nu$       & 300 fb$^{-1}$   & 5.5 M \\
$W t \bar t nj$, $n=0,\dots,4$ & $W \to l \nu$       & 300 fb$^{-1}$   & 50.6 K \\
$Z/\gamma\, nj$, $n=0,1,2$, $m_{ll} < 120$ GeV
                              & $Z \to l^+ l^-$  & 10 fb$^{-1}$    & 54.9 M \\
$Z/\gamma\, nj$, $n=3,\dots,6$, $m_{ll} < 120$ GeV
                              & $Z \to l^+ l^-$  & 30 fb$^{-1}$    & 1.1 M \\
$Z/\gamma\, nj$, $n=0,\dots,6$, $m_{ll} > 120$ GeV
                              & $Z \to l^+ l^-$  & 300 fb$^{-1}$   & 17.3 M \\
$Z b \bar b nj$, $n=0,\dots,4$ & $Z \to l^+ l^-$  & 300 fb$^{-1}$   & 2 M \\
$Z c \bar c nj$, $n=0,\dots,4$ & $Z \to l^+ l^-$  & 300 fb$^{-1}$   & 1.8 M \\
$Z t \bar t nj$, $n=0,\dots,4$ & $Z \to l^+ l^-$  & 300 fb$^{-1}$   & 18.7 K \\
$WWnj$, $n=0,\dots,3$          & $W \to l \nu$       & 300 fb$^{-1}$   & 2.9 M \\
$WZnj$, $n=0,\dots,3$          & $W \to l \nu$, $Z \to l^+ l^-$
                                                       & 300 fb$^{-1}$   & 377 K \\
$ZZnj$,  $n=0,\dots,3$          & $Z \to l^+ l^-$  & 300 fb$^{-1}$   & 37.4 K \\
$WWWnj$, $n=0,\dots,3$         & $2W \to l \nu$      & 300 fb$^{-1}$   & 14.7 K \\
$WWZnj$, $n=0,\dots,3$         & all                    & 300 fb$^{-1}$   & 48.7 K \\
$WZZnj$, $n=0,\dots,3$         & all                    & 300 fb$^{-1}$   & 15.3 K \\
$ZZZnj$, $n=0,\dots,3$         & $2Z \to l^+ l^-$ & 300 fb$^{-1}$   & 114 \\
\end{tabular}
\end{small}
\caption{Background processes considered in the simulations. The second column indicates the decay modes included (where $l=e,\mu,\tau$), and the third column the luminosity equivalent generated. The last column corresponds to the number of events after matching, with K and M standing for $10^3$ and $10^6$ events, respectively.}
\label{tab:bkg}
\end{center}
\end{table}
which are interfaced to {\tt Pythia}
using the MLM prescription \cite{mlm} to perform a matching between the ``soft'' radiation generated by {\tt Pythia} and the ``hard'' jets generated by {\tt Alpgen} avoiding double counting. Background samples are generated with large statistics, often 300 fb$^{-1}$, in order to avoid fluctuations in the final selected samples. This is a demanding computational task, which would take around ten years in a modern single processor system. The backgrounds generated and the corresponding luminosities are collected in Table~\ref{tab:bkg}. For $b \bar b nj$ and $c \bar c nj$ the statistics is small but the samples are still very large, allowing to estimate correctly these backgrounds as described in detail in appendix A of Ref.~\cite{delAguila:2007em}.
Signals are all generated with a statistics of 3000 fb$^{-1}$, and all results are rescaled to 30 fb$^{-1}$ in the Tables.

The procedure used for estimating the statistical significance of a signal is considered case by case. In the absence of any systematic uncertainty on the background, the statistical significance would be $\mathcal{S}_0 \equiv S/\sqrt B$, where $S$ and $B$ are the number of signal and background events, or its analogous from the $P$-number for small backgrounds where Poisson statistics must be applied. Nevertheless, there are systematic uncertainties in the background evaluation from several sources: the theoretical calculation, parton distribution functions (PDFs), the collider luminosity, pile-up, ISR and FSR, etc. as well as some specific uncertainties related to the detector like the energy scale and $b$ tagging efficiency. Provided that the signal manifests as a clear peak in a distribution (as in doubly charged scalar production) or a long tail, it will be possible to normalise the background directly from data, and extract the peak significance. In this case, we give estimators of the sensitivity considering two hypotheses for the background normalisation:
\begin{itemize}
\item[(a)] the SM background normalisation does not have any uncertainty;
\item[(b)] the SM background is normalised directly from data.
\end{itemize}
In some of the cases examined the signal has a wide distribution and the SM background cannot be normalised from data. In such situations we include a 20\% background uncertainty in the significance summed in quadrature, using as estimator $\mathcal{S}_{20} \equiv S/\sqrt{B+(0.2 B)^2}$.

\section{Seesaw I signals} 
\label{sec:4}

\begin{figure}[ht]
\begin{center}
\epsfig{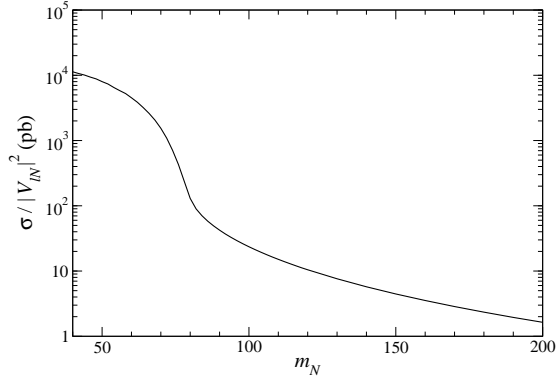}
\caption{Cross section for production of heavy neutrino singlets
$q \bar q \to l^\pm N$ at LHC.}
\label{fig:cross-N}
\end{center}
\end{figure}

For our study we consider the single heavy neutrino production process
\begin{equation}
q \bar q' \to W^{*} \to l^\pm N \,,
\end{equation}
with $l=e,\mu,\tau$.
Its cross section depends on $m_N$ as well as on the small mixing $V_{l N}$, to which the amplitude is proportional. Its dependence on $m_N$ is plotted in Fig.~\ref{fig:cross-N}, normalised to $|V_{l N}|^2 = 1$.\footnote{The total cross section plotted in Fig.~2 of Ref.~\cite{delAguila:2007em} is underestimated by a factor of two, while the cross sections with $N$ decay and with kinematical cuts, as well as the numbers of events included in all Tables, are correct. We thank B. Gavela for bringing this incorrect normalisation into our attention.}
Heavy Majorana neutrino singlets decay to SM leptons plus a gauge or Higgs boson, with partial widths
\begin{eqnarray}
\Gamma(N \to l^- W^+) & = &  \Gamma(N \to l^+ W^-) \nonumber \\
& = & \frac{g^2}{64 \pi} |V_{l N}|^2
\frac{m_N^3}{M_W^2} \left( 1- \frac{M_W^2}{m_N^2} \right) 
\left( 1 + \frac{M_W^2}{m_N^2} - 2 \frac{M_W^4}{m_N^4} \right) \,, \nonumber
\\[0.1cm]
\Gamma(N \to \nu_l Z) & = &  \frac{g^2}{64 \pi c_W^2} |V_{lN}|^2
\frac{m_N^3}{M_Z^2} \left( 1- \frac{M_Z^2}{m_N^2} \right) 
\left( 1 + \frac{M_Z^2}{m_N^2} - 2 \frac{M_Z^4}{m_N^4} \right) \,, \nonumber
\\[0.2cm]
\Gamma(N \to \nu_l H) & = &  \frac{g^2}{64 \pi} |V_{lN}|^2
\frac{m_N^3}{M_W^2} \left( 1- \frac{M_H^2}{m_N^2} \right)^2 \,.
\label{ec:Nwidths}
\end{eqnarray}
For $m_N < M_W$ these two body decays are not possible and $N$ decays into three fermions, mediated by off-shell bosons.
Within any of these four decay modes, the branching fractions for
individual final states $l = e,\mu,\tau$
are in the ratios $|V_{eN}|^2 \,:\, |V_{\mu N}|^2 \,:\, |V_{\tau N}|^2$.
However, as it can be clearly seen from Eqs.~(\ref{ec:Nwidths}),
the total branching ratio for each of the four channels above
(summing over $l$)
is independent of the heavy neutrino mixing and determined only by $m_N$
and the gauge and Higgs boson masses.

Heavy neutrino production cross sections are suppressed by the small mixings
$|V_{eN}|^2 \leq 0.0030$, $|V_{\mu N}|^2 \leq 0.0032$, $|V_{\tau N}|^2 \leq 0.0062$ \cite{delAguila:2008pw}, and then the observability of $lN$ production is limited to masses up to 150 GeV approximately due to the large backgrounds \cite{delAguila:2007em}.\footnote{The mass reach is much larger at $e^+ e^-$
\cite{delAguila:2005pf,del Aguila:2005mf} and $e\gamma$ \cite{Bray:2005wv} colliders, whose environment is cleaner (see Ref.~\cite{del Aguila:2006dx} for a review).}
In this situation, heavy neutrino decay products are not very energetic and SM backgrounds are
important. Among the possible final states given by Eqs.~(\ref{ec:Nwidths}), only charged current decays give final states which may be observable in principle. Likewise, other single production processes like
\begin{align}
& q \bar q \to Z^{*} \to \nu N \,, \notag \\
& gg \to H^{*} \to \nu N
\end{align}
give $\ell^\pm$ and $\ell^+ \ell^-$ final states which are unobservable due to the huge backgrounds. Pair production
\begin{equation}
q \bar q \to Z^* \to N N 
\end{equation}
has its cross section suppressed by $|V_{lN}|^4$, phase space and the $Z$ propagator, and is thus negligible.

In a previous work~\cite{delAguila:2007em} we have studied in great detail the observability of heavy neutrino singlets in the like-sign dilepton final state for $m_N > M_W$ as well as for $m_N < M_W$, performing sophisticated likelihood analyses to effectively suppress the backgrounds. We found that a heavy neutrino coupling only to the electron with $|V_{eN}|^2 = 0.0054$ could be discovered up to $m_N = 145$ GeV, and if it couples to the muon with $|V_{\mu N}|^2 = 0.0096$ it could be discovered up to 200 GeV. (If it couples only to the tau the signals are swamped by the SM background.) For heavy neutrinos lighter than the $W$ boson, we found that, for example, a 60 GeV neutrino coupling to the muon might be discovered up to $|V_{\mu N}|^2 = 4.9 \times 10^{-5}$. These limits, however, are obtained from very optimised analyses which use as input the heavy neutrino mass
to build the probability distributions for the heavy neutrino signal.
In this section we will take the opposite approach, following the philosophy of this paper: we will investigate whether with ``generic'' model-independent cuts the heavy neutrino (as well as seesaw II and seesaw III signals) might be observable. Of course, dedicated experimental searches can be carried out assuming some value for $m_N$ and optimising the kinematical distributions for this mass to achieve the best sensitivity. But, at least in a first step, LHC searches are likely to be performed with general and model-independent event selections.

A major difference between heavy neutrino signals studied in this section and scalar triplet and fermion triplet signals concerns lepton flavour. For the latter, the SM backgrounds involving electrons and muons are alike at large transverse momenta, and it makes sense to perform ``flavour blind'' searches summing electrons and muons. This is also sensible from 
the point of view of the signals, which have the same cross sections if the new states couple to the electron, the muon or both, as it will be argued in sections \ref{sec:5} and \ref{sec:6}. On the other hand, for heavy neutrino production the situation is clearly different.
At low transverse momenta SM backgrounds involving electrons are much larger than those involving muons, as shown in Ref.~\cite{delAguila:2007em}, and searches must be performed independently in order to avoid that a possible signal in muon final states is hidden by electron backgrounds. Moreover, heavy neutrino signals are different if the heavy neutrino couples to the electron, the muon or both:
if it couples to the electron $e^\pm N$ production will take place, if it couples to the muon we will have $\mu^\pm N$ production, and if it couples to both we will have the two processes simultaneously. Therefore, for heavy neutrino searches it is convenient to divide final states by lepton flavour.

In what follows we assume two scenarios: (i) a heavy neutrino coupling only to the electron 
with $|V_{eN}|^2 = 0.0030$, labelled as scenario S1; (ii) coupling only to the muon with 
$|V_{\mu N}|^2 = 0.0032$, labelled as scenario S2.
We will take a mass $m_N = 100$ GeV, between the two cases previously studied. For such a heavy neutrino mass the production cross section is large,
and the kinematics of the signal is completely different from the other cases.
The decay branching ratios are $\text{Br}(N \to l^- W^+) = \text{Br}(N \to l^+ W^-)
= 0.43$, $\text{Br}(N \to \nu Z) = 0.14$.
We will examine final states with (a) three charged leptons $\ell^\pm \ell^\pm \ell^\mp X$; (b) two like-sign dileptons $\ell^\pm \ell^\pm X$, where $X$ denotes possible additional jets and the leptons can have different flavour. Final states with two opposite-sign leptons or only one lepton are unobservable for these small cross sections.


\subsection{Final state $\ell^\pm \ell^\pm \ell^\mp$}
\label{sec:4.1}

Trilepton signals are produced in the two charged current decay channels of the heavy neutrino, with subsequent leptonic decay of the $W$ boson, {\em e.g.}
\begin{align}
& \ell^+ N \to \ell^+ \ell^- W^+ \to \ell^+ \ell^- \ell^+ \bar \nu \,, \nonumber \\
& \ell^+ N \to \ell^+ \ell^+ W^- \to \ell^+ \ell^+ \ell^- \nu \,,
\end{align}
(and small additional contributions from $\tau$ leptonic decays).
This final state is very clean once that $WZnj$ production can be almost eliminated with a simple cut on the invariant mass of opposite charge leptons.

For event pre-selection we require the presence of two like-sign charged leptons $\ell_1$ and $\ell_2$ (ordered by decreasing $p_T$) with transverse momentum larger than 30 GeV, and an additional lepton of opposite charge. The choice of the $p_T$ cut for like-sign leptons is motivated by the need to reduce backgrounds where soft leptons are produced in $b$ decays, for example $t \bar t nj$ in the dilepton channel.
For event selection, in a first step we only require that neither of the two opposite-sign lepton pairs have an invariant mass closer to $M_Z$ than 10 GeV. 
These pre-selection and selection criteria are the same as those applied in the analysis of scalar and fermion triplet signals in the next sections, but in this subsection we split the sample in two disjoint sets: final states with at least two electrons (labelled as ``$2e$'') and with at least two muons  (``$2\mu$'').
The number of events for the signal and the largest backgrounds is given in Table~\ref{tab:nsnb-3lepN} for these two stages of event selection. 

\begin{table}[ht]
\begin{center}
\begin{tabular}{ccccccccc}
      & \multicolumn{2}{c}{Pre-selection} & \quad & \multicolumn{2}{c}{Selection} & \quad & \multicolumn{2}{c}{Impr. selection} \\[1mm]
                & $2e$   & $2\mu$ && $2e$  & $2\mu$ && $2e$ & $2\mu$  \\[1mm]
$N$ (S1)        & 37.1   & 0      && 32.4  & 0      && 28.6 & 0    \\
$N$ (S2)        & 0      & 37.8   && 0     & 33.1   && 0    & 29.6 \\
$t \bar t nj$   & 244.8  & 78.0   && 159.8 & 52.4   && 58.4 & 16.3 \\
$tW$            & 14.8   & 3.0    && 10.5  & 1.7    && 6.5  & 0.6  \\
$W t \bar t nj$ & 25.6   & 19.9   && 20.6  & 14.5   && 3.8  & 2.6  \\
$Z b \bar b nj$ & 17.1   & 16.2   && 1.1   & 0.9    && 0.5  & 0.1  \\
$Z t \bar t nj$ & 82.5   & 69.9   && 10.3  & 6.5    && 2.6  & 1.1  \\
$W Z nj$        & 2166.4 & 1947.3 && 49.2  & 24.3   && 36.8 & 17.8 \\
$Z Z nj$        & 141.0  & 135.0  && 2.8   & 1.4    && 1.6  & 1.2  \\
$WWW nj$        & 10.8   & 12.0   && 7.9   & 8.9    && 4.7  & 5.3 \\
$WWZ nj$        & 23.9   & 18.8   && 1.1   & 0.7    && 0.8  & 0.4 \\
\end{tabular}
\end{center}
\caption{Number of events for $\ell^\pm \ell^\pm \ell^\mp$ signals and main backgrounds with a luminosity of 30 fb$^{-1}$.}
\label{tab:nsnb-3lepN}
\end{table}

\begin{figure}[ht]
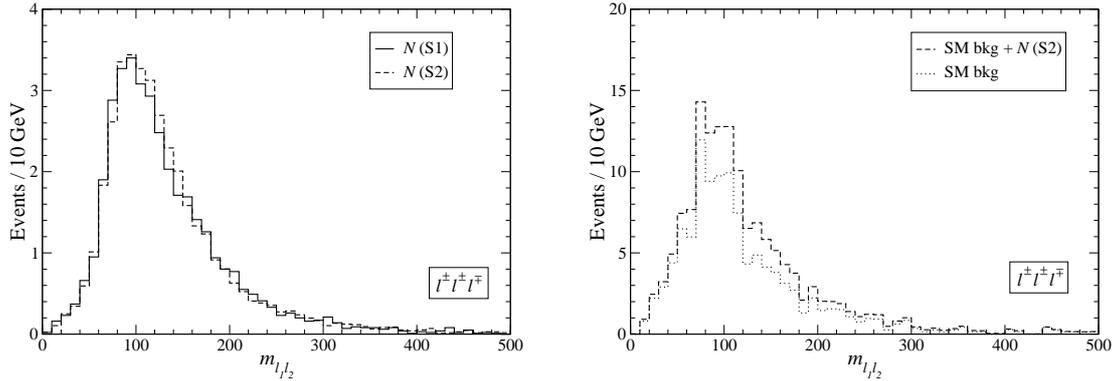

\begin{center}
\begin{tabular}{ccc}
\epsfig{file=Figs/Mll-3lep-N.eps,height=5cm,clip=} & \quad &
\epsfig{file=Figs/Ml1l2-3lep-bkgN2.eps,height=5cm,clip=}
\end{tabular}
\caption{Left: Kinematical distribution at pre-selection of the like-sign dilepton invariant mass for the signals in the two heavy neutrino scenarios. Right: the same, for the SM and the SM plus the signal in scenario S2 at the selection level. The luminosity is 30 fb$^{-1}$.}
\label{fig:3lepN-m12}
\end{center}
\end{figure}

The invariant mass of the like-sign leptons $m_{\ell_1 \ell_2}$ is a good discriminant among different sources of new physics giving $\ell^\pm \ell^\pm \ell^\mp$ signals. In the case 
of heavy neutrino production this distribution, presented in Fig.~\ref{fig:3lepN-m12}, is broad and without a long tail. For larger $N$ masses the $m_{\ell_1 \ell_2}$ tail will be longer, but in this case the cross sections are much smaller. 
In dimuon final states the backgrounds are very small and a signal in scenario S2 might be detected (although with a significance smaller than $5\sigma$) without the need of further improvements in the analysis, provided that the background uncertainties are small. Neglecting them, the excess of events would amount to a statistical significance $\mathcal{S}_0 = 3.1\sigma$, whereas if we consider a 20\% systematic uncertainty in the background the significance is smaller, $\mathcal{S}_{20} = 1.3\sigma$. This excess
is distributed across the $m_{\ell_1 \ell_2}$ range, as it is shown
on the right side of Fig.~\ref{fig:3lepN-m12}, and does not display a peak as it does in scalar triplet production (see next section) nor a long tail as in fermion triplet 
production (see section~\ref{sec:6}). This fact makes it difficult to normalise the background directly from data in a given ``control'' region to extract the significance of an excess in another phase space region, as it will be done in some of the cases analysed in this paper.

Other kinematical distributions, for example the transverse momenta of the like-sign leptons, exhibit analogous behaviour with the event excess distributed in a wide range but without long tails which would be a clear indication of the presence of a new physics signal. This implies that, in order to be detected, heavy neutrino signals require
dedicated analyses, often optimised for a given $m_N$ value, as the one presented in Ref.~\cite{delAguila:2007em}. For this specific heavy neutrino signal there are additional cuts which can be imposed to further reduce backgrounds. For an improved event selection we ask that
\begin{itemize}
\item[(i)] no $b$ jets can be present in the final state;
\item[(ii)] the like-sign leptons must be back-to-back, with their angle in transverse plane larger than $\pi/2$.
\end{itemize}
These selection criteria are convenient for this heavy neutrino singlet signal but rather inadequate for fermion triplet signals in the same final state. The number of events after these additional requirements is given in the last two columns of Table~\ref{tab:nsnb-3lepN}.
The statistical significance does not reach $5\sigma$ in any of the cases:
$\mathrm{S}_{20} = 1.1$ in scenario S1 and $\mathrm{S}_{20} = 2.6$ in S2. The variable selection can still be improved and cuts optimised for this particular value of $m_N$, obtaining $\mathcal{S}_{20} = 1.7$ in scenario S1 and $\mathcal{S}_{20} = 3.7$ in S2 (allowing discovery with 180 fb$^{-1}$), and we expect that much
better results will be obtained with a probabilistic analysis. This is in agreement with our statement that heavy neutrino singlets require dedicated searches, optimised for their detection.


\subsection{Final state $\ell^\pm \ell^\pm$}
\label{sec:4.2}

Heavy neutrino signals in this final state have been widely studied \cite{Datta:1993nm,Almeida:2000pz,Panella:2001wq,Han:2006ip,delAguila:2007em}. They are produced from the LNV neutrino decay and subsequent hadronic $W$ decay, or leptonic decay when the lepton is missed. In this section we investigate whether a search based on simple selection criteria could find such a signal. For event pre-selection we require the presence of two like-sign charged leptons with $p_T > 30$ GeV. Even with this relatively large transverse momentum cut, SM backgrounds are non-negligible, as it has been shown elsewhere~\cite{delAguila:2007em}. The corresponding numbers of events are collected in Table~\ref{tab:nsnb-2likN}. 
We consider independently the $e^\pm e^\pm$ and $\mu^\pm \mu^\pm$ final states for each of the two heavy neutrino scenarios. $e^\pm \mu^\pm$ signals are not generated in any of them (although they are produced if a heavy neutrino simultaneously couples to the electron and muon), and hence they are not considered.

\begin{table}[h]
\begin{center}
\begin{tabular}{cccccc}
      & \multicolumn{2}{c}{Pre-selection} & \quad & \multicolumn{2}{c}{Selection} \\[1mm]
                & $2e$  & $2\mu$ && $2e$ & $2\mu$ \\[1mm]
$N$ (S1)        & 28.1  & 0      && 13.5 & 0    \\
$N$ (S2)        & 0     & 25.6   && 0    & 13.5 \\
$t \bar t nj$   & 620.0 & 8.4    && 36.7 & 0.1  \\
$tW$            & 39.3  & 1.1    && 4.2  & 0.2  \\
$W t \bar t nj$ & 53.7  & 45.1   && 1.1  & 0.7  \\
$W W nj$        & 54.2  & 47.8   && 5.2  & 4.8  \\
$W Z nj$        & 269.9 & 182.6  && 23.7 & 13.8 \\
$WWW nj$        & 21.2  & 22.6   && 1.2  & 1.3  \\
\end{tabular}
\end{center}
\caption{Number of events for the like-sign dilepton signals and main backgrounds for a luminosity of 30 fb$^{-1}$.}
\label{tab:nsnb-2likN}
\end{table}

Despite the larger branching ratio, the number of signal events is smaller
than in the previous trilepton channel because the charged leptons from $N\to \ell W$ decay (with a mass $m_N = 100$ GeV) are not very energetic in general, and the requirement $p_T > 30$ GeV severely reduces the signal.\footnote{In most of the trilepton signal events the like-sign sub-leading lepton results from leptonic $W$ decay, and hence the suppression is smaller.} For larger heavy neutrino masses the efficiency is larger but signal cross sections are smaller.

\begin{figure}[ht]
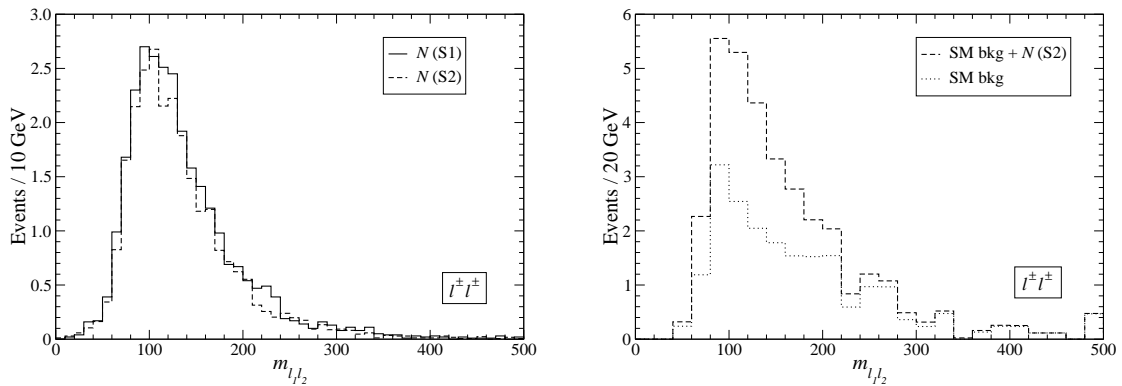

\begin{center}
\begin{tabular}{ccc}
\epsfig{file=Figs/Mll-2lik-N.eps,height=5cm,clip=} & \quad &
\epsfig{file=Figs/Ml1l2-2lik-bkgN2.eps,height=5cm,clip=}
\end{tabular}
\caption{Left: Kinematical distribution at pre-selection of the like-sign dilepton invariant mass for the signals in the two heavy neutrino scenarios. Right: the same, for the SM and the SM plus the signal in scenario S2 at the selection level. The luminosity is 30 fb$^{-1}$.}
\label{fig:2likN-m12}
\end{center}
\end{figure}

For event final selection we also require:
\begin{itemize}
\item[(i)] at least two jets in the final state with $p_T > 20$ GeV, and no $b$-tagged jets;
\item[(ii)] missing energy smaller than 30 GeV;
\item[(iii)] the transverse angle between the two leptons must be larger than $\pi/2$.
\end{itemize}
The number of events passing these cuts is also included in Table~\ref{tab:nsnb-2likN}.
After this simple event selection the heavy neutrino signals in the dielectron channel are not significant, with $\mathcal{S}_0 = 1.5\sigma$, $\mathcal{S}_{20} = 0.7\sigma$ but in the dimuon channel the event excess amounts to $\mathcal{S}_0 = 2.9\sigma$, $\mathcal{S}_{20} = 2.1\sigma$, which would be noticed if the background normalisation is precise enough. The invariant mass distribution of both signals (without background) at pre-selection level is shown in Fig.~\ref{fig:2likN-m12} (left), and for scenario S2 the distribution also including the background is shown in Fig.~\ref{fig:2likN-m12} (right) at selection level. We can observe that again the signal dilepton distribution is very broad but without a long tail, which distinguishes the neutrino singlet from scalar and fermion triplet production.

\begin{figure}[ht]
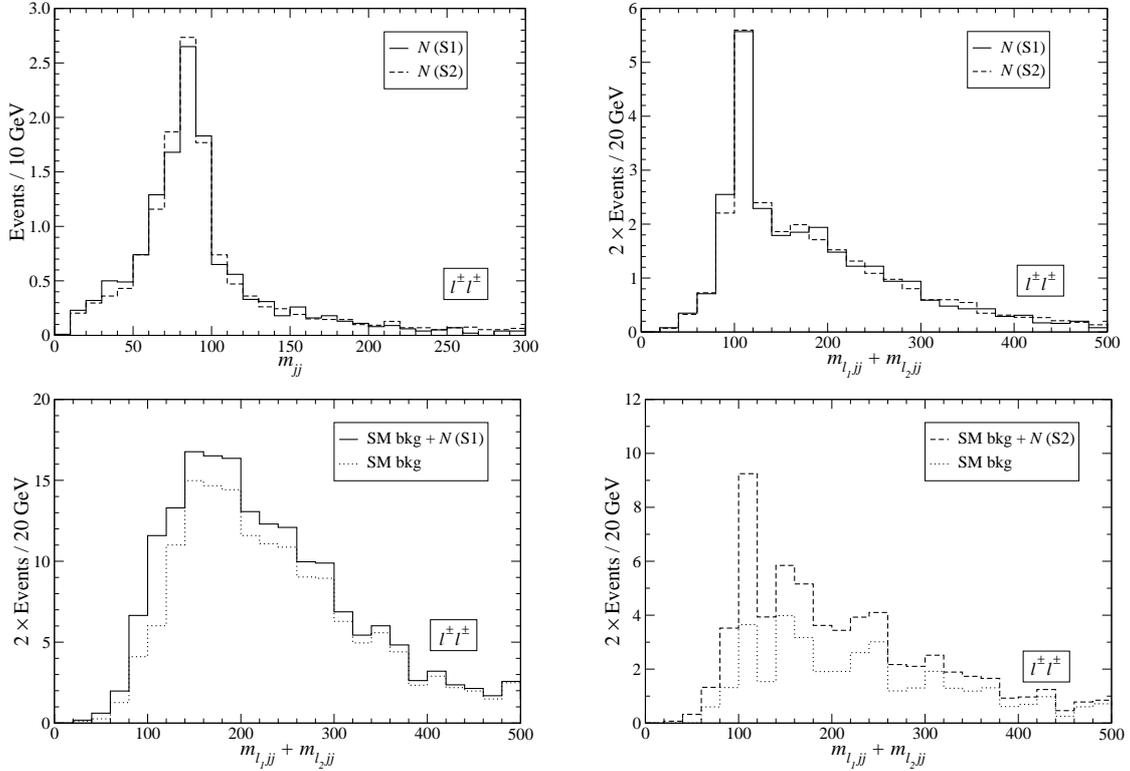

\begin{center}
\begin{tabular}{ccc}
\epsfig{file=Figs/mwrec-2lik-N12.eps,height=5cm,clip=} & \quad &
\epsfig{file=Figs/mN2x-2lik-N12.eps,height=5cm,clip=} \\
\epsfig{file=Figs/mN2x-2lik-bkgN1.eps,height=5cm,clip=} & \quad &
\epsfig{file=Figs/mN2x-2lik-bkgN2.eps,height=5cm,clip=}
\end{tabular}
\caption{Up, left: reconstructed $W$ mass. Up, right: $\ell_1 jj$ + $\ell_2 jj$ invariant mass distribution (two entries per event). Down:
$\ell_1 jj$ + $\ell_2 jj$ distribution for the two scenarios including background. All distributions correspond to selection level and a luminosity of 30 fb$^{-1}$.}
\label{fig:2likN-mrec}
\end{center}
\end{figure}

The reconstruction of the signal may be useful for its identification when large luminosities are available. The $W$ boson decaying hadronically can be reconstructed to some extent from the two jets with largest transverse momentum, as it is shown in Fig.~\ref{fig:2likN-mrec} (up, left). As it happens for larger neutrino masses \cite{delAguila:2007em}, the reconstruction is not very good and the discriminating power against the background is small, so that performing a cut on this variable results in a large signal loss. The reason for this bad reconstruction is that for the heavy neutrino signal the jets from the $W$ decay often have small transverse momentum (once that the charged lepton is required by pre-selection to have $p_T > 30$ GeV), and often one or the two jets selected to reconstruct the $W$ boson are produced from pile-up.

The heavy neutrino mass can be reconstructed from the $W$ boson and one of the charged leptons. In principle, it can be found by taking both possibilities and constructing a plot with two entries per event. The kinematical distribution displays a peak near the true $m_N$, as shown in Fig.~\ref{fig:2likN-mrec} (up, right) which might be visible over the bacground (this Figure, down). The observability of this peak is compromised by the large background in scenario S1, and by the small statistics in both cases. If the heavy neutrino mass is known from other source then invariant mass cuts can be performed, improving the significance to $\mathcal{S}_{20} = 1.2\sigma$, $\mathcal{S}_{20} = 3.1\sigma$ in scenarios S1 and S2, respectively. In the latter, the heavy neutrino signal can be discovered with 180 fb$^{-1}$.
Finally, it is worth remarking again that the results presented here can be improved if, instead of a simple application of kinematical cuts like we have done here, one performs a likelihood analysis as in Ref.~\cite{delAguila:2007em}. But in any case
heavy neutrino signals are very small and difficult to observe.

\subsection{Outlook}
\label{sec:4.3}

Heavy neutrino signals are limited by the small mixing of the heavy neutrino required by precision constraints \cite{delAguila:2008pw}. This fact implies that only masses of the order of 100 GeV are accesible at LHC. For this mass range, SM backgrounds are larger and, since production cross sections are relatively small, heavy neutrino singlets are rather difficult to observe.

In this section we have assumed a mass $m_N = 100$ GeV and examined the possible signals in the like-sign dilepton final state, which is the only channel
considered in many studies, and also in the trilepton final state. To our knowledge, this channel has not been previously studied in the context of heavy neutrino production, possibly due to its smaller cross section. We have found that for this particular heavy neutrino mass the trilepton channel is slightly better than the like-sign dilepton one. Although this fact may well be specific for the heavy neutrino mass assumed, it underlines the importance of
searching for heavy neutrinos in all the channels in which they might give observable signals. Indeed, the trilepton channel allows to discover Dirac neutrino singlets \cite{corto}, which do not give significant like-sign dilepton signals.

Heavy neutrino signals are characterised by low transverse momenta, and by a broad like-sign dilepton invariant mass distribution which does not have peaks nor long tails. 
This allows to distinguish them from scalar triplet (seesaw II) and fermion triplet (seesaw III) signals . But, more importantly, scalar and fermion triplets lead to other final states which are not present in heavy neutrino production, and thus the discrimination should be easy in case that a positive signal is observed.

\section{Seesaw II signals}
\label{sec:5}

\begin{figure}[ht]
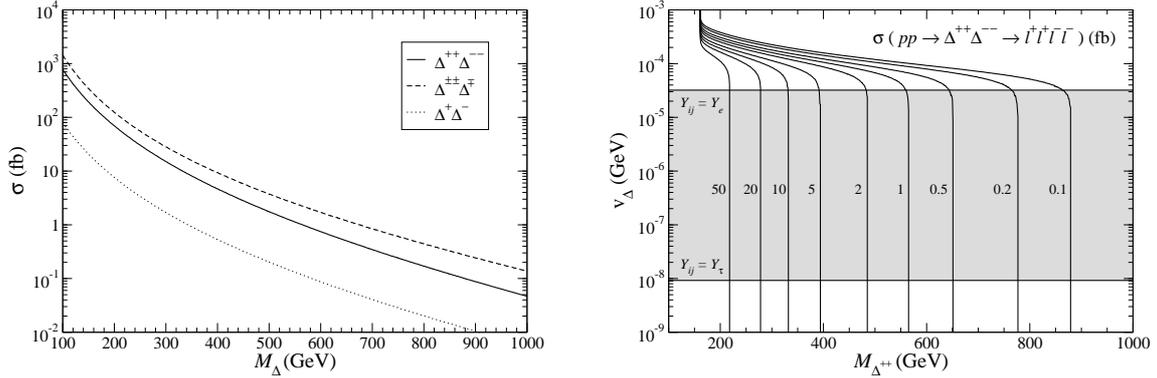

\begin{center}
\begin{tabular}{ccc}
\epsfig{file=Figs/cross-scalar.eps,height=5cm,clip=} & \quad &
\epsfig{file=Figs/contour-Dpp.eps,height=5cm,clip=}
\end{tabular}
\caption{Left: Cross section for production of charged scalar pairs
$\Dpp \Dmm$, $\Dppmm \Delta^\mp$ and $\Dp \Dm$ at LHC. Right: cross section for $\Dpp \Dmm$ decaying into four lepton final states.}
\label{fig:cross-sc}
\end{center}
\end{figure}

We consider three processes in which the members of the scalar triplet can be produced in hadron collisions,
\begin{align}
& q \bar q \to Z^* \,/\, \gamma^* \to \Dpp \Dmm \,, \notag \\
& q \bar q' \to W^* \to \Dppmm \Delta^{\mp} \,, \notag \\
& q \bar q \to Z^* \,/\, \gamma^* \to \Dp \Dm \,.
\label{ec:sc-signals}
\end{align}
Their cross sections only depend on the scalar masses, because the interactions are fixed by the triplet gauge couplings. We assume for simplicity that $\Dpp$
and $\Dp$ are degenerate.\footnote{A term $\lambda_5 (\phi^\dagger \tau^i \phi) (\vec \Delta^\dagger T^i \vec \Delta)$ in the scalar potential
induces a mass splitting between the triplet states $\sim \lambda_5/g^2 \, (M_W^2/M_\Delta)$, which is small enough to neglect scalar decays into other
triplet members if $\lambda _5$ is smaller than 1 and the
Yukawa couplings $Y_{ij}$ are not too small~\cite{Akeroyd:2005gt}.}
 The cross sections are plotted in Fig.~\ref{fig:cross-sc} (left) as a function of the common scalar mass. As emphasised in Ref.~\cite{Akeroyd:2005gt}, the
mixed $\Dppmm \Delta^\mp$ production is the largest source of doubly charged scalars, with a cross section about twice larger than
for $\Dpp \Dmm$, whereas for $\Dp \Dm$ it is smaller. Processes involving $\Delta^0$ production are not included because they do not contribute to the final states studied.

The cross section for four lepton final states in which we are interested
depends on the vev $\vt$ as well, through the decay branching ratios. For the specific case of $\Dpp$ decays, we have
\begin{eqnarray}
\Gamma(\Dpp \to l_i^+ l_j^+) & = & \frac{M_{\Dpp}}{4 \pi (1+\delta_{ij})} |Y_{ij}|^2 \,, \notag \\
\Gamma(\Dpp \to W^+ W^+) & = & \frac{g^4 \vt^2}{32\pi} \, \frac{M_{\Dpp}^3}{M_W^4}
 (1-4 r_W^2)^{1/2} (1-4 r_W^2 + 12 r_W^4) \,,
\end{eqnarray}
where $l_i = e,\mu,\tau$ for $i=1,2,3$, $\delta_{ij}$ is the Kronecker delta and $r_W=M_W/M_{\Dpp}$. The Yukawa couplings $Y_{ij}$ are related to the neutrino masses by Eq.~(\ref{ec:YMNS}), so that the sum of partial widths to dilepton final states is
\begin{equation}
\sum_{ij} \Gamma(\Dpp \to l_i^+ l_j^+) =
\frac{M_{\Dpp}}{8 \pi} \frac{\sum m_{\nu_i}^2}{4 \vt^2} \,,
\end{equation}
independently of the details of light neutrino mixing. The decay of $\Dppmm$ always takes place inside the detector, because when $\vt$ is large and the dilepton channel is suppressed the diboson channel is enhanced~\cite{Perez:2008ha}, and vice versa.
For $\vt$ sufficiently small the decays of the doubly charged scalar are dominated by the dilepton mode. Let us assume for the moment that light neutrino masses saturate the bound
\cite{Amsler:2008zz}
\begin{equation}
\sum m_{\nu_i} < 0.4 ~\text{eV} \,,
\label{ec:nubound}
\end{equation}
in which case they are quasi-degenerate. Then, the dependence of the cross section for $\Dpp \Dmm \to l^+ l^+ l^- l^-$ production on $M_{\Dpp}$ and $\vt$ is as shown on the right side of Fig.~\ref{fig:cross-sc} ($\tau$ leptons are included in the final state but their decay is not taken into account for the moment). For comparison we plot the
$\vt$ band corresponding to Yukawa couplings $Y_{ij}$ of the same
order as the electron and tau Yukawas.
Note, however, that there is no reason in principle to expect that the triplet Yukawa coupling to the charged leptons in
Eq.~(\ref{ec:LLyuk}) are the same as the Dirac coupling to the Higgs doublet.
These values for $\vt$ must be regarded only as a hint, showing that if doublet
and triplet Yukawas are of the same size, then the dilepton decay
mode $\Dpp \to l_i^+ l_j^+$ dominates. For neutrino masses not saturating the bound in Eq.~(\ref{ec:nubound}) the whole plot in Fig.~\ref{fig:cross-sc} (right) scales up or down with the neutrino mass sum, including the band $Y_{ij}=Y_e$, $Y_{ij}=Y_\tau$,
and the same argument applies.
We will then assume that doubly charged scalars
decay to two charged leptons.
Using an analogous argument, it follows that $\Dpm$ predominantly decay 
into a charged lepton and a neutrino, with partial widths
\begin{equation}
\Gamma(\Dp \to l_i^+ \nu_j) = \frac{M_{\Dp}}{8 \pi} |Y_{ij}|^2 \,. 
\end{equation}
Scalar triplets with masses of the order of 1 TeV or lighter are also predicted in Little Higgs models~\cite{ArkaniHamed:2002qy} (see for a review
Ref.~\cite{Perelstein:2005ka}) and some models of grand unification~\cite{Dorsner:2005fq}. If we impose extra symmetries to make the heavy sector
of the model less sensitive to electroweak precision constraints, as for instance in the Littlest Higgs model with T-parity \cite{Cheng:2003ju}, the coupling $\phi^\dagger (\vec{\tau} \cdot \vec{\Delta}) \tilde{\phi}$ can be forbidden and, consequently, a non-zero $\vt$. In this case our analysis fully applies but light neutrino masses are not generated. One can imagine, however, a very weak breaking of T-parity and then a tiny $\vt$, in agreement with our assumption~\cite{Hektor:2007uu}.

The relative abundance of $l = e,\mu,\tau$ in $\Dppmm$ and
$\Dpm$ decays is determined by the light neutrino mixing matrix
\cite{Hektor:2007uu}, including the Dirac and Majorana phases, and in fact it
may be used to determine $V_\text{MNS}$ from branching ratio measurements
\cite{Garayoa:2007fw,Kadastik:2007yd,Akeroyd:2007zv}. Hence,
a crucial consequence of this relation is that the observability of scalar triplets strongly depends on light neutrino mixing parameters. Decays
$\Dppmm \to e^\pm e^\pm / \mu^\pm \mu^\pm / e^\pm \mu^\pm$
are very clean, producing two energetic like-sign charged leptons with an invariant mass close to $M_{\Dpp}$. On the contrary,
decays to tau leptons are more difficult to identify and have much larger backgrounds. Tau leptons can decay leptonically
$\tau \to e \nu \bar \nu$,  $\tau \to \mu \nu \bar \nu$ with a branching ratio around
17\% each, giving electrons and muons less energetic than the parent
$\tau$. Hadronic tau decays 
can only be tagged with a certain efficiency, and always suffer the contamination from SM backgrounds with fake tau tags from jets. (For example,
corresponding to a $\tau$ tag efficiency of 50\%, the fake rate is around
1\%.) The relevant quantity which determines the observability
of $\Dppmm$ is the branching ratio to electrons and muons,
\begin{equation}
\remu \equiv \text{Br}(\Dppmm \to e^\pm e^\pm / \mu^\pm \mu^\pm / e^\pm \mu^\pm) \,.
\end{equation}
From the point of view of the signal, electrons and muons are quite alike, with similar detection efficiencies. From the point of view of SM backgrounds, at high transverse momenta (such as those involved in the decay
of $\Dppmm$ with few hundreds of GeV) like-sign
dielectron and dimuon final states are comparable, in contrast with the
behaviour at lower transverse momenta, where dielectrons are much more abundant
\cite{delAguila:2007em}. In our analysis we will sum over final states with electrons and muons. A detailed examination of the relative number of each is crucial to reconstruct the MNS matrix
\cite{Garayoa:2007fw,Kadastik:2007yd,Akeroyd:2007zv} but hardly affects the observability of doubly charged scalars.

\begin{figure}[ht]
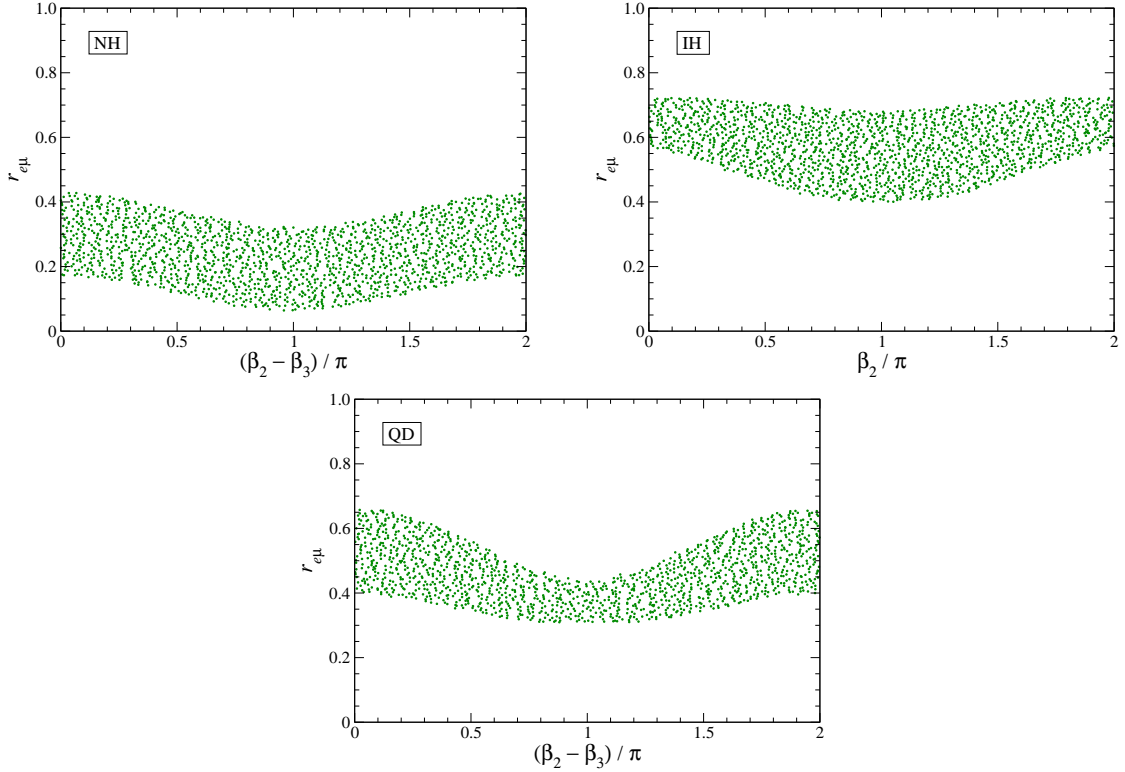

\begin{center}
\begin{tabular}{ccc}
\epsfig{file=Figs/NH-br.eps,height=5cm,clip=} & \quad &
\epsfig{file=Figs/IH-br.eps,height=5cm,clip=} \\
\multicolumn{3}{c}{\epsfig{file=Figs/QD-br.eps,height=5cm,clip=}}
\end{tabular}
\caption{Branching ratio $r_{e\mu}$ to electron and muon final states for normal and inverted hierarchy, and quasi-degenerate neutrinos.}
\label{fig:br-sc}
\end{center}
\end{figure}

In Fig.~\ref{fig:br-sc} we present the 67.3\% CL allowed regions for $\remu$
for normal hierarchy (NH), inverted hierarchy (IH) and quasi-degenerate (QD) neutrino masses. In the first and second cases we assume that the lightest neutrino is massless.
The MNS mixing matrix is parameterised as usual,
\begin{eqnarray}
V_\text{MNS} & = & \left( \! \begin{array}{ccc}
c_{12} c_{13} & s_{12} c_{13} & s_{13} e^{-i\delta} \\
- s_{12} c_{23} - c_{12} s_{23} s_{13} e^{i\delta} & c_{12} c_{23} -s_{12} s_{23} s_{13} e^{i\delta} & s_{23} c_{13} \\
s_{12} s_{23} -c_{12} c_{23} s_{13} e^{i\delta} & -c_{12} s_{23}-s_{12} c_{23}  s_{13} e^{i\delta} & c_{23} c_{13}
\end{array} \! \right) \nonumber \\[2mm]
& & \times \; \text{diag} \, (1 ,\, e^{-i\beta_2/2} ,\, e^{-i\beta_3/2}) \,.
\end{eqnarray}
We use the best fit values of mass differences and mixing angles in
Ref.~\cite{GonzalezGarcia:2007ib} with the errors quoted there, and for the unknown Majorana phases we assume a flat probability.
The 67.3\% CL regions are obtained with the acceptance-rejection method, as described in detail in Ref.~\cite{AguilarSaavedra:2006fy} for the program {\tt TopFit}. The bands show the dependence of $\remu$ on one phase or combination of phases, with the dependence on the rest of parameters (additional phases, the unknown value of $s_{13}$, etc.) reflected in the band width.
For NH $\remu$ mainly depends on the phase difference
$\beta_2-\beta_3$ but the variation is moderate. We observe that
the total branching ratio to electrons and muons is modest, around 30\%, and
for $\beta_2-\beta_3 = \pi$ it can be as low as 5\%, making the doubly charged scalars hard to discover in this case.
For IH $\remu$ is much larger, about 60\%, depending on $\beta_2$. For QD neutrinos $\remu$ depends on both phases and
only the dependence on $\beta_2-\beta_3$ (which is the strongest) is shown.
For this mass hierarchy $\remu \sim 0.45$, between the values obtained for NH and IH. For our simulations we select two benchmark scenarios illustrating the two extreme cases: (i) NH with $s_{13}=0$, $\beta_2-\beta_3 = \pi$, for which $\remu = 0.21$; (ii) IH with $s_{13}=0$, $\beta_2=\beta_3=0$, for which
$\remu = 0.65$. For squared mass differences and mixing angles we take the central values in Ref.~\cite{GonzalezGarcia:2007ib}.

In the rest of this section we study the observability of the scalar triplets in several final states, which we classify according to the number of charged
leptons in the sample:
(a) $\ell^+ \ell^+ \ell^- \ell^- X$; (b) $\ell^\pm \ell^\pm \ell^\mp X$; (c) $\ell^\pm \ell^\pm X$; (d) $\ell^+ \ell^- j_\tau X$; (e) $\ell^\pm
 j_\tau j_\tau j_\tau X$, where $\ell$ only corresponds to electrons and muons (but not necessarily all with the same flavour), $j_\tau$ denotes a jet tagged as a tau jet and $X$ refers to additional jets, tagged or not. We assume a common mass $M_{\Dpp} = M_{\Dp} = 300$ GeV.


\subsection{Final state $\ell^+ \ell^+ \ell^- \ell^-$}
\label{sec:5.1}

This is a very good channel for the observation of $\Dpp \Dmm$ production, because of its practically absent SM background. However, the scalar triplet signals in this decay mode are smaller than in other final states, because
\begin{enumerate}
\item Only $\Dpp \Dmm$ production contributes because $\Dppmm \Dmp$, with a cross section two times larger, gives at most three charged leptons.
\item For NH, requiring the presence of four charged leptons significantly reduces the signal. Four leptons can be produced (a) when both scalars decay $\Dppmm \to
e^\pm e^\pm,\mu^\pm \mu^\pm, e^\pm \mu^\pm$, which has a small total branching ratio $\sim (0.2)^2$ for NH, and; (b) in decays $\Dppmm \to e^\pm \tau^\pm,\mu^\pm \tau^\pm, \tau^\pm \tau^\pm$, when the $\tau$ leptons decay to electrons and muons plus neutrinos, which happens with a branching ratio of 34\%. Final states with a smaller number of leptons have larger branching ratios, which also include combinatorial factors (see next subsection).
\item All four leptons have to be isolated, within the detector acceptance and with transverse momentum above a certain threshold, leading to a lower detection efficiency than in channels with a smaller number of leptons.
\end{enumerate}
As pre-selection we require for signals and backgrounds the presence of four isolated charged leptons, two positively and two negatively charged. Among the four leptons, at least two must have transverse momentum larger than 30 GeV.
We also require the absence of additional non-isolated muons (from now on, this will be implicitly understood).
For event selection we only ask that the event does not have two opposite charge pairs with an invariant mass closer to $M_Z$ than 5 GeV.
Charged leptons are labelled as follows: $\ell_1$ is the one with highest
transverse momentum, $\ell_2$ is the other lepton with the same sign, and
$\ell_3$, $\ell_4$ the remaining two leptons ordered by decreasing $p_T$.
Then, neither the pairs $(\ell_1,\ell_3)$, $(\ell_2,\ell_4)$
nor $(\ell_1,\ell_4)$, $(\ell_2,\ell_3)$ can simultaneously have invariant mass within a 5 GeV interval around $M_Z$.\footnote{A stronger background suppression
can be achieved by demanding that neither of the opposite charge lepton pairs
has an invariant mass consistent with $M_Z$, which eliminates $Z b \bar b nj$ and $Z t \bar t nj$. This slightly decreases the signal and leads to a smaller statistical significance. Moreover, such a cut would suppress a possible fermion triplet signal in this channel (see section~\ref{sec:6.4}).}
This requirement does not affect the signal and is sufficient to suppress $ZZnj$ production below the other backgrounds. The remaining backgrounds, mainly $t \bar t nj$, concentrate at lower invariant masses and are not dangerous.
The number of signal events and main backgrounds at the pre-selection and selection levels is collected in Table~\ref{tab:nsnb-4lep}.

\begin{table}[ht]
\begin{center}
\begin{tabular}{ccc}
            & Pre-selection & Selection  \\[1mm]
$\Dpp \Dmm$ (NH) & 34.9  & 34.9 \\
$\Dpp \Dmm$ (IH) & 120.7 & 120.7 \\
$t \bar t nj$    & 116.0 & 115.7 \\
$Z b \bar b nj$  & 53.1  & 53.1 \\
$Z t \bar t nj$  & 32.9  & 31.5 \\
$Z Z nj$         & 617.7 & 98.7
\end{tabular}
\end{center}
\caption{Number of events for the four lepton signals and main backgrounds for a luminosity of 30 fb$^{-1}$.}
\label{tab:nsnb-4lep}
\end{table}

\begin{figure}[h]
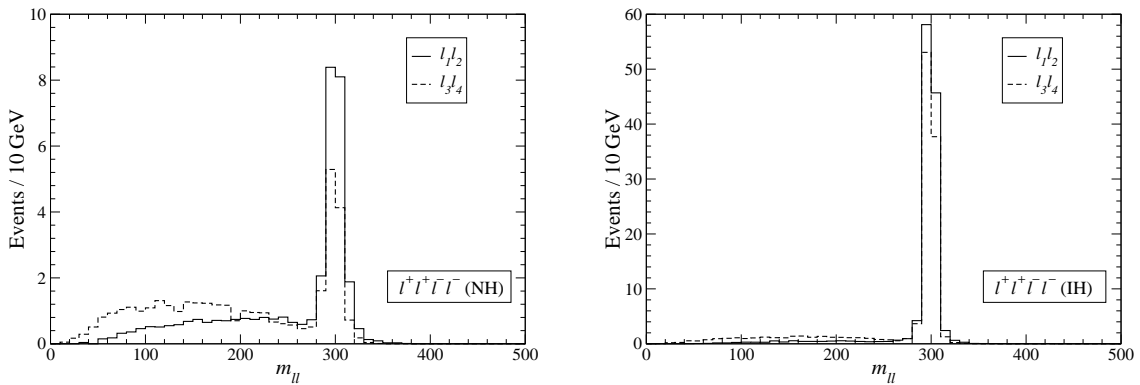

\begin{center}
\begin{tabular}{ccc}
\epsfig{file=Figs/Mll-4lep-sc1.eps,height=5cm,clip=} & \quad &
\epsfig{file=Figs/Mll-4lep-sc2.eps,height=5cm,clip=}
\end{tabular}
\caption{Kinematical distribution at pre-selection of the two like-sign dilepton invariant masses for the NH (left) and IH (right) signals, assuming a luminosity of 30 fb$^{-1}$.}
\label{fig:4lep-mrec}
\end{center}
\end{figure}

There are several interesting points to be learnt from the data in this table.
For NH, the final number of events for four lepton signals at pre-selection is rather small, 34.9 events which correspond to only 7.4\% of the $\Dpp \Dmm$ pairs produced. This fraction is larger than
$\remu^2 = 4.6\%$ due to tau leptonic decays, which give additional four lepton events but with a like-sign dilepton invariant mass smaller than
$M_{\Dpp}$. This can be clearly observed in Fig.~\ref{fig:4lep-mrec} (left).
The peaks correspond to $\Dppmm$ decays into two electrons or muons, while the broad part of the distributions correspond to $\tau$ decays. By construction,
the $\ell_1 \ell_2$ peak is higher because charged leptons from $\tau$ decays are less energetic. The number of events where $m_{\ell_1 \ell_2}$ and
$m_{\ell_3 \ell_4}$ are both in the windows $280-320$ GeV is 7.5, corresponding to only 1.6\% of the $\Dpp \Dmm$ pairs.
We also point out that the broad part of the $m_{\ell \ell}$ distributions behaves as combinatorial background decreasing the height of the peak with respect to the ``flat'' part. 
For IH the number of events at pre-selection is four times larger than for NH, and the peaks are much more pronounced, as it can be observed in Fig.~\ref{fig:4lep-mrec} (right).

\begin{figure}[ht]
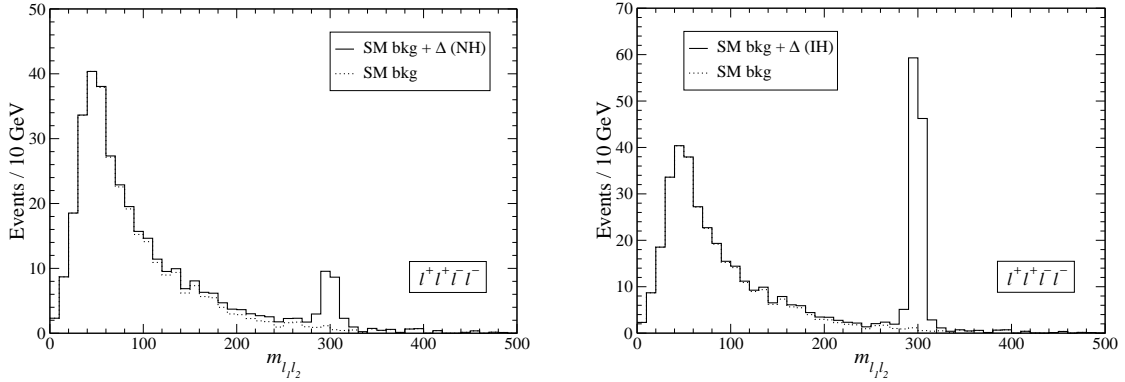

\begin{center}
\begin{tabular}{ccc}
\epsfig{file=Figs/Ml1l2-4lep-sc1.eps,height=5cm,clip=} & \quad &
\epsfig{file=Figs/Ml1l2-4lep-sc2.eps,height=5cm,clip=}
\end{tabular}
\caption{$\ell_1 \ell_2$ invariant mass distribution for the SM and the SM plus the scalar
triplet signal in the cases of NH (left) and IH (right). The luminosity is 30 fb$^{-1}$.}
\label{fig:4lep-mrec2}
\end{center}
\end{figure}

Discovering the $\Dppmm$ does not require to see both dilepton pairs
with masses around $M_{\Dpp}$ (for which the number of events is much smaller), but on the contrary it is enough to identify a clear peak in the $m_{\ell_1 \ell_2}$ distribution, which is plotted in Fig.~\ref{fig:4lep-mrec2} for the
SM backgrounds only and for the SM backgrounds plus the NH signal (left) and the IH signal (right).
In both cases the peaks are clearly visible, although for NH the number of events at the peak is small even for 30 fb$^{-1}$.
In Table~\ref{tab:sig-4lep} we collect the number of signal and background events at the peak, taken as the window
\begin{equation}
280 < m_{\ell_1 \ell_2} < 320~\text{GeV} \,,
\end{equation}
and making the two hypotheses for the background normalisation mentioned in section~\ref{sec:3}:
\begin{itemize}
\item[(a)] The SM background normalisation does not have any uncertainty, so that all the event excess at the peak can be interpreted as signal.
\item[(b)] The SM background must be normalised directly from data, in which case the off-peak signal contributes as combinatorial background, reducing the significance of the peak.
\end{itemize}
The situation in a real experiment will be between these two cases.
We also include in Table~\ref{tab:sig-4lep} the luminosity needed to have $5\sigma$ significance, for which we require to have an event excess not compatible with a background fluctuation at $5\sigma$, and to have at least 10 signal ($\ell^+ \ell^+ \ell^- \ell^-$) events.

\begin{table}[ht]
\begin{center}
\begin{tabular}{ccccccc}
& \multicolumn{3}{c}{Case (a)} & \multicolumn{3}{c}{Case (b)} \\
   & $S$   & $B$ & $L$            & $S$   & $B$ & $L$  \\[1mm]
NH & 20.4  & 3.0 & 14.7 fb$^{-1}$ & 18.1  & 5.3 & 18.6 fb$^{-1}$ \\
IH & 110.4 & 3.0 & 2.7 fb$^{-1}$  & 107.3 & 6.1 & 2.8 fb$^{-1}$
\end{tabular}
\end{center}
\caption{Number of signal ($S$) and background ($B$) events in the $m_{\ell_1 \ell_2}$ peak
for 30 fb$^{-1}$ in cases (a) and (b) explained in the text, and luminosity $L$ required to have a $5\sigma$ discovery in the $\ell^+ \ell^+ \ell^- \ell^-$ final state.}
\label{tab:sig-4lep}
\end{table}

\begin{figure}[htb]
\begin{center}
\epsfig{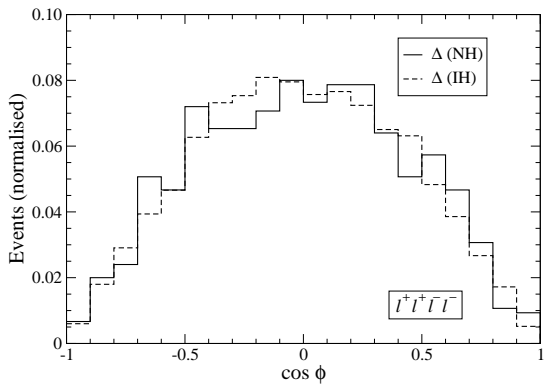}
\caption{Normalised $\Dpp$ opening angle distribution at the $m_{\ell \ell}$ peaks for NH and IH, for the four lepton signals.}
\label{fig:open-4lep}
\end{center}
\end{figure}

\begin{figure}[htb]
\begin{center}
\begin{tabular}{ccc}
\epsfig{file=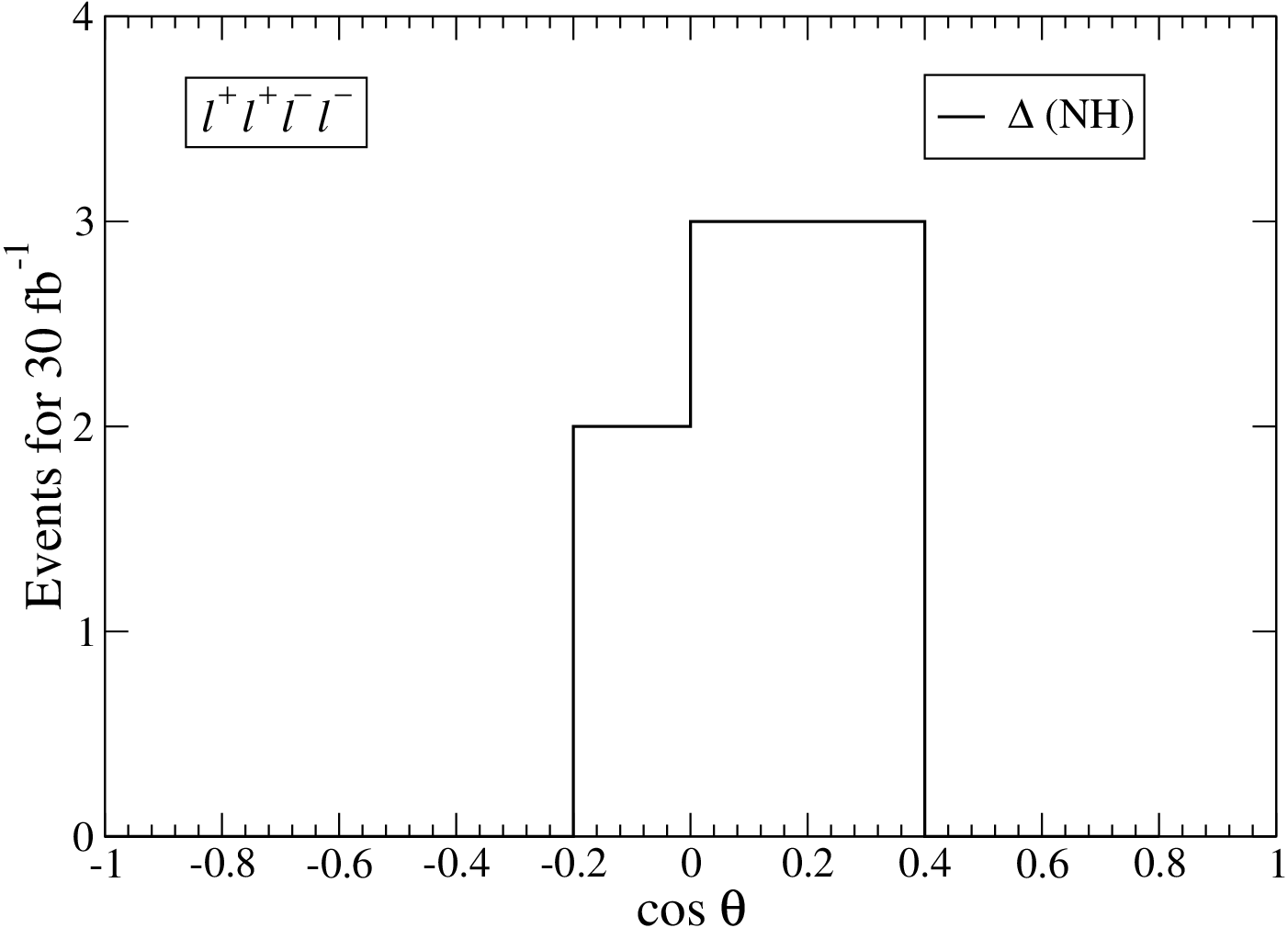,height=5cm,clip=} & \quad &
\epsfig{file=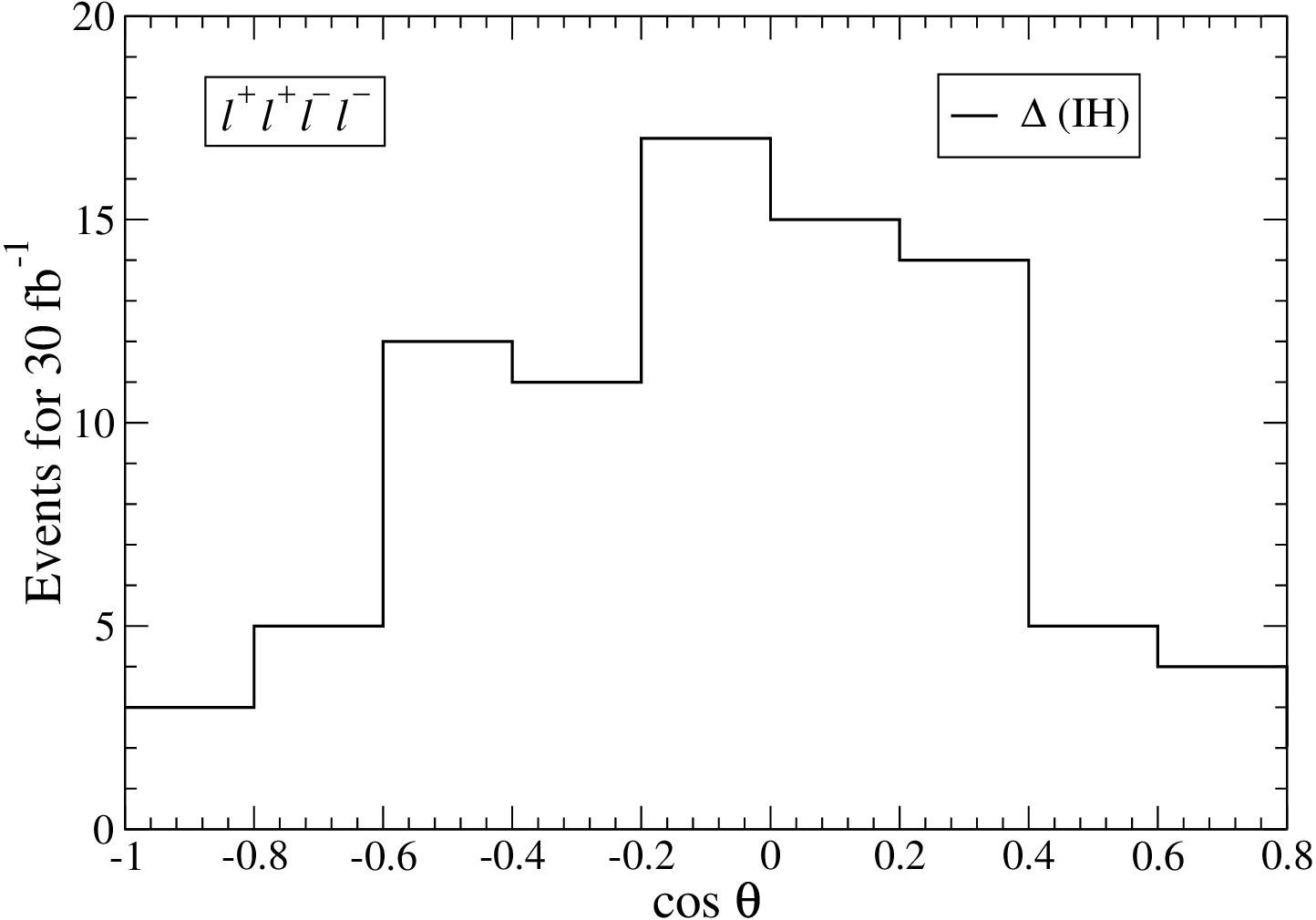,height=5cm,clip=}
\end{tabular}
\caption{Possible experimental results for the $\Dpp$ opening angle distribution in the case of NH (left) and IH (right), for a luminosity of 30 fb$^{-1}$. The number of events is 8 and 88 for NH and IH, respectively.}
\label{fig:openX-4lep}
\end{center}
\end{figure}

We finally investigate if the scalar nature of $\Dppmm$ can be established. We examine the opening angle distribution, defined in terms of the angle $\theta$ between the momenta of $\Dpp$ 
and the estimated direction of the incoming quark (positive $z$ if the $\Dpp \Dmm$ system moves in this direction or negative $z$ otherwise) in the $\Dpp \Dmm$ centre of mass (CM) frame. In order to ensure a correct reconstruction of this frame we require that both dilepton pairs have a mass close to the peak, between 280 and 320 GeV.
The dependence of the peak cross section on the opening angle is presented in Fig.~\ref{fig:open-4lep} for both NH and IH scenarios. We observe that the reconstruction is very good even without introducing correction functions to account for the detector effects, and
refinements such as using the Collins-Soper angle \cite{Collins:1977iv} are not necessary either.  The shape of the distributions obtained, proportional to
$1-\cos^2 \theta$, corresponds to the production of scalar particles.
However, the number of events at the peaks, which is 7.5 for NH and 88.4 for IH with a luminosity of 30 fb$^{-1}$, is too small to observe these distributions except for relatively large integrated luminosities. (In Fig.~\ref{fig:open-4lep} the signal is simulated using 3000 fb$^{-1}$.) In Fig.~\ref{fig:openX-4lep} we show the possible results of an experiment with 30 fb$^{-1}$. For NH one has 
some hints pointing to a $1-\cos^2 \theta$ distribution, although nothing can be concluded with the small number of events observed. For IH the distribution seems sufficiently good so as to establish
the scalar nature of $\Dppmm$, but we do not address here this issue quantitatively.

\subsection{Final state $\ell^\pm \ell^\pm \ell^\mp$}
\label{sec:5.2}

This final state can be considered as the ``golden channel'' for scalar triplet discovery. It has very small SM backgrounds as the four lepton channel, and the kinematical reconstruction of the missing particles can be achieved. Moreover, an important advantage over the former
is that three lepton final states receive contributions (which are actually dominant) from $\Dppmm \Dmp$ production, giving much larger signals and allowing for an earlier discovery of $\Dppmm$.
For pre-selection we require two like-sign leptons $\ell_1$ and $\ell_2$ with transverse momentum larger than 30 GeV and an additional charged lepton of opposite charge.
The number of events for the signal and main backgrounds is gathered in Table~\ref{tab:nsnb-3lep}. For selection we require that neither of the two opposite-sign lepton pairs have an invariant mass closer to $M_Z$ than 10 GeV. As expected, this requirement significantly reduces the backgrounds involving $Z$ boson production. The numbers of events after selection are also listed in
Table~\ref{tab:nsnb-3lep}. 

\begin{table}[ht]
\begin{center}
\begin{tabular}{cccccccc}
            & Pre-selection & Selection & \quad & & Pre-selection & Selection \\[1mm]
$\Dpp \Dmm$ (NH)   & 86.5   & 79.2  & & $Z b \bar b nj$    & 33.3   & 2.0  \\
$\Dppmm \Dmp$ (NH) & 97.6   & 89.9  & & $Z t \bar t nj$    & 152.5  & 16.8 \\
$\Dpp \Dmm$ (IH)   & 141.6  & 133.2 & & $W Z nj$           & 4113.8 & 73.4 \\
$\Dppmm \Dmp$ (IH) & 276.1  & 260.8 & & $Z Z nj$           & 276.1  & 4.2 \\
$t \bar t nj$      & 322.8  & 212.2 & & $WWW nj$           & 22.7   & 16.8\\
$tW$               & 17.8   & 12.2  & & $WWZ nj$           & 42.7   & 1.7  \\
$W t \bar t nj$    & 45.5   & 35.1 \\
\end{tabular}
\end{center}
\caption{Number of events for the three-lepton signals and main backgrounds with a luminosity of
30 fb$^{-1}$.}
\label{tab:nsnb-3lep}
\end{table}

\begin{figure}[ht]
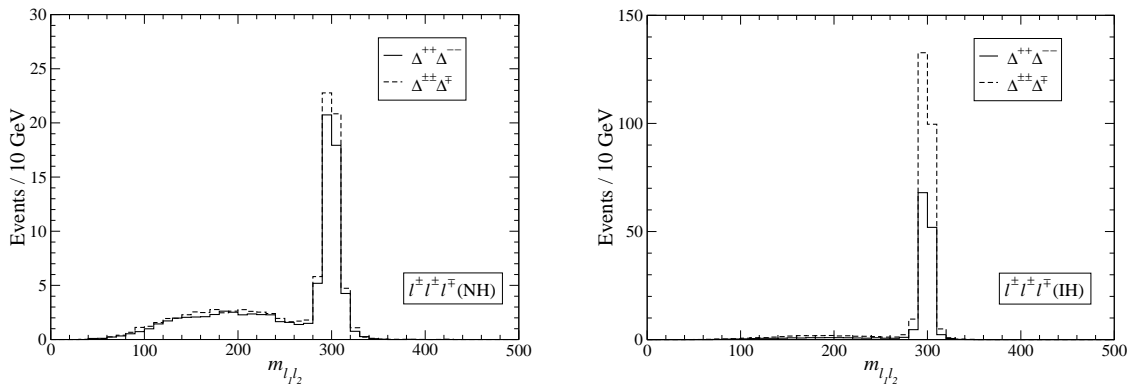

\begin{center}
\begin{tabular}{ccc}
\epsfig{file=Figs/Mll-3lep-sc1.eps,height=5cm,clip=} & \quad &
\epsfig{file=Figs/Mll-3lep-sc2.eps,height=5cm,clip=}
\end{tabular}
\caption{Kinematical distribution at pre-selection of the like-sign dilepton invariant mass for
the $\Dpp \Dmm$ and $\Dppmm \Dmp$ signals, for NH (left) and IH (right). The luminosity is
30 fb$^{-1}$.}
\label{fig:3lep-mrec}
\end{center}
\end{figure}

\begin{figure}[ht]
\begin{center}
\epsfig{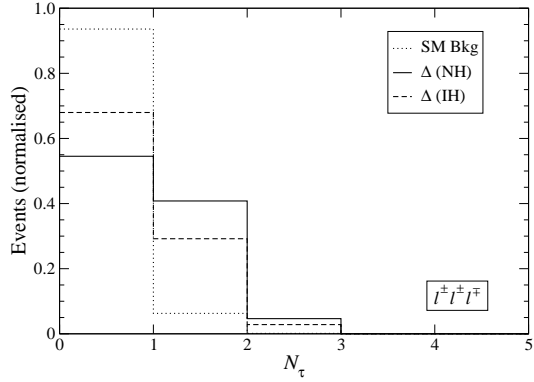} 
\caption{Multiplicity of $\tau$-tagged jets for the SM background and the NH, IH signals in trilepton final states at pre-selection level.}
\label{fig:3lep-tmult}
\end{center}
\end{figure}

Comparing with the four lepton final state we see that for NH the $\Dpp \Dmm$ signal is 2.5 times larger, mainly because of the larger branching ratios due to combinatorial factors. The additional contribution from $\Dppmm \Dmp$ makes the trilepton signal more than five times larger than the four lepton one in the previous subsection. In the case of IH the enhancement is mainly due to the $\Dppmm \Dmp$ process, and gives a trilepton signal 3.5 times larger than the four lepton one.
The signals have a sizeable contribution in which $\Dppmm$ decays give $\tau$ leptons, as it can be observed in the like-sign dilepton invariant mass
distribution, presented in Fig.~\ref{fig:3lep-mrec} for both NH and IH. The contributions of $\Dpp \Dmm$ and $\Dppmm \Dmp$ are separated for convenience. The behaviour is completely analogous to the one in Fig.~\ref{fig:4lep-mrec} for the four lepton final state. We also point out that around 40\% of the total number of signal events (which in this case correspond to $\Dpp \Dmm$ production) have jets in the final state which are tagged as tau jets. In Fig.~\ref{fig:3lep-tmult} we plot the $\tau$ multiplicity for the background and the NH and IH signals at pre-selection (notice that the trilepton signals can have at most one tau jet, but a second one can appear due to mistags).
Although the SM backgrounds rarely have tau leptons, it is not convenient to ask for one $\tau$ jet in event selection, since it decreases the signal considerably. On the other hand, separate analyses for each multiplicity $N_\tau=0,1,2$ can be performed increasing the total sensitivity, but for brevity we do not present them here.

\begin{figure}[ht]
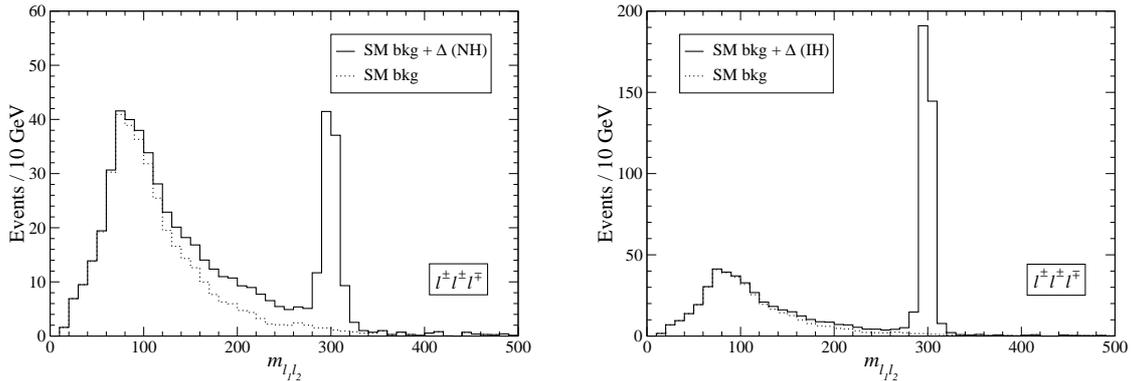

\begin{center}
\begin{tabular}{ccc}
\epsfig{file=Figs/Ml1l2-3lep-sc1.eps,height=5cm,clip=} & \quad &
\epsfig{file=Figs/Ml1l2-3lep-sc2.eps,height=5cm,clip=}
\end{tabular}
\caption{$\ell_1 \ell_2$ invariant mass distribution for the SM and the SM plus the scalar triplet signal in the cases of NH (left) and IH (right). The luminosity is 30 fb$^{-1}$.}
\label{fig:3lep-mrec2}
\end{center}
\end{figure}

After event selection, trilepton SM backgrounds are almost four times larger than those involving four leptons, but again they concentrate at low $m_{\ell_1 \ell_2}$ invariant masses. The doubly charged scalars can be discovered as a peak in the $\ell_1 \ell_2$ invariant mass, whose distribution is plotted in
Fig.~\ref{fig:3lep-mrec2} for the
SM backgrounds only and for the SM backgrounds plus the NH signal (left) and the IH signal (right)
after event selection criteria.
The peaks are much more pronounced than in the four lepton final state, making the discovery of the $\Dppmm$ signal in this final state much easier. The number of signal and background events at the peak
\begin{equation}
280 < m_{\ell_1 \ell_2} < 320~\text{GeV}
\end{equation}
is collected in Table~\ref{tab:sig-3lep}, together with the luminosity necessary for a $5\sigma$ discovery. We distinguish the two cases: (a) if the SM background can be predicted with negligible uncertainty and (b) if it is normalised from data. For NH the luminosity needed to discover $\Dppmm$ is $4-5$ times smaller than
in the four lepton final state, and for IH three times smaller. This improvement is very significant, making the three lepton final state the best one for the discovery of the doubly charged scalars at LHC.

\begin{table}[ht]
\begin{center}
\begin{tabular}{ccccccc}
& \multicolumn{3}{c}{Case (a)} & \multicolumn{3}{c}{Case (b)} \\
   & $S$   & $B$ & $L$            & $S$   & $B$ & $L$  \\[1mm]
NH & 94.5  & 5.1 & 3.2 fb$^{-1}$  & 84.4  & 15.2 & 3.6 fb$^{-1}$ \\
IH & 353.0 & 5.1 & 0.85 fb$^{-1}$ & 343.5 & 14.4 & 0.87 fb$^{-1}$
\end{tabular}
\end{center}
\caption{Number of signal ($S$) and background ($B$) events at the $m_{\ell_1 \ell_2}$ peak
for 30 fb$^{-1}$ in cases (a) and (b), and luminosity $L$ needed to have a $5\sigma$ discovery in the $\ell^\pm \ell^\pm \ell^\mp$ final state.}
\label{tab:sig-3lep}
\end{table}

We finally address the identification of the scalar nature of $\Dppmm$. The reconstruction of the final state is more involved due to the presence of two signal contributions with different kinematics. Signal events involve two scalars, one of them (labelled as $\Delta_1$) decays to the like-sign pair and the other (labelled as $\Delta_2$) produces the third lepton $\ell_3$ plus an additional missed charged lepton or $\tau$ jet (if it is doubly charged) or a light neutrino (if it is a $\Dpm$). Both possibilities must be disentangled on an event by event basis. We identify events corresponding to $\Dpp \Dmm$ and
$\Dppmm \Dmp$ production using these criteria:
\begin{enumerate}
\item If the event has a $\tau$-tagged jet $j_\tau$, it is assumed that it corresponds to $\Dpp \Dmm$ production and it is reconstructed accordingly.
\item If the event does not have $\tau$-tagged jets but has additional energetic jets, it is taken as $\Dpp \Dmm$
if the transverse momentum of the hardest jet
(which is then regarded as coming from a $\tau$ decay, albeit not tagged)
is larger than the missing energy of the event. Otherwise the event is reconstructed as $\Dppmm \Dmp$.
\item If the event does not have additional jets, it is reconstructed as $\Dppmm \Dmp$.
\end{enumerate}
$\Dpp \Dmm$ events are reconstructed as follows.
The third charged lepton $\ell_3$ may have been directly produced in a $\Dppmm$ decay or may be a secondary charged lepton from a leptonic $\tau$ decay, in which case it is produced together with two neutrinos, of combined momentum $p_{\nu_1}$, taken collinear to $p_{\ell_3}$.
The neutrino associated to the hadronic $\tau$ has momentum $p_{\nu_2}$ collinear to the jet. If the like-sign dilepton pair has an invariant mass close to the $M_{\Dpp}$ peak (a fact which is enforced using a suitable kinematical cut), then all the missing energy of the event corresponds to these neutrinos, whose momenta $p_{\nu_1} = t_1 p_{\ell_3}$, $p_{\nu_2} = t_2 p_{j_\tau}$ can be determined using the equations
\begin{eqnarray}
t_1 (p_{\ell_3})_x + t_2 (p_{j_\tau})_x & = & p_x\!\!\!\!\!\!\!\!\not\,\,\,\,\,\,\,\, \,, \notag \\
t_1 (p_{\ell_3})_y + t_2 (p_{j_\tau})_y & = & p_y\!\!\!\!\!\!\!\!\not\,\,\,\,\,\,\,\, \,,
\end{eqnarray}
where $p_x\!\!\!\!\!\!\!\!\not\,\,\,\,\,\,$ and $p_y\!\!\!\!\!\!\!\!\not\,\,\,\,\,\,$ are the two components of the missing momentum $p_T\!\!\!\!\!\!\!\!\not\,\,\,\,\,\,\,$. Both $t_1$ and $t_2$ must be positive, otherwise the event is discarded.
The reconstructed momenta of the two scalars are then
\begin{eqnarray}
p_{\Delta_1} & = & p_{\ell_1} + p_{\ell_2} \,, \notag \\
p_{\Delta_2} & = & p_{\ell_3} + p_{j_\tau} + p_{\nu_1} + p_{\nu_2} \,.
\label{ec:3lep-rec1}
\end{eqnarray}

Reconstruction of events classified as $\Dppmm \Dmp$ is done neglecting the possible missing momentum associated to $\ell_3$ and using the equations
\begin{eqnarray}
(p_{\nu_2})_x & = & p_x\!\!\!\!\!\!\!\!\not\,\,\,\,\,\,\,\, \,, \notag \\
(p_{\nu_2})_y & = & p_y\!\!\!\!\!\!\!\!\not\,\,\,\,\,\,\,\, \,, \notag \\
(p_{\ell_3} + p_{\nu_2})^2 & = & (p_{\ell_1} + p_{\ell_2})^2 \,,
\end{eqnarray}
plus the on-shell condition $p_{\nu_2}^2 = 0$. The reconstructed momenta of $\Delta_1$ and $\Delta_2$ are
\begin{eqnarray}
p_{\Delta_1} & = & p_{\ell_1} + p_{\ell_2} \,, \notag \\
p_{\Delta_2} & = & p_{\ell_3} + p_{\nu_2} \,.
\end{eqnarray}
The quality of the reconstruction is ensured by setting cuts
\begin{align}
& 280~\text{GeV} \leq m_{\ell_1 \ell_2} \leq 320~\text{GeV} \,, \notag \\  
& 280~\text{GeV} \leq \sqrt{p_{\Delta_2}^2} \leq 320~\text{GeV} \,.
\end{align}
With these cuts, the number of events at the peaks is 71.1 and 281.8 for the NH and IH signals, respectively, classified as shown in Table~\ref{tab:ns-3lep}. We can observe that the discrimination method is good, although it may eventually be improved with a kinematical fit.
\begin{table}[htb]
\begin{center}
\begin{tabular}{ccccc}
& \multicolumn{2}{c}{NH} & \multicolumn{2}{c}{IH} \\
& $\Dpp \Dmm$          & $\Dppmm \Dmp$ & $\Dpp \Dmm$ & $\Dppmm \Dmp$ \\[1mm]
Total                  & 21.7 & 49.4 & 52.4  & 229.4 \\
$\geq 1~j_\tau$        & 4.5  & 0.0  & 7.4  & 0.0 \\
$1~j$, $p_T > \ptmiss$ & 2.6  & 0.0  & 6.6  & 0.0 \\
$1~j$, $p_T < \ptmiss$ & 7.6  & 6.3  & 11.7  & 34.9 \\
$0j$                   & 7.0  & 43.1 & 26.7  & 194.5
\end{tabular}
\end{center}
\caption{Number of signal events at the $m_{\ell_1 \ell_2}$ peak for each of the classes considered in the reconstruction }
\label{tab:ns-3lep}
\end{table}

\begin{figure}[htb]
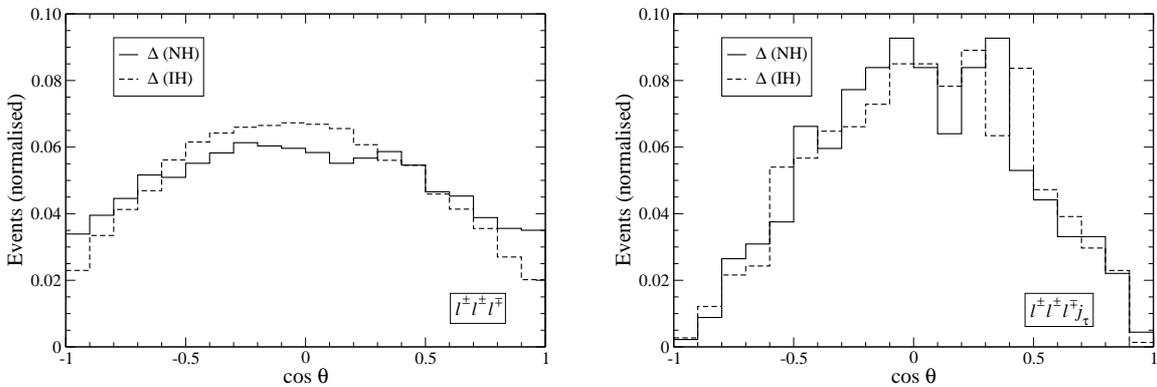

\begin{center}
\begin{tabular}{ccc}
\epsfig{file=Figs/cProd-3lep.eps,height=5cm,clip=} & \quad &
\epsfig{file=Figs/cProd-3leptau.eps,height=5cm,clip=}
\end{tabular}
\caption{Normalised opening angle distribution at the $m_{\ell \ell}$ peak for NH and IH. On the right side we plot the same distribution but only for events with a tagged $\tau$ jet.}
\label{fig:open-3lep}
\end{center}
\end{figure}

\begin{figure}[htb]
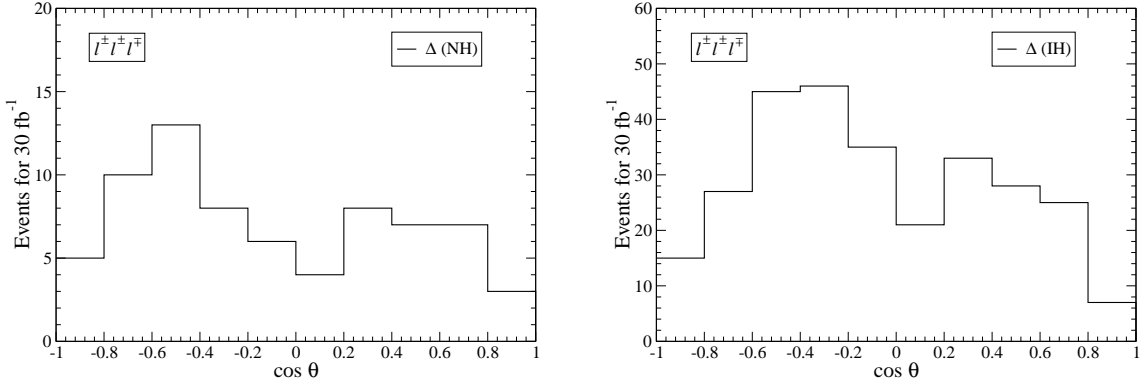

\begin{center}
\begin{tabular}{ccc}
\epsfig{file=Figs/cProd-3lep-sc1X.eps,height=5cm,clip=} & \quad &
\epsfig{file=Figs/cProd-3lep-sc2X.eps,height=5cm,clip=}
\end{tabular}
\caption{Possible experimental results for the opening angle distribution in the case of NH (left) and IH (right), for a luminosity of 30 fb$^{-1}$. The number of events is 71 and 282 for NH and IH, respectively.}
\label{fig:openX-3lep}
\end{center}
\end{figure}

The opening angle $\theta$ is defined as the angle between the momentum of the $\Dpp$ ($\Dmm$) and the momentum of the initial quark (antiquark) in the CM frame. The latter is estimated as in the case of  the four lepton signal in the previous subsection, because the improvement found using the Collins-Soper angle is very small. The resulting distribution is presented in Fig.~\ref{fig:open-3lep} (left). The shape is similar to the ``true'' one although the extreme bins have a sizeable fraction of events, and correction functions must be used in order to unfold the effect of the detector and reconstruction (see for example Refs.~\cite{Hubaut:2005er,AguilarSaavedra:2007rs}).
Requiring the presence of a tagged $\tau$ jet reduces the signals to 4.5 and 7.4 events for NH and IH, respectively, but improves the quality of the reconstruction.
As it can be observed in
Fig.~\ref{fig:open-3lep} (right), the distribution in this case is very similar to the one found in the four lepton final state, but includes a smaller number of events. Possible experimental results corresponding to Fig.~\ref{fig:open-3lep} (left) are shown in Fig.~\ref{fig:openX-3lep}, taking a luminosity of 30 fb$^{-1}$. The distributions seem to indicate that
the cross section is proportional to $1-\cos^2 \theta$, especially for the IH case, although the results must be corrected for detector effects in order to draw a quantitative conclusion.

Finally, it is worth remarking that the presence of reconstructed trilepton events with large missing energy is a signature of $\Dppmm \Dmp$ production, providing evidence of the non-singlet nature of $\Delta$.


\subsection{Final state $\ell^\pm \ell^\pm$}
\label{sec:5.3}

Scalar triplet production gives like-sign dilepton signals when one doubly charged scalar decays into two charged leptons while the accompanying scalar does into tau jets, neutrinos or charged leptons missed by the detector. Like-sign dilepton signals are common to the three types of seesaw mechanism but in the case of the scalar triplet seesaw the like-sign dilepton invariant mass spectrum exhibits a peak at $M_{\Dpp}$, produced when the doubly charged scalars directly decay to light charged leptons (electrons and muons). SM backgrounds in this channel are larger than in the previous two modes, but the signal significance is still comparable to the one achieved in the four lepton channel.

\begin{figure}[ht]
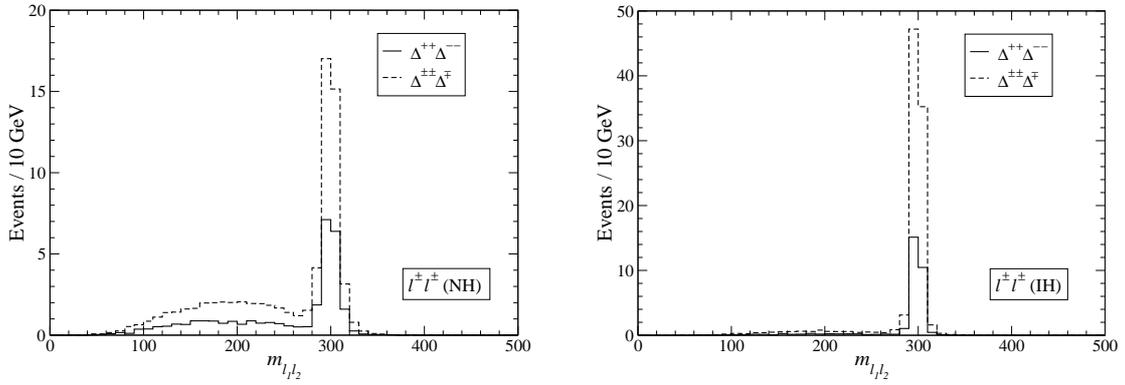

\begin{center}
\begin{tabular}{ccc}
\epsfig{file=Figs/Mll-2lik-sc1.eps,height=5cm,clip=} & \quad &
\epsfig{file=Figs/Mll-2lik-sc2.eps,height=5cm,clip=}
\end{tabular}
\caption{Kinematical distribution at pre-selection of the like-sign dilepton invariant mass for
the $\Dpp \Dmm$ and $\Dppmm \Dmp$ signals, in NH (left) and IH (right). The luminosity is
30 fb$^{-1}$.}
\label{fig:2lik-mrec}
\end{center}
\end{figure}

\begin{figure}[ht]
\begin{center}
\epsfig{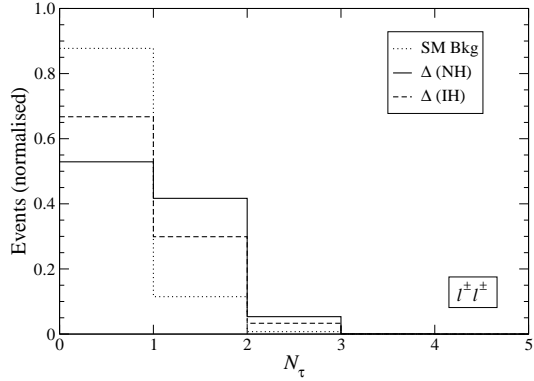} 
\caption{Multiplicity of $\tau$-tagged jets for the SM background and the NH, IH signals in like-sign dilepton final states at pre-selection level.}
\label{fig:2lik-tmult}
\end{center}
\end{figure}

For event pre-selection we require two like-sign charged leptons $\ell_1$, $\ell_2$ with transverse momentum larger than 30 GeV and no additional leptons (otherwise events correspond to the channels studied in the previous sections).
The number of signal events are collected in Table~\ref{tab:nsnb-2lik}, together with the main backgrounds.  
\begin{table}[h]
\begin{center}
\begin{tabular}{cccccccc}
            & Pre-selection & \quad & & Pre-selection \\[1mm]
$\Dpp \Dmm$ (NH)   & 30.6   & & $W t \bar t nj$    & 194.0 \\
$\Dppmm \Dmp$ (NH) & 72.2   & & $W W nj$           & 205.7 \\
$\Dpp \Dmm$ (IH)   & 30.3   & & $W Z nj$           & 892.2 \\
$\Dppmm \Dmp$ (IH) & 97.4   & & $WWW nj$           & 86.9  \\
$t \bar t nj$      & 1193.6 & &  \\
\end{tabular}
\end{center}
\caption{Number of events for the like-sign dilepton signals and main backgrounds for a luminosity of 30 fb$^{-1}$.}
\label{tab:nsnb-2lik}
\end{table}
These pre-selection criteria are sufficient to observe the signals, and the improvement achieved with further cuts ({\em e.g.} requiring that the leptons are not back-to-back) is small.
The $m_{\ell_1 \ell_2}$ distribution for the separate $\Dpp \Dmm$ and
$\Dppmm \Dmp$ signals is presented in Fig.~\ref{fig:2lik-mrec}, for NH (left) and IH (right). The shape of the distributions is as in the two previous subsections, but in this case the combinatorial background from $\tau$ decays is less significant compared to the SM background.
Like-sign dilepton signals from scalar triplet production benefit from the presence of $\tau$-tagged jets in the final state, as it is shown in Fig.~\ref{fig:2lik-tmult}. Therefore, the sensitivity can be improved by splitting the like-sign dilepton sample by the $\tau$ jet multiplicity $N_\tau = 0,1,2$ and performing an analysis for each subsample. For brevity we do not present such a study here.

\begin{figure}[ht]
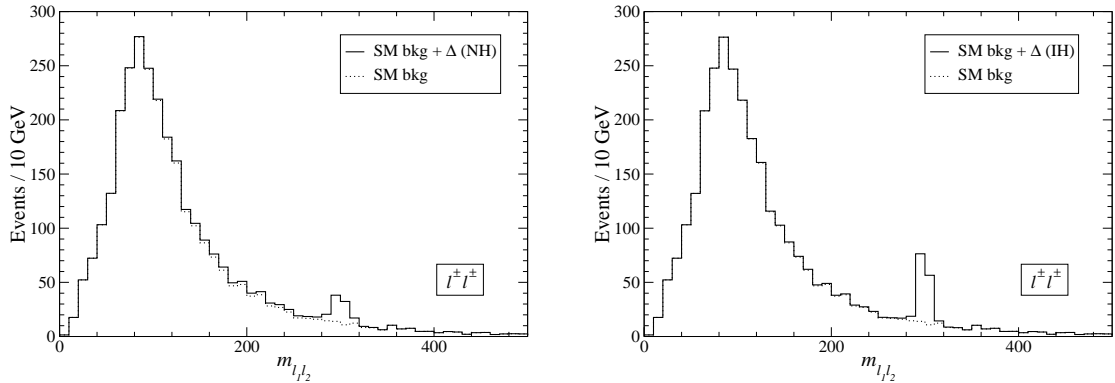

\begin{center}
\begin{tabular}{ccc}
\epsfig{file=Figs/Ml1l2-2lik-sc1.eps,height=5cm,clip=} & \quad &
\epsfig{file=Figs/Ml1l2-2lik-sc2.eps,height=5cm,clip=}
\end{tabular}
\caption{$\ell_1 \ell_2$ invariant mass distribution for the SM and the SM plus the triplet signal in the cases of NH (left) and IH (right). The luminosity is 30 fb$^{-1}$.}
\label{fig:2lik-mrec2}
\end{center}
\end{figure}

In this channel SM backgrounds are much larger than the signals but, as it happens with trilepton and four lepton final states, they concentrate at low
dilepton invariant masses. Hence, even with the loose pre-selection cuts used here, the presence of a $\Dppmm$ resonance can be spotted with the examination of the $m_{\ell_1 \ell_2}$ distribution, shown in Fig.~\ref{fig:2lik-mrec2} for the SM background alone and also including the NH and IH signals.
The $\Dppmm$ peaks are less pronounced than in the three and four lepton final states.
Despite the larger backgrounds at the peak region
\begin{equation}
280 < m_{\ell_1 \ell_2} < 320~\text{GeV}
\end{equation}
(see Table~\ref{tab:sig-2lik}), the larger number of signal events provides a signal significance very similar to the one in the four lepton final state, and the luminosities required for $5\sigma$ discovery in both NH and IH scenarios, listed in Table~\ref{tab:sig-2lik}, are comparable to the four lepton channel.
Nevertheless, a disadvantage of the $\ell^\pm \ell^\pm$ final state is that the full event reconstruction, with two competing signal processes and several missing particles, is much more involved. The opening angle distribution obtained in this case is very distorted from the theoretical value and a background subtraction must also be performed. This study is beyond the scope of the present work.

\begin{table}[ht]
\begin{center}
\begin{tabular}{ccccccc}
& \multicolumn{3}{c}{Case (a)} & \multicolumn{3}{c}{Case (b)} \\
   & $S$   & $B$ & $L$            & $S$  & $B$ & $L$  \\[1mm]
NH & 56.5  & 51.7 & 15 fb$^{-1}$  & 53.4 & 54.7 & 17.4 fb$^{-1}$ \\
IH & 114.3 & 51.7 & 4.4 fb$^{-1}$ & 114.3 & 51.7 & 4.4 fb$^{-1}$
\end{tabular}
\end{center}
\caption{Number of signal ($S$) and background ($B$) events at the $m_{\ell_1 \ell_2}$ peak
for 30 fb$^{-1}$ and luminosity $L$ required to have a $5\sigma$ discovery in the $\ell^\pm \ell^\pm$ final state.}
\label{tab:sig-2lik}
\end{table}


\subsection{Final state $\ell^+ \ell^- j_\tau$}
\label{sec:5.4}

Opposite-sign dilepton backgrounds are huge at LHC, mainly coming from $t \bar t nj$ and $Z^* / \gamma^*\,nj$ production, and make the observation of scalar triplet signals in the $\ell^+ \ell^-$ channel virtually impossible. However, the requirement of an energetic $\tau$ jet, which is often present in the signals (except in $\Dp \Dm$) makes the backgrounds manageable. The main objective of 
the study in this section is to show that scalar triplet signals are observable in this difficult channel too.
A likelihood analysis taking advantage of the differences in the kinematical distributions of signals and backgrounds will certainly improve the results.
We select the events with:
\begin{itemize}
\item[(i)] two oppositely charged leptons with invariant mass larger than 200 GeV, and no additional leptons;
\item[(ii)] at least one jet tagged as $\tau$ jet, with transverse momentum larger than 20 GeV;
\item[(iii)] not more than 2 additional untagged jets with $p_T > 20$ GeV, and no $b$-tagged jets.
\end{itemize}
For the scalar triplet signals the two charged leptons have a very broad
invariant mass distribution because they are produced in the decay of different particles. The jet $j_\tau$ and one of the charged leptons (typically, the most energetic one) have an invariant mass distribution which concentrates at $M_{\Dpp}$ and below. Thus, the presence of the signal can be detected as a bump in the $m_{\ell_1 j_\tau}$ distribution. Nevertheless, the mass reconstruction is not very good because of the missing energy from the $\tau$ decay, and we will skip its presentation here. A better discriminating variable is the transverse momentum of the leading charged lepton $p_T^{\ell_1}$, whose distribution exhibits a long tail once that SM backgrounds are conveniently reduced. The kinematical cuts applied with this purpose are:
\begin{itemize}
\item[(i)] the missing energy $p_T\!\!\!\!\!\!\!\!\not\,\,\,\,\,\,\,$ must be larger than 50 GeV;
\item[(ii)] at least one of the $\tau$-tagged jets must have transverse momentum
$p_T > 50$ GeV;
\item[(iii)] the angle $\phi$ between the two charged leptons in transverse plane has to be larger than $\pi/2$.
\end{itemize}
The first requirement eliminates $Z^*/\gamma^* nj$ production. The remaining backgrounds involve charged leptons from $W$ decays, so the number of
$e^+ e^-$, $\mu^+ \mu^-$, $e^+ \mu^-$ and $\mu^+ e^-$ events is similar.
The number of signal and background events at the pre-selection and selection levels can be read in Table~\ref{tab:nsnb-2opp}.

\begin{table}[t]
\begin{center}
\begin{tabular}{cccccccc}
            & Pre-selection & Selection & \quad & & Pre-selection & Selection \\[1mm]
$\Dpp \Dmm$   (NH) & 26.4  & 16.2 & & $t \bar t nj$      & 486.0 & 55.2 \\
$\Dppmm \Dmp$ (NH) & 38.1  & 28.3 & & $tW$               & 98.4  & 9.0 \\
$\Dpp \Dmm$ (IH)   & 12.3  & 6.6  & & $WW nj$            & 216.9  & 7.0 \\
$\Dppmm \Dmp$ (IH) & 27.3  & 19.5 & & $Z^*/\gamma^* nj$  & 2424.5 & 0.7
\end{tabular}
\end{center}
\caption{Number of events for the $\ell^+ \ell^- j_\tau$ signals and main backgrounds for a luminosity of 30 fb$^{-1}$.}
\label{tab:nsnb-2opp}
\end{table}

\begin{figure}[ht]
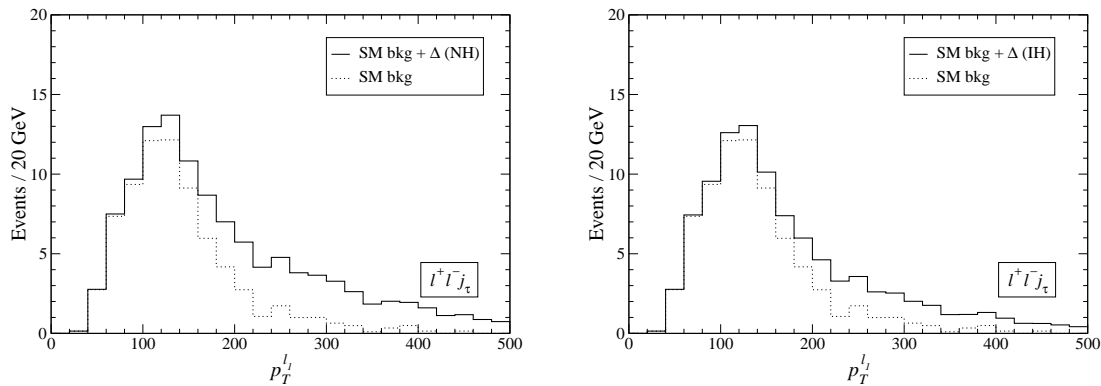

\begin{center}
\begin{tabular}{ccc} 
\epsfig{file=Figs/ptlep1-2opp-sc1.eps,height=5cm,clip=} & \quad &
\epsfig{file=Figs/ptlep1-2opp-sc2.eps,height=5cm,clip=}
\end{tabular}
\caption{$p_T^{\ell_1}$ distributions for the SM and the SM plus the triplet signal in the cases of NH (left) and IH (right). The luminosity is 30 fb$^{-1}$.}
\label{fig:2opp-mpt}
\end{center}
\end{figure}

The contribution of scalar triplets to the $\ell^+ \ell^- j_\tau$ final state can be detected as a long tail in the transverse momentum distribution for the
leading charged lepton, presented in Fig.~\ref{fig:2opp-mpt} for the cases of NH and IH. The signal contributions are spread across a wide $p_T^{\ell_1}$ range, and normalising the SM background seems non-trivial. In order to estimate the signal significance in this channel we assume a 20\% uncertainty in the backgrounds, incorporated in the calculation of the statistical significance. Requiring
\begin{equation}
p_T^{\ell_1} > 200~\text{GeV}
\end{equation}
the number of background events is 10.1, while most of the signal survives,
34.4 events for NH and 20.2 for IH, giving significances $\mathcal{S}_{0} = 10.6\sigma$,
$\mathcal{S}_{20} = 9.0\sigma$ (NH) and $\mathcal{S}_{0} = 6.2\sigma$,
$\mathcal{S}_{20} = 5.3\sigma$ (IH).

\subsection{Final state $\ell^\pm j_\tau j_\tau j_\tau$}
\label{sec:5.5}

The huge cross section for $Wnj$ production makes the signals in Eq.~(\ref{ec:sc-signals}) unobservable in final states with only one charged lepton. In order to reduce this and the rest of backgrounds we require three tagged $\tau$ jets with transverse momentum larger than 20 GeV, in addition to a charged lepton with
$p_T > 30$ GeV. The number of events for a luminosity of 30 fb$^{-1}$ is gathered in Table~\ref{tab:nsnb-1l3t} for the most relevant processes. It is clear that
even requiring three $\tau$ jets, which imply a background rejection factor $\sim 10^{-6}$, is not enough to make the signals observable.

\begin{table}[ht]
\begin{center}
\begin{tabular}{cc}
            & Pre-selection \\[1mm]
$\Dpp \Dmm$ (NH)   & 3.0 \\
$\Dppmm \Dmp$ (NH) & 0.3 \\
$\Dpp \Dmm$ (IH)   & 0.5 \\
$\Dppmm \Dmp$ (IH) & 0.1 \\
$t \bar t nj$      & 3069.8 \\
$b \bar b nj$      & 1200 \\
$Wnj$              & 72740
\end{tabular}
\end{center}
\caption{Number of events for the $\ell^\pm j_\tau j_\tau j_\tau$ signals and main backgrounds for a luminosity of 30 fb$^{-1}$.}
\label{tab:nsnb-1l3t}
\end{table}

\subsection{Outlook}
\label{sec:5.6}

In this section we have examined the scalar triplet signals in the case of small vev $\vt$, such that triplet decays are dominated by the leptonic channels. Our approach has been different from recent studies \cite{Perez:2008zc,Perez:2008ha}. 
Instead of classifying signals by the particles produced ({\em e.g.} light charged leptons, taus, neutrinos) we have classified them by the signatures actually seen. We believe that the latter is more adequate because most final states (except the one with four leptons) receive contributions from $\Dpp \Dmm$ and $\Dppmm \Dmp$ production, although these two processes can be separated to some extent with an adequate reconstruction, as the one performed for the
$\ell^\pm \ell^\pm \ell^\mp$ channel.

We have devoted special attention to $\tau$ lepton decays. Indeed, the invariant mass distribution of like-sign dileptons resulting from $\Dppmm$ decays has, in addition to a clear peak from direct $\Dppmm \to \ell^\pm
\ell^\pm$ decays, a broad bump originated when $\Dppmm$ decays into one or two taus, which subsequently decay leptonically. This bump constitutes a ``combinatorial background'', which in the cleanest $\ell^+ \ell^+ \ell^- \ell^-$ and $\ell^\pm \ell^\pm \ell^\mp$ channels is actually larger than
the SM background and decreases the relative height of the peaks.
If the SM trilepton and four lepton backgrounds have to be normalised with data,\footnote{This is a pessimistic hypothesis, but perhaps it will be the case in the first months of LHC running, when a 300 GeV scalar triplet would be discovered.} then this
combinatorial background decreases the observability of the scalar triplet signals. The effect is not very dramatic, however.

Comparing the several channels with two, three and four leptons we have concluded that the trilepton channel is by far the best one for both NH and IH, and could give $5\sigma$ evidence of doubly charged scalars with a luminosity five (for NH) or three (for IH) times  smaller than the one required in the four lepton channel. This is due not only to the additional
contribution of $\Dppmm \Dmp$ to trilepton final states but also to the
larger branching ratio in $\Dpp \Dmm$ production. The like-sign dilepton and four lepton channels have similar sensitivities. The channel with an opposite charge lepton pair and a tagged $\tau$ jet is more difficult, although positive signals can be observed in the tail of the leading lepton momentum distribution.
Channels with only one charged lepton suffer from huge backgrounds and triplet signals are unobservable.

Finally, we have addressed the identification of the scalar nature of the $\Dppmm$. We have examined the opening angle distribution in the three and four lepton channels, finding that the detector effects do not alter significantly the distributions (which can be eventually corrected, anyway) and
they are compatible with the hypotesis of a spin 0 particle. In the
$\ell^\pm \ell^\pm \ell^\mp$ final state, the presence of reconstructed events with large missing energy indicates $\Dppmm \Dmp$ production, thus giving evidence of the non-singlet nature of $\Delta$.

\section{Seesaw III signals}
\label{sec:6}

\begin{figure}[ht]
\begin{center}
\epsfig{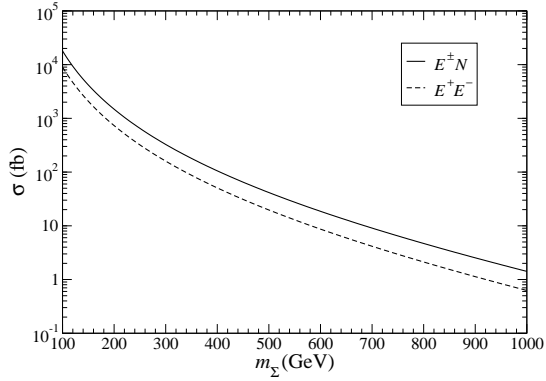}
\caption{Cross section for production of heavy lepton pairs
$E^\pm N$, $E^+ E^-$ at LHC.}
\label{fig:cross-fermion}
\end{center}
\end{figure}

The charged and neutral members of the fermion triplet can be produced in the partonic processes
\begin{align}
& q \bar q \to Z^* \,/\, \gamma^* \to E^+ E^- \,, \notag \\
& q \bar q' \to W^* \to E^\pm N \,.
\label{ec:f-signals}
\end{align}
Neutral lepton pairs are not produced because $N$ has $T_3=0$, $Y=0$ and thus they do not couple to the $Z$ boson.\footnote{Similar analyses can be performed to investigate the discovery prospects for heavy leptons transforming
as electroweak doublets and charged singlets~\cite{delAguila:1989rq}.} The production cross section only depends on the $E$, $N$ masses, since the triplet interactions are fixed by the gauge symmetry. The triplet splitting is expected to be very small, and the mass differences are irrelevant for production. The dependence of the cross sections on the common mass $m_E = m_N \equiv m_\Sigma$ is represented in Fig.~\ref{fig:cross-fermion}.

Except for tiny mixings
$V_{l N} \sim 10^{-8}$,
the new heavy leptons decay almost exclusively to SM leptons plus a gauge or Higgs boson. (For mixings of this size and the mass assumed here $m_{E,N} = 300$ GeV the branching ratios to decays between triplet members, allowed by the small mass splitting, are below 10\%  \cite{Franceschini:2008pz}.) The partial widths for $E$ decays are
\begin{eqnarray}
\Gamma(E^+ \to \bar \nu W^+ ) & = & \frac{g^2}{32 \pi} |V_{lN}|^2
\frac{m_E^3}{M_W^2} \left( 1- \frac{M_W^2}{m_E^2} \right) 
\left( 1 + \frac{M_W^2}{m_E^2} - 2 \frac{M_W^4}{m_E^4} \right) \,, \nonumber
\\[0.1cm]
\Gamma(E^+ \to l^+ Z) & = &  \frac{g^2}{64 \pi c_W^2} |V_{lN}|^2
\frac{m_E^3}{M_Z^2} \left( 1- \frac{M_Z^2}{m_E^2} \right) 
\left( 1 + \frac{M_Z^2}{m_E^2} - 2 \frac{M_Z^4}{m_E^4} \right) \,, \nonumber
\\[0.2cm]
\Gamma(E^+ \to l^+ H) & = &  \frac{g^2}{64 \pi} |V_{lN}|^2
\frac{m_E^3}{M_W^2} \left( 1- \frac{M_H^2}{m_E^2} \right)^2 \,,
\label{ec:Ewidths}
\end{eqnarray}
while the widths for $N$ decays are the same as for the heavy neutrino singlet in Eqs.~(\ref{ec:Nwidths}). The total branching ratios for the $W$, $Z$ and $H$ modes, summing over light charged lepton and neutrino species,
do not depend on the value of the mixings $V_{lN}$ but only on the heavy masses $m_{E,N}$. The partial widths to different flavours are in the ratios
\begin{equation}
\frac{\Gamma(E^+ \to \bar \nu_{l_1} W^+ )}{\Gamma(E^+ \to \bar \nu_{l_2} W^+)} = 
\frac{\Gamma(E^+ \to l_1 Z)}{\Gamma(E^+ \to l_2 Z)} = 
\frac{\Gamma(E^+ \to l_1 H)}{\Gamma(E^+ \to l_2 H)} = 
\frac{|V_{l_1 N}|^2}{|V_{l_2 N}|^2} \,.
\end{equation}
The observability of the new states $E$, $N$ strongly depends on their coupling to the SM leptons. A triplet coupling to the electron and/or muon leads to final states with very energetic electrons and/or muons, which give clean signals. Since at high transverse momenta the backgrounds involving electrons and muons have roughly the same size, the observability of the new signals is similar in these cases. On the other hand, a triplet coupling to the third generation gives $\tau$ leptons as decay products, which are much more difficult to observe. We will then select two scenarios for our analysis. In scenario T1 we assume that $E$ and $N$ only couple to $(e,\nu_e)$, so that their decays give $e$, $\nu_e$ in the final states. As we have argued, the observability in case that $E$, $N$ couple to the second generation as well (or only to it) is similar. Scenario T2 assumes that $E$, $N$ only couple to $(\tau,\nu_\tau)$, and it is the most pessimistic one.

Even with $E$, $N$ coupling to one lepton flavour, there is a plethora of possible final states resulting from $E^+ E^-$ and $E^\pm N$ production, given by the several decay possibilities of each heavy lepton.
They are collected in Table~\ref{tab:trbr}, not including the decay of the $W$, $Z$ and $H$ bosons.
\begin{table}[htb]
\begin{center}
\begin{tabular}{|c|c|c|c|}
\hline
                          & $E^+ \to \bar \nu W^+$ (0.54) & $E^+ \to l^+ Z$ (0.27) & $E^+ \to l^+ H$ (0.19) \\
\hline
$E^- \to \nu W^-$ (0.54)  & $\nu \bar \nu W^+ W^-$ (0.29) & $l^+ \nu Z W^-$ (0.15) & $l^+ \nu H W^-$ (0.10) \\
\hline
$E^- \to l^- Z$ (0.27) & $l^- \bar\nu Z W^+$ (0.15) & $l^+ l^- ZZ$ (0.07) & $l^+ l^- ZH$ (0.05) \\
\hline
$E^- \to l^- H$ (0.19) & $l^- \bar\nu H W^+$ (0.10) & $l^+ l^- ZH$ (0.05) & $l^+ l^- HH$ (0.04) \\
\hline \hline
$N \to l^- W^+$ (0.27) & $l^-\bar\nu W^+ W^+$ (0.14) & $l^+ l^- Z W^+$ (0.07) & $l^+ l^- HW^+$ (0.05) \\
\hline
$N \to l^+ W^-$ (0.27) & $l^+\bar\nu W^+ W^-$ (0.14) & $l^+ l^+ Z W^-$ (0.07) & $l^+ l^+ HW^-$ (0.05) \\
\hline
$N \to \nu Z$ (0.27)      & $\nu \bar \nu Z W^+$ (0.15)    & $l^+ \nu ZZ$ (0.07)       & $l^+ \nu ZH$ (0.05) \\
\hline
$N \to \nu H$ (0.19)      & $\nu \bar \nu H W^+$ (0.10)    & $l^+ \nu ZH$ (0.05)       & $l^+ \nu HH$ (0.04) \\
\hline
\end{tabular}
\end{center}
\caption{Final states for $E^+ E^-$ and $E^+ N$ production, with their approximate branching ratio for $m_{E,N} = 300$ GeV.}
\label{tab:trbr}
\end{table}
These decay channels lead to many final state signatures which can be conveniently classified by their charged lepton multiplicity. A realistic analysis must take into account all possible contributions to a given signal. Indeed, in our analysis we find that for most final states there are several possible competing contributions from different decay chains. Additionally, final states with smaller lepton multiplicity receive contributions from final states with more leptons when one or more are missed by the detector. In our study we consider all decay channels generated by \tria\ and passed through a parton shower Monte Carlo and a detector simulation. The final states studied are: (a) six leptons; (b) five leptons; (c) $\ell^\pm \ell^\pm \ell^\pm \ell^\mp X$; (d) $\ell^+ \ell^+ \ell^- \ell^-X$; (e) $\ell^\pm \ell^\pm \ell^\pm X$; (f) $\ell^\pm \ell^\pm \ell^\mp X$; (g) $\ell^\pm \ell^\pm X$; (h) $\ell^+ \ell^- jjjjX$; (i) $\ell^\pm jjjj X$. As before, the charged leptons are $\ell=e,\mu$, including all flavour combinations. Signals with four like-sign leptons are in principle possible but they are found to be negligible. We take a common mass $m_{E,N} = 300$ GeV, for which the decay branching ratios are
\begin{align}
& \mathrm{Br}(E^+ \to \bar \nu W^+) = 0.537 \,, \nonumber \\
& \mathrm{Br}(E^+ \to l^+ Z) = 0.271 \,, \nonumber \\
& \mathrm{Br}(E^+ \to l^+ H) = 0.192 \,, \nonumber \\
& \mathrm{Br}(N \to l^- W^+) = \mathrm{Br}(N \to l^+ W^-) = 0.269
  \,, \nonumber \\
& \mathrm{Br}(N \to \nu Z) = 0.271 \,, \nonumber \\
& \mathrm{Br}(N \to \nu H) = 0.192 \,.
\end{align}
We give results for both scenarios T1 and T2, summing backgrounds with electrons and muons.


\subsection{Six lepton final states}
\label{sec:6.1}

These final states are the cleanest ones but the signal cross sections are tiny. Six lepton final states can only be produced in the channel
\begin{align}
& E^+ E^-  \to \ell^+ Z \, \ell^- Z \,,
  && \quad Z \to \ell^+ \ell^-
  & (\mathrm{Br} = 3.5 \times 10^{-4})\,.
\label{ec:ch6l}
\end{align}
The overall branching ratio is calculated for
$m_{E} = 300$ GeV and includes $Z$ decay.
For an $E^+ E^-$ cross section of 160 fb, this corresponds to 1.7 events in 30 fb$^{-1}$, further reduced by detection efficiencies.
Minor additional contributions are also present from $Z \to \tau^+ \tau^-$ with the subsequent $\tau$ leptonic decay.

We require for event selection six isolated charged leptons, at least two of them having transverse momentum larger than 30 GeV and all of them with $p_T > 15$ GeV (if they are electrons) and $p_T > 10$ GeV (if they are muons). No SM background events survive these selection criteria,
while 0.6 signal events are found in scenario T1 and 0.1 events in scenario T2.
Therefore, this final state is only relevant for very high integrated luminosities, well above 300 fb$^{-1}$.


\subsection{Five lepton final states}
\label{sec:6.2}

Fermion triplet production can give five lepton final states in several decay channels,
\begin{align}
& E^+ N \to \ell^+ Z \, \ell^\pm W^\mp \,,
  && \quad Z \to \ell^+ \ell^-, W \to \ell \nu
  & (\mathrm{Br} = 2.2 \times 10^{-3}) \,,  \notag \\
& E^+ N \to \ell^+ Z \, \nu Z \,,
  && \quad Z \to \ell^+ \ell^-
  & (\mathrm{Br} = 3.5 \times 10^{-4})\,, 
\label{ec:ch5l}
\end{align}
and similarly for $E^- N$ production. 
Small additional contributions from $W$, $Z$ decay to $\tau$ leptons are also present.
Additionally, five leptons can be produced in
the channel of Eq.~(\ref{ec:ch6l}) when a lepton is missed by the detector.
Clearly, this is a general feature: final states in which a given number of leptons are produced contribute to final states with a smaller lepton multiplicity when one or more of them are missed by the detector.
Five lepton signals have branching ratios 7 times larger than six lepton signals, and so they are expected to be more significant, because they still have
tiny backgrounds.

For event selection we require, analogously to the previous case, five charged leptons, at least two of them having transverse momentum larger than 30 GeV and all of them with $p_T > 15$ GeV (for electrons) and $p_T > 10$ GeV (for muons). The number of signal and background events is collected in Table~\ref{tab:nsnb-5l}.
We do not set any additional selection criteria, since the background is sufficiently small. For the scenario T1, $5\sigma$ discovery can be reached with 28 fb$^{-1}$, while for scenario T2 the signal is too small to be observed even with 300 fb$^{-1}$. It is important to remark that neither neutrino singlet nor scalar triplet production produce five lepton final states, so this mode can signal fermion triplet production, although only for relatively large luminosities.

\begin{table}[ht]
\begin{center}
\begin{tabular}{ccccc}
            & Selection & \quad & & Selection \\[1mm]
$E^+ E^-$ (T1)   & 0.9 & & $t \bar t nj$   & 0.1 \\
$E^\pm N$ (T1)   & 9.7 & & $WZZ nj$        & 0.7 \\
$E^+ E^-$ (T2)   & 0.3 & & $ZZZ nj$        & 0.1 \\
$E^\pm N$ (T2)   & 1.2
\end{tabular}
\end{center}
\caption{Number of events for the five lepton signals and main backgrounds for a luminosity of 30 fb$^{-1}$.}
\label{tab:nsnb-5l}
\end{table}


\subsection{Final state $\ell^\pm \ell^\pm \ell^\pm \ell^\mp$}
\label{sec:6.3}

The $E^\pm N$ production process with decay
\begin{align}
& E^+ N \to \ell^+ Z \, \ell^+ W^- \,,
  && \quad Z \to \ell^+ \ell^-, W \to q \bar q'
  & (\mathrm{Br} = 3.4 \times 10^{-3})\,, 
\label{ec:ch4q2}
\end{align}
(and its charge conjugate)
can produce this interesting final state with three like-sign leptons and a fourth one of opposite sign. An additional contribution approximately five times smaller arises from $N \to \ell^- W^+$ when $\ell^-$ is missed and the $W$ boson decays leptonically. The $\ell^\pm \ell^\pm \ell^\pm \ell^\mp$ final state cannot be produced in seesaw I and II scenarios, so it constitutes a very characteristic signature for fermion triplet production. Additionally, this final state provides a clean measurement of the heavy $E$ mass, as we will show in this subsection.

Signal and background events are selected by requiring four isolated
charged leptons with a total charge of $\pm 2$, two of them with transverse momentum larger than 30 GeV. We do not apply veto cuts on the invariant mass of 
opposite charge lepton pairs, since the signal itself involves $Z$ boson decays. The number of signal and background events fulfilling these requirements is collected in Table~\ref{tab:nsnb-4Q2}. The SM background 
is small enough so as to allow the observation of the signal (in scenario T1) by the analysis of kinematical distributions, and further kinematical cuts are not necessary.

\begin{table}[ht]
\begin{center}
\begin{tabular}{ccccc}
            & Pre-selection & \quad & & Pre-selection \\[1mm]
$E^+ E^-$ (T1)   &  0.7 & & $t \bar t nj$     & 26.7 \\
$E^\pm N$ (T1)   & 24.2 & & $Z b \bar b nj$   & 18.8 \\
$E^+ E^-$ (T2)   &  0.4 & & $Z t \bar t nj$   & 3.5 \\
$E^\pm N$ (T2)   &  4.5
\end{tabular}
\end{center}
\caption{Number of events for the $\ell^\pm \ell^\pm \ell^\pm \ell^\mp$ signals and main backgrounds with a luminosity of 30 fb$^{-1}$.}
\label{tab:nsnb-4Q2}
\end{table}

\begin{figure}[ht]
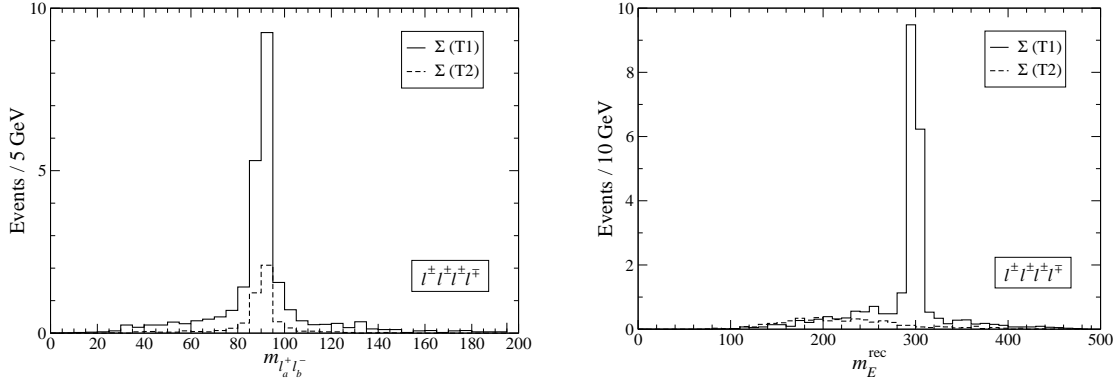

\begin{center}
\begin{tabular}{ccc}
\epsfig{file=Figs/mzrec-4Q2.eps,height=5cm,clip=} & \quad &
\epsfig{file=Figs/mErec-4Q2.eps,height=5cm,clip=}
\end{tabular}
\caption{Left: $\ell_a^+ \ell_b^-$ invariant mass distribution for the fermion triplet signal in the two scenarios T1 and T2. Right: reconstructed $E$ mass distribution (see the text).
The luminosity is 30 fb$^{-1}$.}
\label{fig:4Q2-mzme}
\end{center}
\end{figure}

There are several kinematical distributions in which the presence of the fermion triplet signals may manifest. Among these, the mass reconstruction of the heavy states deserves special attention. We follow a procedure adapted for the channel in Eq.~(\ref{ec:ch4q2}), which gives the main contribution to the signal:
\begin{enumerate}
\item First, the two charged leptons coming from the $Z$ boson decay are identified, selecting among the three possibilities the opposite sign pair $\ell_a^+ \ell_b^-$ which has an invariant mass closest to $M_Z$. As the plot in Fig.~\ref{fig:4Q2-mzme} (left) shows, there are long tails in this distribution caused by signal channels different from that in Eq.~(\ref{ec:ch4q2}). The background has not been included in this plot for clarity.
\item Assuming for the moment that the true heavy charged lepton mass $m_E$ is known, we can determine which leptons are its decay products choosing between the two possibilities $\ell_a^+ \ell_b^- \ell_c$ and $\ell_a^+ \ell_b^- \ell_d$, the one giving an invariant mass closest to $m_E$. The reconstructed $E$ mass $m_E^\text{rec}$ is then the three-lepton invariant mass.
This distribution is presented in Fig.~\ref{fig:4Q2-mzme} (right) for the signal alone, in order to see how the reconstruction procedure works.
\end{enumerate}

\begin{figure}[ht]
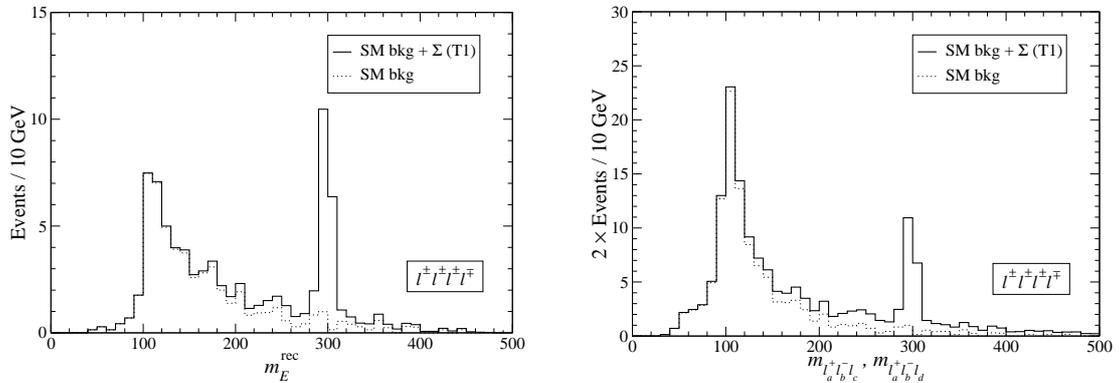

\begin{center}
\begin{tabular}{ccc}
\epsfig{file=Figs/mErec-4Q2-Ts1.eps,height=5cm,clip=} & \quad &
\epsfig{file=Figs/mE2x-4Q2-Ts1.eps,height=5cm,clip=}
\end{tabular}
\caption{Left: $m_E^\text{rec}$ distribution for the SM and the SM plus the fermion triplet signal in scenario T1. Right:  $\ell_a^+ \ell_b^- \ell_c$ + $\ell_a^+ \ell_b^- \ell_d$ distribution (two entries per event). 
The luminosity is 30 fb$^{-1}$.}
\label{fig:4Q2-merec}
\end{center}
\end{figure}

The reconstructed $E$ mass distribution is presented in Fig.~\ref{fig:4Q2-merec} (left) for the SM background and the background plus the heavy triplet signal in scenario T1 (in scenario T2 the peak is unobservable, as shown in Fig.~\ref{fig:4Q2-mzme}). We notice that the background is not significantly biased by the selection of the charged lepton giving $m_E^\text{rec}$ closest to $m_E$.
The input value of $m_E$ necessary for the reconstruction can be found by plotting the invariant mass of $\ell_a^+ \ell_b^-$ with each of the two remaining leptons,
obtaining a plot with two entries per event shown in Fig.~\ref{fig:4Q2-merec} (right). This plot displays a clear peak from which $m_E$ can be determined,
although the low statistics may compromise the determination for low luminosities.
Note that we do not reconstruct the $N$ mass at this stage because we do not require extra jets (which are produced in $N \to \ell W \to \ell q \bar q'$) in event pre-selection to keep the signal as large as possible.

\begin{figure}[ht]
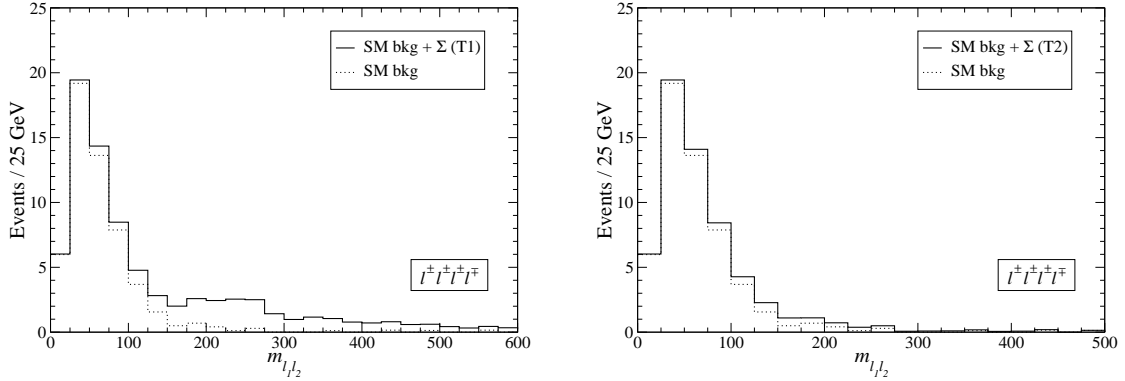

\begin{center}
\begin{tabular}{ccc}
\epsfig{file=Figs/Ml1l2-4Q2-Ts1.eps,height=5cm,clip=} & \quad &
\epsfig{file=Figs/Ml1l2-4Q2-Ts2.eps,height=5cm,clip=}
\end{tabular}
\caption{$\ell_1 \ell_2$ invariant mass distribution for the SM and the SM plus the fermion triplet signal in the scenarios T1 (left) and T2 (right). The luminosity is 30 fb$^{-1}$.}
\label{fig:4Q2-ml12}
\end{center}
\end{figure}

A more adequate variable for low statistics is the mass distribution of the leading and sub-leading like-sign leptons, labelled
as $\ell_1$ and $\ell_2$, respectively. The variable $m_{\ell_1 \ell_2}$ is a very good discriminator, as it can be seen in Fig.~\ref{fig:4Q2-ml12}. Since the signal almost always has two very energetic leptons, their invariant mass is large.
The number of signal (in scenario T1) and background events when cuts are applied on the reconstructed $E$ mass or in $m_{\ell_1 \ell_2}$ are collected in Table~\ref{tab:sig-4Q2}. In the first case, we select a 20 GeV interval around the peak, and in the second we require $m_{\ell_1 \ell_2} > 150$ GeV.
We assume that (a) the SM background is perfectly known, and (b) it is normalised from data (for $m_{\ell_1 \ell_2}$ we use the region $m_{\ell_1 \ell_2} < 150$). The luminosity needed for $5\sigma$ discovery is also given.
For the scenario T2, the event selection done here is not sufficient to discover a signal with a luminosity of 300 fb$^{-1}$. Moreover, the small size of this signal and its distribution across all the $m_E^\text{rec}$ and $m_{\ell_1 \ell_2}$ range implies that the systematic uncertainty in background normalisation has to be investigated in detail in order to draw a definite prediction for the sensitivity.

\begin{table}[ht]
\begin{center}
\begin{tabular}{lcccccc}
& \multicolumn{3}{c}{Case (a)} & \multicolumn{3}{c}{Case (b)} \\
  & $S$   & $B$ & $L$ & $S$   & $B$ & $L$            \\[1mm]
T1 (cut on $m_E^\text{rec}$)
  & 17.4  & 2.5 & 17.4 fb$^{-1}$ & 16.6 & 3.3 & 18.3 fb$^{-1}$  \\
T1 (cut on $m_{\ell_1 \ell_2}$)
  & 20.9  & 2.6 & 14.4 fb$^{-1}$ & 20.8 & 2.7 & 14.7 fb$^{-1}$  
\end{tabular}
\end{center}
\caption{Number of signal ($S$) and background ($B$) events for 30 fb$^{-1}$
in the regions 
$280 < m_E^\text{rec} < 320$ GeV (upper row) and
 $m_{\ell_1 \ell_2} > 150$ GeV (lower row), and luminosity $L$ required to have a $5\sigma$ discovery in the $\ell^\pm \ell^\pm \ell^\pm \ell^\mp$ final state.}
\label{tab:sig-4Q2}
\end{table}

\begin{figure}[ht]
\begin{center}
\epsfig{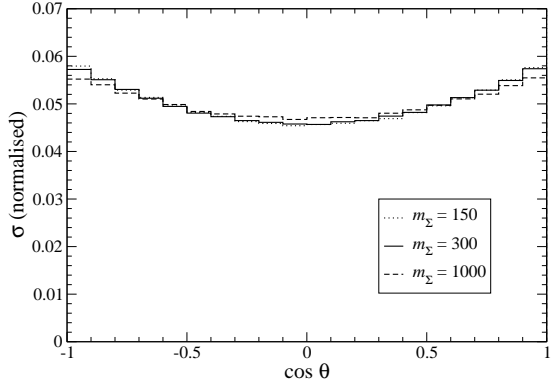} 
\caption{Dependence of the $E^\pm N$ cross section on the CM polar angle $\theta$, at the generator level, for different values of $m_\Sigma$.}
\label{fig:open-4Q2-gen}
\end{center}
\end{figure}

Once that a positive signal is discovered, it can be
investigated whether the opening angle distribution corresponds to the expectation for $E^\pm N$ production. For this process,
the dependence of the partonic cross section
on the polar angle $\theta$ in the CM system is given by
\begin{equation} 
\frac{d\sigma}{d \cos \theta} \propto 1 +
\frac{E^2-m_\Sigma^2}{E^2+m_\Sigma^2} \cos^2 \theta \,,
\end{equation}
where $E=\sqrt{\hat s}/2$, with $\hat s$ the partonic CM energy. After weighting with phase space factors and PDFs, the resulting distribution depends little on the common mass $m_\Sigma$, as it is shown in
Fig.~\ref{fig:open-4Q2-gen}. The distribution
approximately corresponds to a dependence
$d\sigma \propto 1+0.28 \cos^2 \theta$, rather flat compared to the
one for massless final state fermions,
$d\sigma \propto 1+ \cos^2 \theta$.

\begin{figure}[htb]
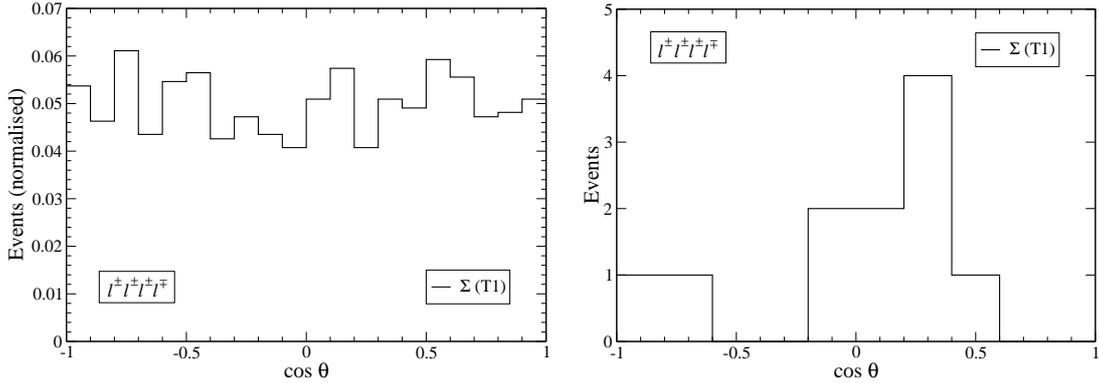

\begin{center}
\begin{tabular}{ccc}
\epsfig{file=Figs/cProd-4Q2-T1.eps,height=5cm,clip=} &
\epsfig{file=Figs/cProd-4Q2-T1X.eps,height=5cm,clip=}
\end{tabular}
\caption{Left: Normalised $E^\pm$ opening angle distribution for the
$\ell^\pm \ell^\pm \ell^\pm \ell^\mp$ signal in scenario T1. Right: Possible experimental result for a luminosity of 30 fb$^{-1}$ (with 11 events).}
\label{fig:open-4Q2}
\end{center}
\end{figure}

In the simulations we require for the reconstruction of the CM system the presence of two jets, as produced in the channel which gives the largest contribution to the signal,
$E^+ N \to \ell^+ Z (\to \ell^+ \ell^-) \ell^\pm W^\mp (\to q \bar q')$. Both jets must have transverse momentum larger than 20 GeV. The reconstruction of the $E$ momentum proceeds as before, identifying the three leptons resulting from the $E$ decay. The heavy neutrino is reconstructed from the two jets and the remaining lepton:
\begin{enumerate}
\item First, the $W$ boson momentum is reconstructed using the two jets with larger transverse momentum, and rescaled so that their invariant mass is $M_W$.
\item Then, the momentum of the remaining charged lepton is summed, to form the $N$ momentum.
\end{enumerate}
The quality of the reconstruction is ensured by requiring that $m_E^\text{rec}$ lies between 280 and 320 GeV, after which 10.4 signal events survive.
Cuts are not applied on $m_N^\text{rec}$ in order to keep the signal as large as possible.
The opening angle distribution after event selection and reconstruction for the scenario T1 is shown in
Fig.~\ref{fig:open-4Q2} (left). The result seems compatible with the theoretical expectation even without using any correction function to parameterise the detector effects, although the Monte Carlo statistics is insufficient even with 3000 fb$^{-1}$ generated. Clearly, this distribution will be useful only for very large integrated luminosities. In
Fig.~\ref{fig:open-4Q2} (right), we present a possible experimental result for 30 fb$^{-1}$, generated from the distribution in the left side of this figure, and corresponding to 11 signal events. This hypothetical ``experimental result'' has little resemblance with the theoretical expectation due precisely to the small number of events.

\subsection{Final state $\ell^+ \ell^+ \ell^- \ell^-$}
\label{sec:6.4}

In contrast with the previous $Q=\pm 2$ four lepton final state, the $\ell^+ \ell^+ \ell^- \ell^-$ signal is common to scalar and fermion triplet production. In the latter case, four leptons can result from many decay channels,
namely
\begin{align}
& E^+ E^- \to \ell^+ Z \, \ell^- Z  \,,
  && \quad ZZ \to \ell^+ \ell^- q \bar q / \nu \bar \nu
  & (\mathrm{Br} = 9.1 \times 10^{-3}) \,, \nonumber \\
& E^+ E^- \to \ell^+ Z \, \ell^- H / \ell^+ H \, \ell^- Z \,,
  && \quad Z \to \ell^+ \ell^- , H \to q \bar q 
  & (\mathrm{Br} = 6.9 \times 10^{-3}) \,, \nonumber \\
& E^+ E^- \to \nu W^+ \ell^- Z / \ell^+ Z \nu W^- \,,
  && \quad Z \to \ell^+ \ell^- , W \to \ell \nu 
  & (\mathrm{Br} = 4.5 \times 10^{-3}) \,, \nonumber \\
& E^\pm N \to \ell^\pm Z \, \ell^- W^+ \,,
  && \quad Z \to \ell^+ \ell^- , W \to q \bar q' 
  & (\mathrm{Br} = 3.4 \times 10^{-3}) \,.
\label{ec:ch4q0}
\end{align}
Additional channels with $\tau$ leptons, or more charged leptons which are missed, also contribute. This final state is crucial in order to establish the production of the heavy charged lepton $E$, which is seen as a sharp peak in a trilepton invariant mass distribution. This distribution also can be used to experimentally measure $m_E$.

In order to better compare with the scalar triplet production in section \ref{sec:5.1} we use the same pre-selection and selection criteria: for pre-selection we require four isolated charged leptons with total charge $Q=0$, at least two of them with transverse momentum larger than 30 GeV, and
for event selection we ask that opposite charge leptons cannot be paired in such a way that both pairs have a mass closer to $M_Z$ than 5 GeV. The number of events at pre-selection and selection levels are collected in Table~\ref{tab:nsnb-4lepT} (backgrounds are the same as in Table~\ref{tab:nsnb-4lep} but quoted for convenience).

\begin{table}[ht]
\begin{center}
\begin{tabular}{ccccccc}
                 & Pre-selection & Selection & \quad & & Pre-selection & Selection \\[1mm]
$E^+ E^-$ (T1)   & 43.6 & 42.9 & & $t \bar t nj$     & 116.0 & 115.7 \\
$E^\pm N$ (T1)   & 27.3 & 26.8 & & $Z b \bar b nj$   & 53.1  & 53.1 \\
$E^+ E^-$ (T2)   & 8.2  &  7.7 & & $Z t \bar t nj$   & 32.9  & 31.5 \\
$E^\pm N$ (T2)   & 8.6  &  7.6 & & $ZZ nj$           & 617.7 & 98.7
\end{tabular}
\end{center}
\caption{Number of events for the four-lepton signals and main backgrounds for a luminosity of 30 fb$^{-1}$.}
\label{tab:nsnb-4lepT}
\end{table}

\begin{figure}[h]
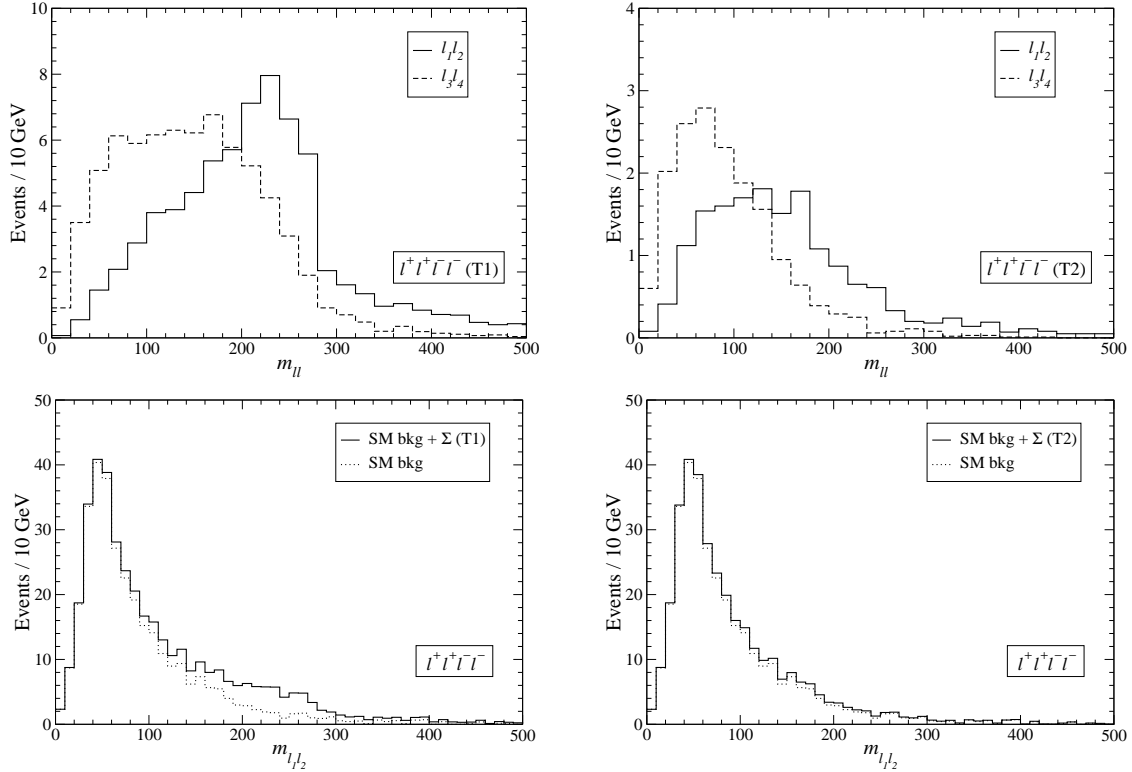

\begin{center}
\begin{tabular}{ccc}
\epsfig{file=Figs/Mll-4lep-T1.eps,height=5cm,clip=} & \quad &
\epsfig{file=Figs/Mll-4lep-T2.eps,height=5cm,clip=} \\
\epsfig{file=Figs/Ml1l2-4lep-Ts1.eps,height=5cm,clip=} & \quad &
\epsfig{file=Figs/Ml1l2-4lep-Ts2.eps,height=5cm,clip=}
\end{tabular}
\caption{Up: Kinematical distribution at pre-selection of the two like-sign dilepton invariant masses $m_{\ell_1 \ell_2}$ and $m_{\ell_3 \ell_4}$ for the fermion triplet signal in scenarios T1 (left) and T2 (right). Down: $m_{\ell_1 \ell_2}$ distribution for the SM and the SM plus the fermion triplet signals in the scenarios T1 (left) and T2 (right). The luminosity in all plots is 30 fb$^{-1}$.}
\label{fig:4lepT-mll}
\end{center}
\end{figure}

The processes in Eqs.~(\ref{ec:ch4q0}) originate broad like-sign dilepton
invariant mass distributions, shown in Fig.~\ref{fig:4lepT-mll} at the pre-selection level. The distributions are much wider and completely different from the peaks found for the scalar triplet, plotted in Fig.~\ref{fig:4lep-mrec} of section~\ref{sec:5.1}, and the discrimination between both possibilities should be possible already with a small number of events. To see this,
in Fig.~\ref{fig:4lepT-mll} we plot the $m_{\ell_1 \ell_2}$ distribution
for the SM background and in the presence of the fermion triplet signals in scenarios T1 (left) and T2 (right), after event selection. 

\begin{figure}[ht]
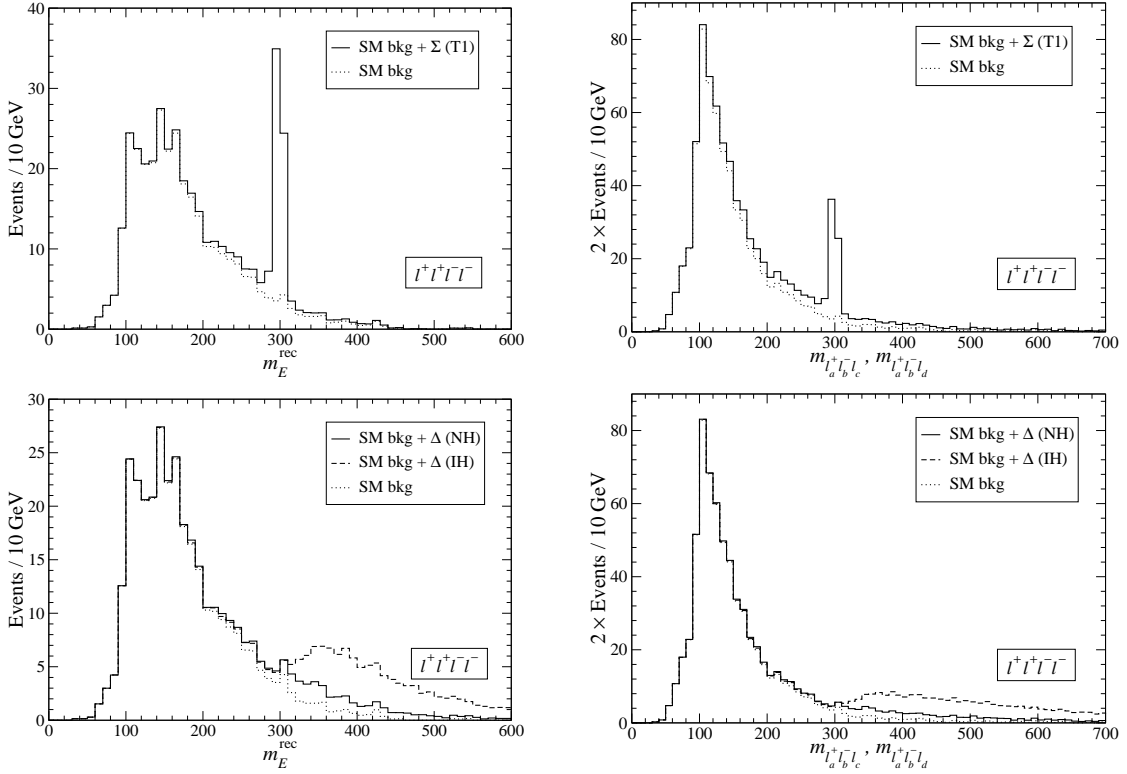

\begin{center}
\begin{tabular}{ccc}
\epsfig{file=Figs/mErec-4lep-Ts1.eps,height=5cm,clip=} & \quad &
\epsfig{file=Figs/mE2x-4lep-Ts1.eps,height=5cm,clip=} \\
\epsfig{file=Figs/mErec-4lep-S12.eps,height=5cm,clip=} & \quad &
\epsfig{file=Figs/mE2x-4lep-S12.eps,height=5cm,clip=} 
\end{tabular}
\caption{Up: $m_E^\text{rec}$ distribution for the SM and the SM plus the fermion triplet signal in scenario T1 (left) and $\ell_a^+ \ell_b^- \ell_c$ + $\ell_a^+ \ell_b^- \ell_d$ invariant mass distribution (right). Down: the same for the scalar triplet
$\Delta$, for NH and IH. The luminosity in all cases is 30 fb$^{-1}$.}
\label{fig:4lepT-m3l}
\end{center}
\end{figure}

The heavy $E$ mass can be reconstructed in a way analogous to that for the
previous $\ell^\pm \ell^\pm \ell^\pm \ell^\mp$ final state. Notice that all channels in Eqs.~(\ref{ec:ch4q0}) involve the decay $E \to \ell Z \to
\ell \ell^+ \ell^-$. The $Z$ boson can be identified selecting among the four possibilities the opposite sign pair $\ell_a^+ \ell_b^-$ with an invariant mass closest to $M_Z$. Then, the charged lepton produced in the $E$ decay is chosen among the remaining ones $\ell_c$, $\ell_d$ as the one giving a three-lepton invariant mass closest to $m_E$. The resulting signal distributions are very similar to the ones in Fig.~\ref{fig:4Q2-mzme}, and are not shown for brevity. The $m_E^\text{rec}$ and $\ell_a^+ \ell_b^- \ell_c$ + $\ell_a^+ \ell_b^- \ell_d$ distributions for scenario T1 (the latter with two entries per event) are shown  in Fig.~\ref{fig:4lepT-m3l}, including the background. For scenario T2 the number of events is much smaller (see Table~\ref{tab:nsnb-4lepT}) and widely distributed (see Fig.~\ref{fig:4Q2-mzme}), so the signal is not visible in this channel.
For comparison, in the lower half of Fig.~\ref{fig:4lepT-m3l} we show the same variables for scalar triplet production in NH and IH. Clearly, this distribution together with the $m_{\ell_1 \ell_2}$ distribution can serve to discriminate among the two seesaw scenarios. The number of events at the peak
\begin{equation}
280 < m_E^\text{rec} < 320~\text{GeV}
\end{equation}
is collected in Table~\ref{tab:sig-4lepT} for the cases (a) and (b) used in the previous analysis, with the luminosity needed for discovery in scenario T1. For scenario T2 the signal significance is smaller than $1\sigma$ for 30 fb$^{-1}$, and discovery cannot be accomplished unless very large luminosities are collected and a very precise background normalisation is achieved.

\begin{table}[ht]
\begin{center}
\begin{tabular}{ccccccc}
& \multicolumn{3}{c}{Case (a)} & \multicolumn{3}{c}{Case (b)} \\
   & $S$   & $B$ & $L$ & $S$   & $B$ & $L$            \\[1mm]
T1 & 55.7  & 14.3 & 6 fb$^{-1}$  & 53.7 & 16.3 & 6.9 fb$^{-1}$
\end{tabular}
\end{center}
\caption{Number of signal ($S$) and background ($B$) events at the
$m_E^\text{rec}$ peak for 30 fb$^{-1}$, and luminosity $L$ required to have a $5\sigma$ discovery in the $\ell^+ \ell^+ \ell^- \ell^-$ final state.}
\label{tab:sig-4lepT}
\end{table}

\begin{figure}[htb]
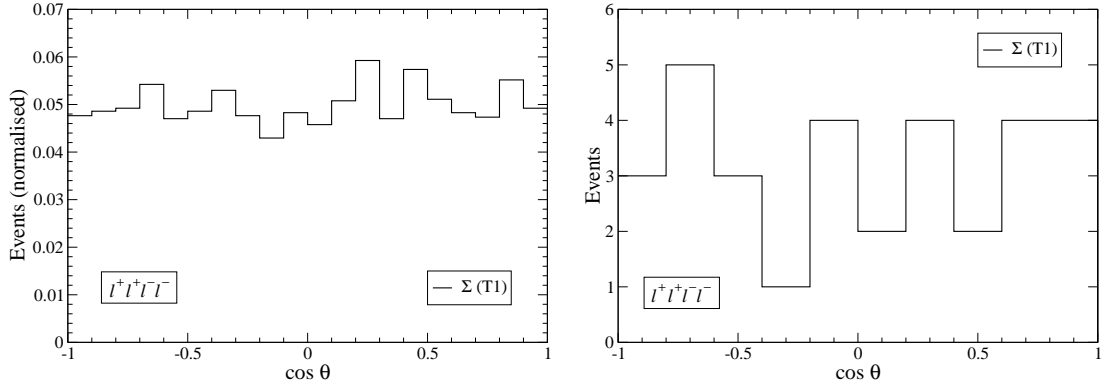

\begin{center}
\begin{tabular}{ccc}
\epsfig{file=Figs/cProd-4lep-T1.eps,height=5cm,clip=} &
\epsfig{file=Figs/cProd-4lep-T1X.eps,height=5cm,clip=}
\end{tabular}
\caption{Left: Normalised $E^\pm$ opening angle distribution for the
$\ell^+ \ell^+ \ell^- \ell^-$ signal in scenario T1. Right: Possible experimental result for a luminosity of 30 fb$^{-1}$ (with 32 events).}
\label{fig:open-4lepT}
\end{center}
\end{figure}

Finally, in this decay channel the full reconstruction of the final state kinematics can be done as in the previous subsection, requiring two jets with $p_T > 20$ GeV.
We point out that the distribution is the same for $E^+ E^-$ and $E^\pm N$ production. The result is shown in Fig.~\ref{fig:open-4lepT}. Although the distribution after detector simulation and reconstruction is flatter than in the $\ell^\pm \ell^\pm \ell^\pm \ell^\mp$ final state the statistics is three times better, and with an adequate parameterisation of the detector effects this channel may be more useful to determine the opening angle distribution.

\subsection{Final state $\ell^\pm \ell^\pm \ell^\pm$}
\label{sec:6.5}

This conspicuous final state is produced when one or several charged leptons are missed by the detector, or from a decay $Z \to \tau^+ \tau^-$ with one $\tau$ decaying hadronically and the other one leptonically. Both the signal and its SM backgrounds are small. Pre-selection criteria are analogous to the other channels studied, and in this case they involve the presence of three like-sign charged leptons, two of them with $p_T > 30$ GeV. The number of like-sign trilepton signal and background events at pre-selection can be read in Table~\ref{tab:nsnb-3Q3}. Backgrounds can be further suppressed by raising the $p_T$ threshold of the leading and sub-leading leptons to 50 GeV, which reduces the total background to 1.9 events.  Since in this case the discovery potential mainly depends on the size of the signal itself,
we can neglect the SM background normalisation uncertainty. Then, the signal in scenario T1 has $5\sigma$ significance for a luminosity of 30 fb$^{-1}$, while in scenario T2 the statistical significance is of only $1.5\sigma$.

\begin{table}[ht]
\begin{center}
\begin{tabular}{ccccccc}
                 & Pre-selection & Selection & \quad & & Pre-selection & Selection \\[1mm]
$E^+ E^-$ (T1)   & 0.1  & 0.0  & & $t \bar t nj$     & 5.3 & 0.1 \\
$E^\pm N$ (T1)   & 11.0 & 10.2 & & $W b \bar b nj$   & 0.4 & 0.0 \\
$E^+ E^-$ (T2)   & 0.1  & 0.0  & & $W t \bar t nj$   & 3.6 & 1.8 \\
$E^\pm N$ (T2)   & 1.6  & 0.9  & & $Z t \bar t nj$   & 0.3 & 0.0 
\end{tabular}
\end{center}
\caption{Number of events for the like-sign trilepton signals and main backgrounds with a luminosity of 30 fb$^{-1}$.}
\label{tab:nsnb-3Q3}
\end{table}


\subsection{Final state $\ell^\pm \ell^\pm \ell^\mp$}
\label{sec:6.6}

This is an excellent final state for the discovery of
fermion triplets (as it is also for scalar triplets), due to the relatively high signal rate and the small background. It provides the same signal significance as the like-sign dilepton channel studied in the next section without the need of event reconstruction, and has the advantage that it serves to establish the production of the heavy neutrino $N$, which is seen as a peak in the invariant mass distribution of two opposite charge leptons plus the missing momentum.

The $\ell^\pm \ell^\pm \ell^\mp$ signal receives contributions from many $E$, $N$ decay channels. Final states with three leptons include
\begin{align}
& E^+ N \to \ell^+ Z \, \ell^\pm W^\mp \,,
  && \quad Z \to q \bar q / \nu \bar \nu, W \to \ell \nu 
  & (\mathrm{Br} = 2.8 \times 10^{-2}) \,, \nonumber \\
& E^+ N \to \ell^+ H \, \ell^\pm W^\mp \,,
  && \quad H \to q \bar q , W \to \ell \nu 
  & (\mathrm{Br} = 2.2 \times 10^{-2}) \,, \nonumber \\
& E^+ N \to \bar \nu W^+ \, \ell^\pm W^\mp \,,
  && \quad W \to \ell \nu 
  & (\mathrm{Br} = 7.2 \times 10^{-3}) \,,
\label{ec:ch3l}
\end{align}
with similar channels for $E^- N$, and additional decay modes where two of the charged leptons result from decays $Z \to \ell^+ \ell^-$. The latter are less important, because they have smaller branching ratios and are suppressed by the selection criteria used for this final state. On the other hand, important additional contributions arise from decay modes with additional charged leptons missed by the detector, in particular from $E^+ E^-$ production. Despite the lower charged lepton multiplicity, this three lepton signal is quite clean and has a cross section much larger than those with more leptons.

In order to compare the scalar and fermion triplet signals, we use the same pre-selection and selection criteria as in section~\ref{sec:5.2}: for pre-selection we demand the presence of two like-sign leptons $\ell_1$ and $\ell_2$ with transverse momentum larger than 30 GeV and an additional lepton of opposite sign, and for selection we ask that neither of the two opposite-sign lepton pairs has an invariant mass closer to $M_Z$ than 10 GeV. This cut reduces the signal by about one fifth, because some of the decay channels involve two charged leptons from $Z$ decays, but it is crucial in order to remove the large background from $WZnj$ production, as it is clearly seen in Table~\ref{tab:nsnb-3lepT}, where 
the number of events for each process at the two stages of event selection is collected. 
\begin{table}[ht]
\begin{center}
\begin{tabular}{cccccccc}
            & Pre-selection & Selection & \quad & & Pre-selection & Selection \\[1mm]
$E^+ E^-$ (T1)  & 62.7   & 21.2  & & $Z b \bar b nj$    & 33.3   & 2.0  \\
$E^\pm N$ (T1)  & 406.2  & 298.5 & & $Z t \bar t nj$    & 152.5  & 16.8 \\
$E^+ E^-$ (T2)  & 26.5   & 4.5   & & $W Z nj$           & 4113.8 & 73.4 \\
$E^\pm N$ (T2)  & 78.1   & 34.4  & & $Z Z nj$           & 276.1  & 4.2 \\
$t \bar t nj$   & 322.8  & 212.2 & & $WWW nj$           & 22.7   & 16.8\\
$tW$            & 17.8   & 12.2  & & $WWZ nj$           & 42.7   & 1.7  \\
$W t \bar t nj$ & 45.5   & 35.1 \\
\end{tabular}
\end{center}
\caption{Number of events for $\ell^\pm \ell^\pm \ell^\mp$ signals and main backgrounds for a luminosity of 30 fb$^{-1}$.}
\label{tab:nsnb-3lepT}
\end{table}

\begin{figure}[htb]
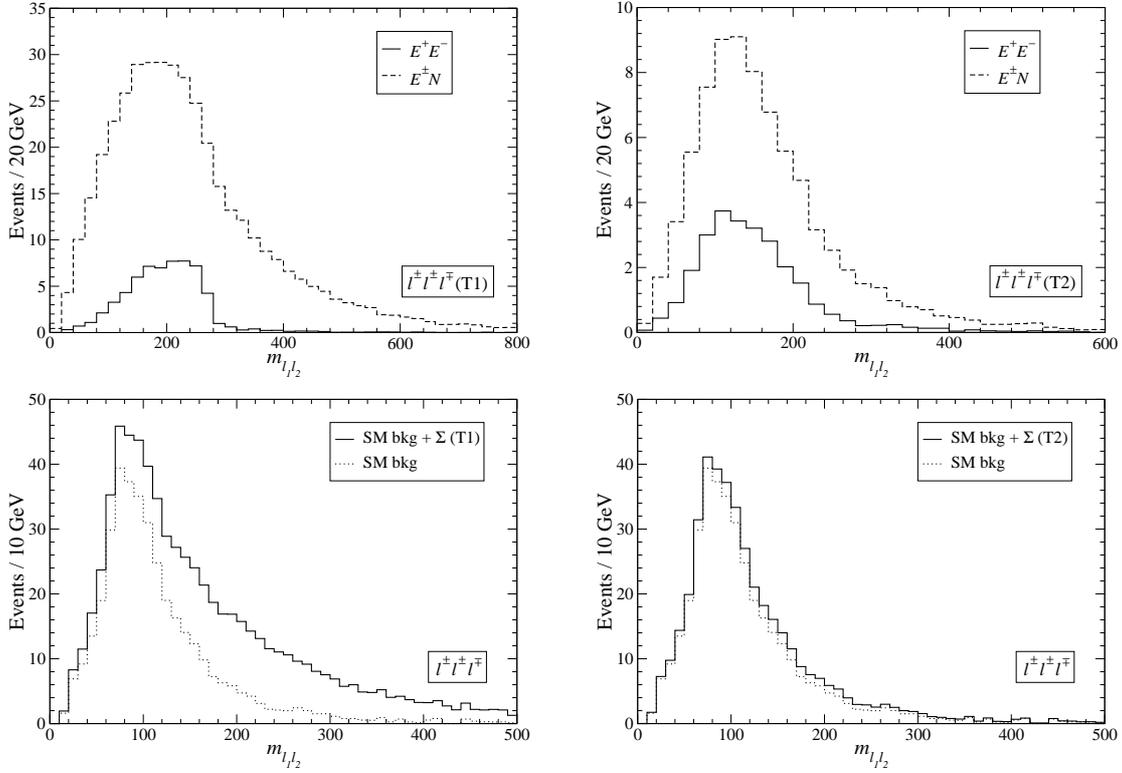

\begin{center}
\begin{tabular}{ccc}
\epsfig{file=Figs/Mll-3lep-T1.eps,height=5cm,clip=} & \quad &
\epsfig{file=Figs/Mll-3lep-T2.eps,height=5cm,clip=} \\
\epsfig{file=Figs/Ml1l2-3lep-T1.eps,height=5cm,clip=} & \quad &
\epsfig{file=Figs/Ml1l2-3lep-T2.eps,height=5cm,clip=}
\end{tabular}
\caption{Up: Kinematical distribution at pre-selection of the like-sign dilepton invariant mass $m_{\ell_1 \ell_2}$ for the $E^+ E^-$ and $E^\pm N$ signals in scenarios T1 (left) and T2 (right). Down: $\ell_1 \ell_2$ invariant mass distribution for the SM and the SM plus the fermion triplet signal in the scenarios T1 (left) and T2 (right), at the selection level. The luminosity is 30 fb$^{-1}$ in all cases.}
\label{fig:3lep-mrecT}
\end{center}
\end{figure}

\begin{figure}[ht]
\begin{center}
\epsfig{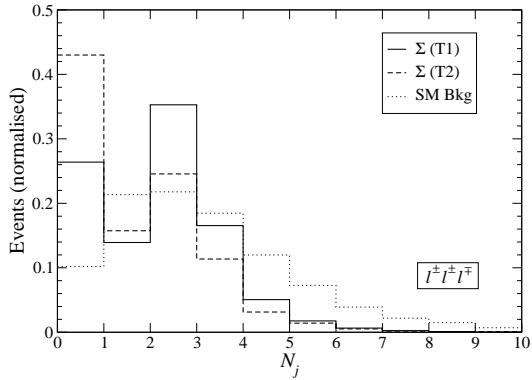} 
\caption{Jet multiplicity of the SM background and the fermion triplet signals in $\ell^\pm \ell^\pm \ell^\mp$ final states at selection level.}
\label{fig:3lep-mult}
\end{center}
\end{figure}

The simplest discriminating variable between the scalar and fermion triplet signals is the like-sign dilepton invariant mass $m_{\ell_1 \ell_2}$. In Fig.~\ref{fig:3lep-mrec2} of section \ref{sec:5.2} we observed that the scalar triplet signals produce a sharp peak in this distribution, while for the fermion triplet signals the distribution is very broad and has a long tail at large $m_{\ell_1 \ell_2}$, as shown in Fig.~\ref{fig:3lep-mrecT}.
For the production of a relatively light neutrino singlet the excess of events is broad, concentrating at intermediate $m_{\ell_1 \ell_2}$ values and without a tail.
Thus, an excess of $\ell^\pm \ell^\pm \ell^\mp$ events with a broad $m_{\ell_1 \ell_2}$ distribution and a long tail points towards $E^+ E^-$ and/or $E^\pm N$ production, a fact which can be confirmed by the event reconstruction performed below.
If the trilepton backgrounds can be accurately predicted, the analysis of this distribution can already signal the presence of new physics with a high significance. We set the cut
\begin{equation}
m_{\ell_1 \ell_2} > 150~\text{GeV}
\end{equation}
and give in Table~\ref{tab:sig-3lepT1} the number of signal and background events in this region, with the luminosity needed to achieve $5\sigma$ significance if the background uncertainty is neglected. Since in the $m_{\ell_1 \ell_2}$ distribution the signal does not appear as a clear peak, it is difficult to estimate a priori to what extent the background can be normalised from data in case that a signal is found. We will assume anyway that the background is normalised in the region $m_{\ell_1 \ell_2} < 150$ GeV to give an approximate estimate of the discovery limit. In scenario T1 the triplet signal can be quicky seen with few fb$^{-1}$, while in scenario T2 the size of the signal is much smaller. However, the signal significance in scenario T2 is better in this final state than in the other ones studied, and this signal
could be seen with sufficient luminosity provided that the background is well understood.
\begin{table}[ht]
\begin{center}
\begin{tabular}{ccccccc}
& \multicolumn{3}{c}{Case (a)} & \multicolumn{3}{c}{Case (b)} \\
   & $S$   & $B$ & $L$ & $S$   & $B$ & $L$            \\[1mm]
T1 (cut on $m_{\ell_1 \ell_2}$) 
  & 228.6  & 80.1 & 1.7 fb$^{-1}$  & 204.6 & 104.1 & 2.5 fb$^{-1}$ \\
T2 (cut on $m_{\ell_1 \ell_2}$) 
  & 18.1  & 80.1 & 186 fb$^{-1}$  & 12.5 & 85.7 & $>300$ fb$^{-1}$
\end{tabular}
\end{center}
\caption{Number of signal ($S$) and background ($B$) events with $m_{\ell_1 \ell_2} > 150$ GeV for 30 fb$^{-1}$, and luminosity $L$ required to have a $5\sigma$ discovery in the $\ell^\pm \ell^\pm \ell^\mp$ final state, without event reconstruction.}
\label{tab:sig-3lepT1}
\end{table}
We also point out that requiring one or more jets in the final state does not improve the signal observability. In Eqs.~(\ref{ec:ch3l}) we observe that several decay channels do not involve the production of extra quarks in the final state, so that requiring the presence of extra jets reduces the signal. This is easily observed in the multiplicity distribution, shown in Fig.~\ref{fig:3lep-mult} for the signals and the backround. Note that the dip at $N_j = 1$ in the signal distributions is not a statistical fluctuation but the effect
of summing different channels with different quark multiplicities in the final state.

\begin{figure}[htb]
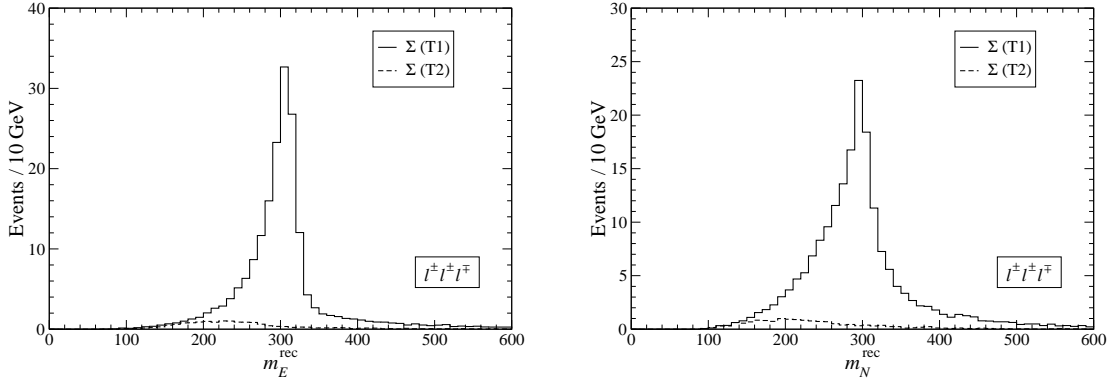

\begin{center}
\begin{tabular}{ccc}
\epsfig{file=Figs/mErec-3lep-T12.eps,height=5cm,clip=} & \quad &
\epsfig{file=Figs/mNrec-3lep-T12.eps,height=5cm,clip=}
\end{tabular}
\caption{Reconstructed $E$ (left) and $N$ (right) masses for the fermion triplet signals in scenarios T1 and T2, without background,
in the $\ell^\pm \ell^\pm \ell^\mp$ final state. The luminosity is 30 fb$^{-1}$.}
\label{fig:3lepT-m3rec}
\end{center}
\end{figure}

Although the trilepton excess at high $m_{\ell_1 \ell_2}$ indicates the presence of new physics, its source must be identified. The confirmation that the excess is due to $E^\pm N$ production is given by the mass reconstruction of the heavy charged lepton $E$ and the neutrino $N$.
In order to fully reconstruct the event kinematics we must specify further selection criteria to restrict ourselves to some of the channels in Eqs.~(\ref{ec:ch3l}). We ask the presence of at least two jets with transverse momentum larger than 20 GeV, as produced in the first two channels listed with hadronic $Z/H$ decay. This additional requirement reduces the signals to 190.8 events (T1) and 16.1 events (T2), and the background to 258.0 events. With three leptons and two jets identified, plus missing energy, the kinematics can be reconstructed as follows:
\begin{enumerate}
\item The momentum of the $Z$ or $H$ boson decaying hadronically is reconstructed as the sum of the momenta of the leading and sub-leading jets.
\item The heavy charged lepton $E$ can be reconstructed from this boson and one of the two like-sign leptons, and the heavy neutrino $N$ from the two remaining leptons and the missing neutrino momentum (the longitudinal component of the neutrino momentum is neglected for the moment, and the transverse component is taken as the missing energy). There are two possibilities for this pairing, and we choose the one giving closest invariant masses for the reconstructed $E$ and $N$.
In scenario T1 this selection procedure gives the correct choice in most of the cases, as it can be observed in the kinematical distribution of the reconstructed $E$ mass in Fig.~\ref{fig:3lepT-m3rec}. In scenario T2, due to the missing energy from the $\tau$ leptonic decay, the distribution does not exhibit a peak.
\item The $N$ reconstruction can be refined by including the longitudinal neutrino momentum. We select among the two charged leptons the least energetic one $\ell_\text{s}$, and require that its invariant mass with the neutrino is $M_W$,
\begin{equation}
(p_{\ell_\text{s}} + p_\nu)^2 = M_W^2 \,,
\label{ec:pnurec3l}
\end{equation}
taking the transverse components of $p_\nu$ as the missing energy.
This quadratic equation determines the longitudinal neutrino momentum up to a twofold ambiguity, which is resolved selecting the solution with smaller $(p_\nu)_z$. In case that no real solution exists, the transverse neutrino momentum 
used in Eq.~(\ref{ec:pnurec3l}) is decreased until a real solution is found.
The resulting $N$ invariant mass distribution is presented in Fig.~\ref{fig:3lepT-m3rec}.
\end{enumerate}

\begin{figure}[htb]
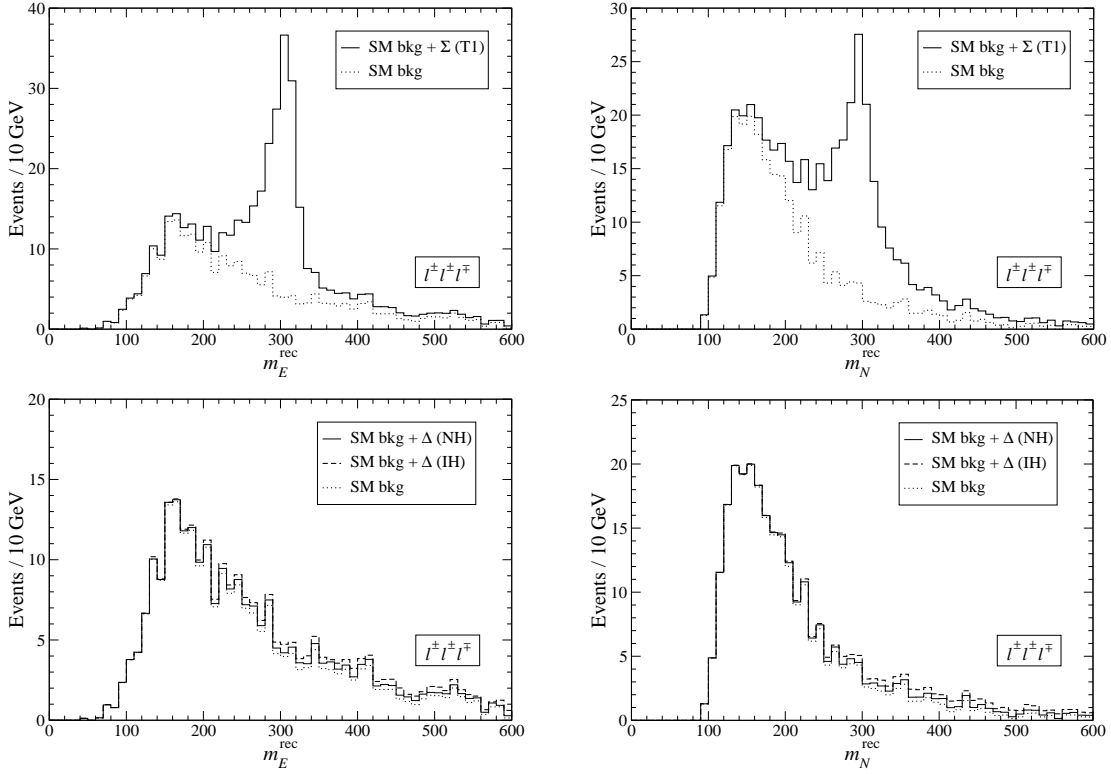

\begin{center}
\begin{tabular}{ccc}
\epsfig{file=Figs/mErec-3lep-bkgT1.eps,height=5cm,clip=} & \quad &
\epsfig{file=Figs/mNrec-3lep-bkgT1.eps,height=5cm,clip=} \\
\epsfig{file=Figs/mErec-3lep-bkgS12.eps,height=5cm,clip=} & \quad &
\epsfig{file=Figs/mNrec-3lep-bkgS12.eps,height=5cm,clip=}
\end{tabular}
\caption{Up: $m_E^\text{rec}$ (left) and $m_N^\text{rec}$ (right) distributions for the SM and the SM plus the fermion triplet signal in scenario T1. Down: the same for the scalar triplet
$\Delta$, for NH and IH. The luminosity is 30 fb$^{-1}$ in all cases.}
\label{fig:3lepT-m3rec2}
\end{center}
\end{figure}

Fig.~\ref{fig:3lepT-m3rec2} displays how the fermon triplet signal in scenario T1 would appear in the presence of background: the signal produces clear peaks over the background in both distributions. We do not include the corresponding distributions for scenario T2, where the signal is much smaller and difficult to see.
For comparison, we also show in Fig.~\ref{fig:3lepT-m3rec2} the distributions corresponding to scalar triplet production. In the latter case the signals are much smaller after the requirement of two extra jets, which are not produced at the partonic level.
The statistical significance of the peaks is very high, as it can readily be observed from the corresponding distributions. We define the peaks as
\begin{align}
240 < m_E^\text{rec} < 360~\text{GeV} \,, \notag \\
240 < m_N^\text{rec} < 360~\text{GeV} \,.
\end{align}
The number of events at both peaks can be found in Table~\ref{tab:sig-3lepT2},
and the luminosity needed for $5\sigma$ discovery. We neglect the background uncertainty, because the discovery significance in this case is mainly controlled by the size of the signal itself.

\begin{table}[ht]
\begin{center}
\begin{tabular}{ccccccc}
& \multicolumn{3}{c}{Case (a)} \\
   & $S$   & $B$ & $L$      \\[1mm]
T1 (cut on $m_E^\text{rec}$, $m_N^\text{rec}$) 
  & 110.3  & 15.9 & 2.7 fb$^{-1}$ \\
\end{tabular}
\end{center}
\caption{Number of signal ($S$) and background ($B$) events in the $m_E^\text{rec}$ and $m_N^\text{rec}$ peaks (defined in the text) for 30 fb$^{-1}$, and luminosity $L$ required to have a $5\sigma$ discovery in the $\ell^\pm \ell^\pm \ell^\mp$ plus two jet final state.}
\label{tab:sig-3lepT2}
\end{table}

Finally, although the statistics in this channel is very good, the reconstructed opening angle distribution shows large deviations from the theoretical expectation due to detector effects, and the use of correction functions is compulsory in order to make a meaningful comparison. This issue will not be addressed here.


\subsection{Final state $\ell^\pm \ell^\pm$}
\label{sec:6.7}

This final state has a similar discovery potential as the previous one $\ell^\pm \ell^\pm \ell^\mp$. Decay channels giving like-sign dileptons include
\begin{align}
& E^+ N \to \ell^+ Z \, \ell^+ W^- \,,
  && \quad Z \to q \bar q / \nu \bar \nu, W \to q \bar q' 
  & (\mathrm{Br} = 4.3 \times 10^{-2}) \,, \nonumber \\
& E^+ N \to \ell^+ H \, \ell^+ W^- \,,
  && \quad H \to q \bar q , W \to q \bar q'
  & (\mathrm{Br} = 3.3 \times 10^{-2}) \,, \nonumber \\
& E^+ N \to \bar \nu W^+ \, \ell^+ W^- \,,
  && \quad W^+ \to \ell \nu, W^- \to q \bar q' 
  & (\mathrm{Br} = 2.1 \times 10^{-3}) \,,
\label{ec:ch2lik}
\end{align}
and analogous channels for $E^- N$ decays.
The contributions with final state neutrinos have not been included in Ref.~\cite{Franceschini:2008pz}, which focuses on four jet final states. As usual, additional contributions arise from decays with larger lepton multiplicities when one or more of them are not detected. For event pre-selection we use the same requirement as for the scalar triplet analysis, that is, two like-sign leptons with $p_T > 30$ GeV. The number of signal and background events are gathered in Table~\ref{tab:nsnb-2likT}.

\begin{table}[h]
\begin{center}
\begin{tabular}{cccccccc}
            & Pre-selection & Selection & \quad & & Pre-selection & Selection \\[1mm]
$E^+ E^-$ (T1) & 13.0   & 0.4   & & $W t \bar t nj$    & 194.0 & 19.3 \\
$E^\pm N$ (T1) & 678.1  & 226.4 & & $W W nj$           & 205.7 & 15.2 \\
$E^+ E^-$ (T2) & 6.8    & 0.1   & & $W Z nj$           & 892.2 & 32.3 \\
$E^\pm N$ (T2) & 114.7  & 1.8   & & $WWW nj$           & 86.9  & 4.3 \\
$t \bar t nj$  & 1193.6 & 166.0 & &  \\
\end{tabular}
\end{center}
\caption{Number of events for the like-sign dilepton signals and main backgrounds for a luminosity of 30 fb$^{-1}$.}
\label{tab:nsnb-2likT}
\end{table}

\begin{figure}[htb]
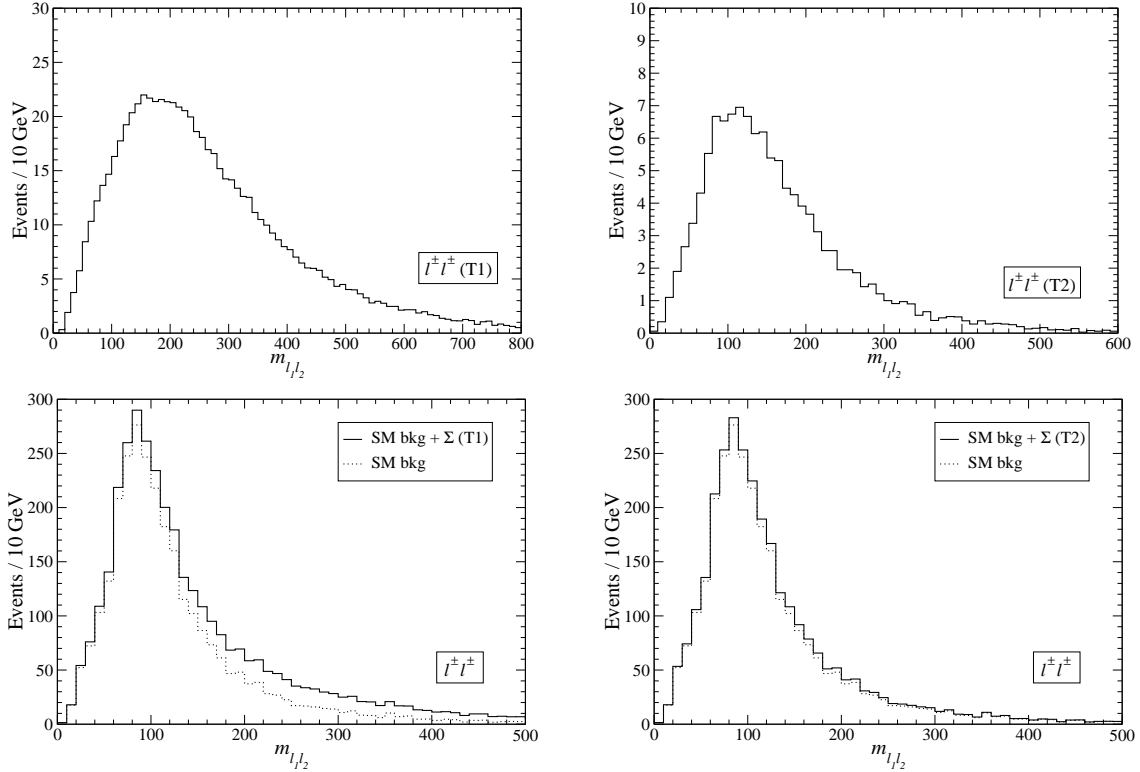

\begin{center}
\begin{tabular}{ccc}
\epsfig{file=Figs/Mll-2lik-T1.eps,height=5cm,clip=} & \quad &
\epsfig{file=Figs/Mll-2lik-T2.eps,height=5cm,clip=} \\
\epsfig{file=Figs/Ml1l2-2lik-T1.eps,height=5cm,clip=} & \quad &
\epsfig{file=Figs/Ml1l2-2lik-T2.eps,height=5cm,clip=}
\end{tabular}
\caption{Up: Kinematical distribution at pre-selection of the like-sign dilepton invariant mass $m_{\ell_1 \ell_2}$ for the fermion triplet signals in scenarios T1 (left) and T2 (right). Down: $\ell_1 \ell_2$ invariant mass distribution for the SM and the SM plus the signal in the scenarios T1 (left) and T2 (right). The luminosity is 30 fb$^{-1}$ in all cases.}
\label{fig:2lik-mllT}
\end{center}
\end{figure}

Already at the pre-selection level,
the differences between the scalar and fermion triplet signals are evident by merely considering the dilepton invariant mass distribution, which displays a sharp peak in the former case and a non-localised excess in the latter, with a long tail. The distributions for the fermion triplet signal in scenarios T1 and T2 are shown in Fig.~\ref{fig:2lik-mllT}, without and with the SM background.
This distribution can already be used to discover the fermion triplet signal, which shows up as a dilepton excess at large invariant mass. We will use, as in the previous subsection, the cut
\begin{equation}
m_{\ell_1 \ell_2} > 150~\text{GeV} \,.
\end{equation}
The number of signal and background events is given in Table~\ref{tab:sig-2likT1} with the luminosity needed for $5\sigma$ discovery, if the background uncertainty is neglected (a) or the background is normalised from data in the region $m_{\ell_1 \ell_2} < 150$ GeV (b).

\begin{table}[ht]
\begin{center}
\begin{tabular}{ccccccc}
& \multicolumn{3}{c}{Case (a)} & \multicolumn{3}{c}{Case (b)} \\
   & $S$   & $B$ & $L$ & $S$   & $B$ & $L$            \\[1mm]
T1 (cut on $m_{\ell_1 \ell_2}$) 
  & 525.3  & 692.7 & 2.2 fb$^{-1}$  & 471.5 & 746.5 & 2.9 fb$^{-1}$ \\
T2 (cut on $m_{\ell_1 \ell_2}$) 
  & 56.2  & 692.7 & 165 fb$^{-1}$  & 35.1 & 713.8 & $>300$ fb$^{-1}$
\end{tabular}
\end{center}
\caption{Number of signal ($S$) and background ($B$) events with $m_{\ell_1 \ell_2} > 150$ GeV for 30 fb$^{-1}$, and luminosity $L$ required to have a $5\sigma$ discovery in the $\ell^\pm \ell^\pm$ final state, at the pre-selection level without event reconstruction.}
\label{tab:sig-2likT1}
\end{table}

\begin{figure}[ht]
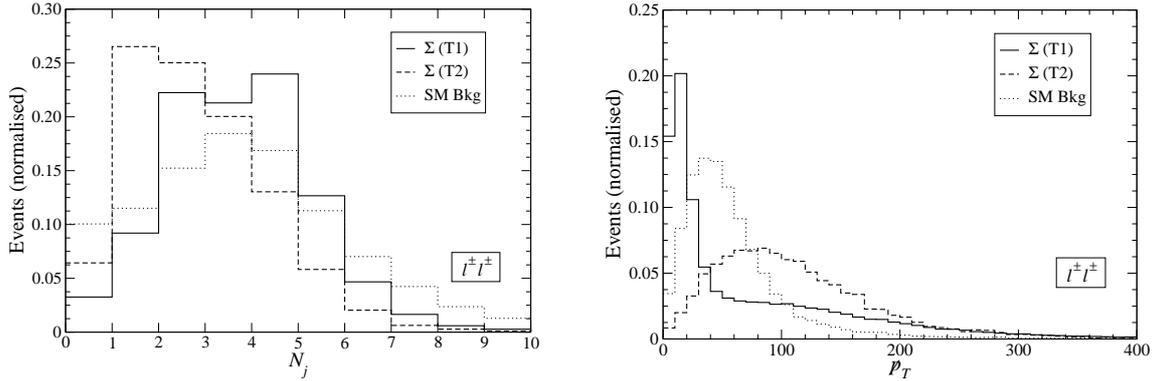

\begin{center}
\begin{tabular}{ccc}
\epsfig{file=Figs/mult-2lik-T12.eps,height=5cm,clip=} & \quad &
\epsfig{file=Figs/ptmiss-2lik-T12.eps,height=5cm,clip=}
\end{tabular}
\caption{Left: Jet multiplicity of the SM background and the fermion triplet signals in like-sign dilepton final states at pre-selection level. Right: missing energy distribution.}
\label{fig:2lik-mult}
\end{center}
\end{figure}

The signal significance can be improved with signal selection criteria and invariant mass reconstruction. For event selection we require:
\begin{itemize}
\item[(i)] missing energy $\ptmiss < 30$ GeV;
\item[(ii)] the presence of four jets with $p_T > 20$ GeV.
\end{itemize}
This event selection is slightly different from the one chosen for the heavy neutrino
singlet ($\ptmiss < 30$, four jets with no $b$ tags and back-to-back leptons) because the signals are quite different and the mass reconstruction
in the triplet case requires four jets.  
The number of events at the selection level is given in Table~\ref{tab:nsnb-2likT}.
Given the fact that some of the decay channels in Eqs.~(\ref{ec:ch2lik}) do not involve the production of extra quarks but undetected neutrinos, the signal is reduced. This can already be seen by examining the multiplicity distribution in Fig.~\ref{fig:2lik-mult} (left), which has been normalised for convenience. The requirement of no missing energy, whose normalised distribution is displayed in Fig.~\ref{fig:2lik-mult} (right) also suppresses the signal channels involving final state neutrinos but does not significantly affect final states which already have four jets.

Although not strictly necessary to have evidence of new physics, the kinematical reconstruction is convenient to identify the type of new physics
and to gain a higher signal significance. In order to do this,
we reconstruct the kinematics as corresponds to the decay channels with production of four quarks:
\begin{enumerate}
\item We associate each charged lepton to a pair of jets in all possible ways, using the four jets with larger $p_T$.
\item Among the six possibilities, we choose the one minimising the difference between the two $jj$ and the two $\ell jj$ invariant masses,
\begin{equation}
(m_{j_1 j_2} - m_{j_3 j_4})^2 + (m_{\ell_1 j_1 j_2} - m_{\ell_2 j_3 j_4})^2 \,.
\label{ec:rechad}
\end{equation}
Note that for the leading signal contributions two of the jets in principle correspond to a hadronic $W$ decay and the other two to a $Z$ or Higgs boson decay. However, if a wrong assignment is made, it is expected that the invariant mass differences will be larger.
\end{enumerate}
Both heavy states $E$, $N$ give as decay products one charged lepton and two jets,
and they cannot be easily distinguished. Then, we label the heavy states as $\Sigma_1$ and $\Sigma_2$, corrsponding to the leading and sub-leading lepton. A discrimination based on the two-jet invariant mass (it should be smaller in average for the two jets coming from the $N \to \ell W \to \ell q \bar q'$ decay than for $E \to \ell Z/H \to \ell q \bar q$)
is not very significant. We point out that, unlike other final states studied, here no information is gained about the nature of the two resonances because the jet charges are not measured.

\begin{figure}[htb]
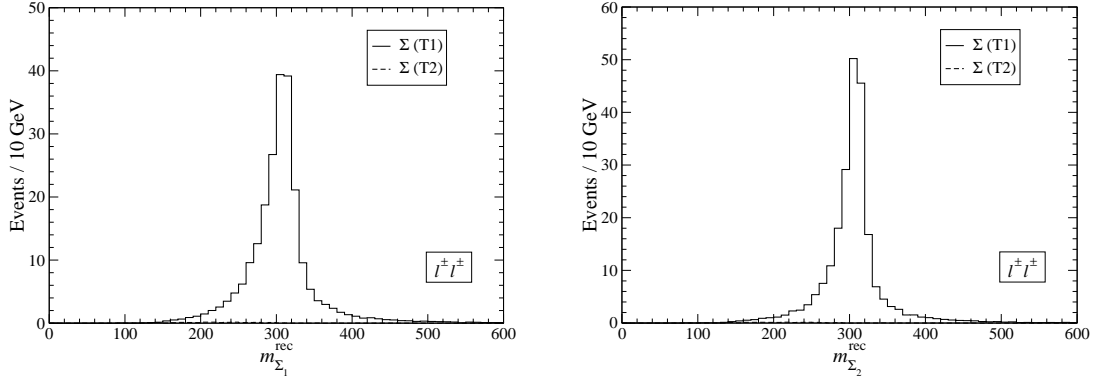

\begin{center}
\begin{tabular}{ccc}
\epsfig{file=Figs/mS1rec-2lik-T12.eps,height=5cm,clip=} & \quad &
\epsfig{file=Figs/mS2rec-2lik-T12.eps,height=5cm,clip=}
\end{tabular}
\caption{Reconstructed triplet masses in scenarios T1 and T2, without background, in the $\ell^\pm \ell^\pm$ final state at the selection level. The luminosity is 30 fb$^{-1}$.}
\label{fig:2likT-m3rec}
\end{center}
\end{figure}

\begin{figure}[htb]
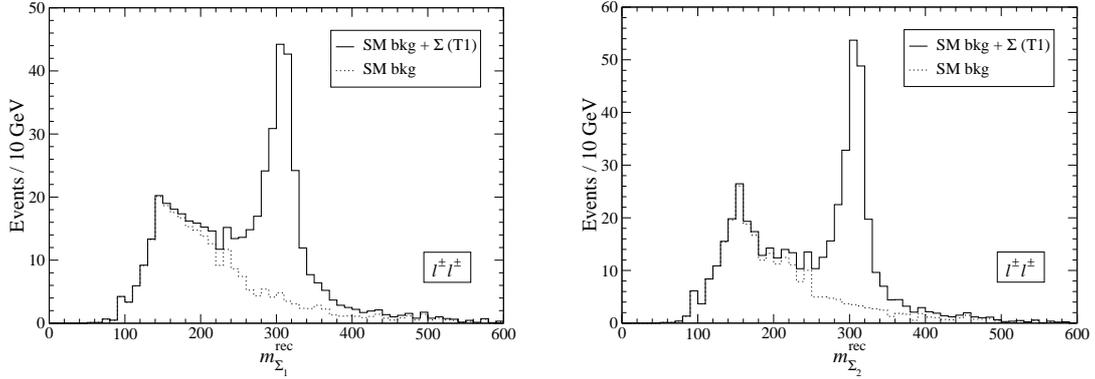

\begin{center}
\begin{tabular}{ccc}
\epsfig{file=Figs/mS1rec-2lik-bkgT1.eps,height=5cm,clip=} & \quad &
\epsfig{file=Figs/mS2rec-2lik-bkgT1.eps,height=5cm,clip=}
\end{tabular}
\caption{$m_{\Sigma_1}^\text{rec}$ (left) and $m_{\Sigma_2}^\text{rec}$ (right) distributions for the SM and the SM plus the fermion triplet signal in scenario T1 at the selection level. The luminosity is 30 fb$^{-1}$.}
\label{fig:2lik-m3rec2}
\end{center}
\end{figure}

The mass reconstruction is excellent for scenario T1, and both distributions display clear peaks at $m_\Sigma = 300$ GeV and small tails, as it can be observed in Fig. \ref{fig:2likT-m3rec}. On the other hand, for scenario T2 the distribution is very broad, and with very few events.
We present in Fig.~\ref{fig:2lik-m3rec2} the two distributions for scenario T1 including the background, in order to show how the presence of the signal would manifest. For scalar triplet signals less than one event survives the selection criteria adopted here, because jets are not produced in the hard interaction. Neutrino singlet signals also have a smaller jet multiplicity, which makes them unobservable if one requires four jets in the final state.

The signal significance is estimated for the peak regions
\begin{align}
250 < m_{\Sigma_1}^\text{rec} < 350~\text{GeV} \,, \notag \\
250 < m_{\Sigma_2}^\text{rec} < 350~\text{GeV}
\end{align}
neglecting the uncertainty in the background normalisation (which is not very important since the background itself it is very small for the discovery luminosity). The results are included in Table~\ref{tab:sig-2likT3}.

\begin{table}[ht]
\begin{center}
\begin{tabular}{ccccccc}
& \multicolumn{3}{c}{Case (a)}  \\
   & $S$   & $B$ & $L$ &     \\[1mm]
T1 (cut on $m_{\Sigma_1}^\text{rec}$, $m_{\Sigma_2}^\text{rec}$) 
  & 177.8  & 19.5 & 1.7 fb$^{-1}$  \\
\end{tabular}
\end{center}
\caption{Number of signal ($S$) and background ($B$) events in the $m_{\Sigma_1}^\text{rec}$ and $m_{\Sigma_2}^\text{rec}$ peaks (defined in the text) for 30 fb$^{-1}$, and luminosity $L$ required to have a $5\sigma$ significance for the peaks in the $\ell^\pm \ell^\pm$ plus four jet final state.}
\label{tab:sig-2likT3}
\end{table}

Finally, as it happens in the $\ell^\pm \ell^\pm \ell^\mp$ final state, the reconstructed opening angle distribution shows large deviations from the theoretical expectation due to detector effects, and the use of correction functions is required.


\subsection{Final state $\ell^+ \ell^- jjjj$}
\label{sec:6.8}

The large SM backgrounds make final states with two opposite charge leptons very demanding, even when the charged leptons have different flavour. Nevertheless, these final states must be studied to check the consistency of a possible discovery in the like-sign dilepton channels. Opposite sign pairs can be produced for example in the decay modes
\begin{align}
& E^+ N \to \ell^+ Z \, \ell^- W^+
  && \quad Z \to q \bar q / \nu \bar \nu, W \to q \bar q' 
  & (\mathrm{Br} = 4.3 \times 10^{-2}) \,, \nonumber \\
& E^+ N \to \ell^+ H \, \ell^- W^+
  && \quad H \to q \bar q , W \to q \bar q'
  & (\mathrm{Br} = 3.3 \times 10^{-2}) \,, \nonumber \\
& E^+ N \to \bar \nu W^+ \, \ell^- W^+
  && \quad W W\to \ell \nu q \bar q' 
  & (\mathrm{Br} = 4.3 \times 10^{-3}) \,, \nonumber \\
& E^+ E^- \to \ell^+ Z \, \ell^- Z
  && \quad Z \to q \bar q / \nu \bar \nu
  & (\mathrm{Br} = 5.9 \times 10^{-2}) \,, \nonumber \\
& E^+ E^- \to \ell^+ Z \, \ell^- H / \ell^+ H \, \ell^- Z
  && \quad Z \to q \bar q / \nu \bar \nu , H \to q \bar q 
  & (\mathrm{Br} = 8.9 \times 10^{-2}) \,, \nonumber \\
& E^+ E^- \to \ell^+ H \, \ell^- H
  && \quad H \to q \bar q 
  & (\mathrm{Br} = 3.4 \times 10^{-2}) \,.
\label{ec:ch2opp}
\end{align}
Performing an inclusive opposite-sign dilepton search of these signals is hopeless, because they are produced at an invariant mass which is not sufficient to separate them from the background.\footnote{Note that a $Z'$ boson decaying to $\ell^+ \ell^-$ produces opposite charge leptons with invariant mass $m_{\ell^+ \ell^-} \simeq M_{Z'}$ with a cross section enhancement from the on-shell $Z'$ propagator, and thus the signal is quite easy to see. On the contrary, $E^+ E^-$, $E^\pm N$ production is different because (i) the cross section does not have such an enhancement, and (ii) the dilepton invariant mass distribution is broader, and dileptons produced at high invariant mass are rare.} Thus, we concentrate on final states with four jets.

Contrarily to what has been recently claimed~\cite{Franceschini:2008pz}, we find that $e^\pm \mu^\mp$ final states with charged lepton flavour violation provide little advantage for the detection of the heavy triplet signal, due to the large LFV backgrounds. This can be easily understood with a simple counting argument. As we will find in the following, the SM background after selection cuts is mainly 
constituted by $t \bar t nj$ (and similar processes involving $WW$ decay) and $Z^* / \gamma^* \, nj$ production, with the latter being two times larger than the former. $WW$ decays give $e^+ e^-$, $\mu^+ \mu^-$ and $e^\pm \mu^\mp$ final states in the ratio $1\,:\,1\,:\,2$, while $Z^* / \gamma^* \, nj$ gives approximately an equal number of $e^+ e^-$ and $\mu^+ \mu^-$ final states (up to different detection efficiencies). Then, the $e^+ e^+$, $\mu^+ \mu^-$ and $e^\pm \mu^\mp$ components of the SM background are approximately
\begin{equation}
B_{ee} = \frac{5}{12} B \,,\quad B_{\mu \mu} = \frac{5}{12} B \,,\quad
B_{e\mu} = \frac{1}{6} B \,.
\end{equation}
On the other hand, $e^\pm \mu^\mp$ signals are only possible if the triplet simultaneously couples to the electron and muon. LFV signals are largest when $V_{eN} = V_{\mu N}$, so that the $e^+ e^-$, $\mu^+ \mu^-$ and $e^\pm \mu^\mp$ signals are produced with cross sections
in the ratio $1\,:\,1\,:\,2$, the total cross section being independent of the triplet mixing. Then, in a maximally LFV scenario
\begin{equation}
S_{ee} = \frac{1}{4} S \,,\quad S_{\mu \mu} = \frac{1}{4} S \,,\quad
S_{e\mu} = \frac{1}{2} S \,,
\end{equation}
where $S$ is also the signal in the lepton flavour conserving (LFC) case, which is the one considered here. A straightforward calculation shows that after combining the statistical significance $S/\sqrt B$ of the different channels,
in a LFV scenario the signal can be seen with a significance at most 1.34 times larger than in a LFC one.\footnote{In the LFC case we conservatively sum all dilepton backgrounds to take all possible triplet couplings into account. A triplet with a definite coupling will not contribute to all final states, and in this case the statistical significance of the signals will be slightly better.}
As we have done for the analysis of other channels, we will sum final states with electrons and muons.
Our pre-selection criteria are:
\begin{itemize}
\item[(i)] two opposite charge leptons with $p_T > 30$ GeV and an invariant mass larger than 200 GeV;
\item[(ii)] four jets with $p_T > 20$ GeV.
\end{itemize}
The number of events at this level of event selection
is collected in Table~\ref{tab:nsnb-2oppT}.
\begin{table}[ht]
\begin{center}
\begin{tabular}{cccccccc}
            & Pre-selection & Selection & \quad & & Pre-selection & Selection \\[1mm]
$E^+ E^-$ (T1) & 184.9 & 98.9 & & $t \bar t nj$      & 18017 & 762.4 \\
$E^\pm N$ (T1) & 214.2 & 112.7 & & $tW$              & 1180  & 53.8 \\
$E^+ E^-$ (T2) & 5.0   & 0.2   & & $WW nj$           & 784.5 & 40.6 \\
$E^\pm N$ (T2) & 7.7   & 0.4   & & $Z^*/\gamma^* nj$ & 6330  & 1966.8
\end{tabular}
\end{center}
\caption{Number of events for the $\ell^+ \ell^- jjjj$ signals and main backgrounds for a luminosity of 30 fb$^{-1}$.}
\label{tab:nsnb-2oppT}
\end{table}
\begin{figure}[ht]
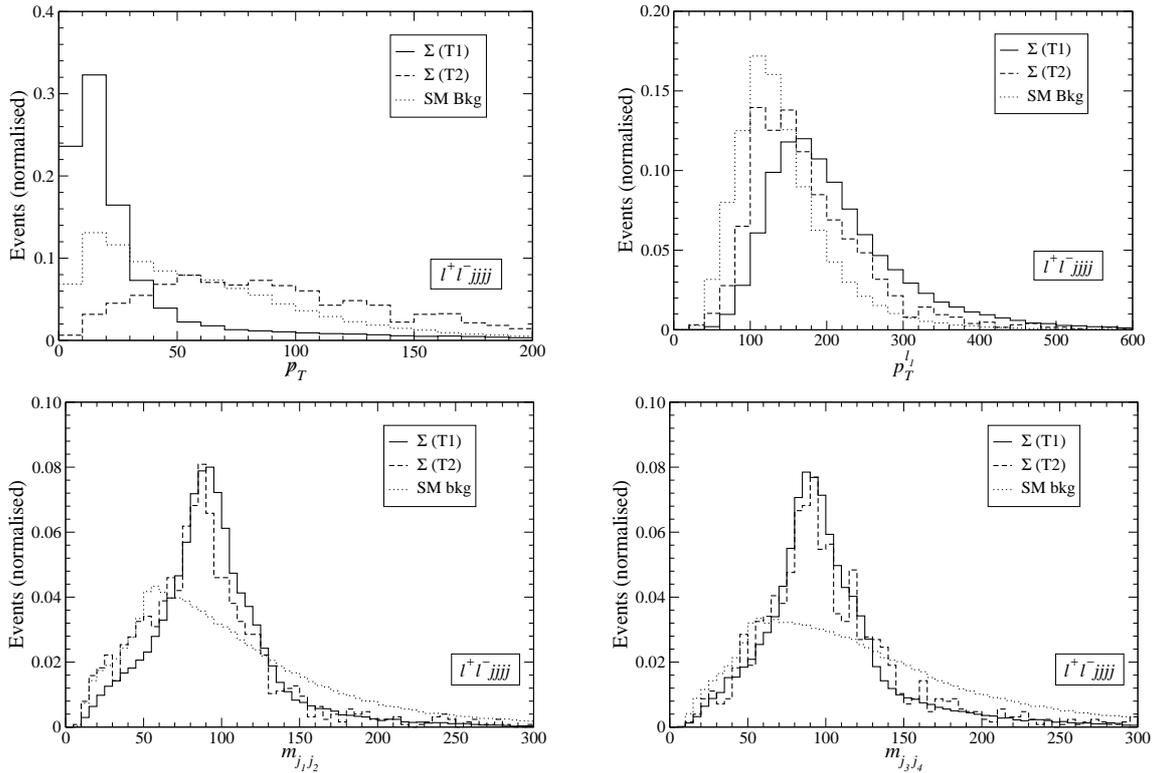

\begin{center}
\begin{tabular}{ccc}
\epsfig{file=Figs/ptmiss-2opp-T12.eps,height=5cm,clip=} & \quad &
\epsfig{file=Figs/ptlep1-2opp-T12.eps,height=5cm,clip=} \\
\epsfig{file=Figs/mB1rec-2opp-T12.eps,height=5cm,clip=} & \quad &
\epsfig{file=Figs/mB2rec-2opp-T12.eps,height=5cm,clip=} \\
\end{tabular}
\caption{Up, left: normalised missing energy distribution of the SM background and the fermion triplet signals in $\ell^+ \ell^- jjjj$ final states at pre-selection level. Up, right: transverse momentum of the leading charged lepton.
Down: dijet invariant masses reconstructing the bosons decaying hadronically.}
\label{fig:2opp-ptmiss}
\end{center}
\end{figure}
As in like-sign dilepton final states, backgrounds involving extra neutrinos can be reduced by requiring small missing energy. Additionally, 
the reconstruction of the two heavy states can be done in the same way as in the like-sign dilepton final state of the previous subsection, and the reconstructed invariant masses for the signal are very similar to those obtained there.
The selection criteria used for the analysis are:
\begin{itemize}
\item[(i)] missing energy $\ptmiss < 30$ GeV;
\item[(ii)] the leading charged lepton $\ell_1$ must have transverse momentum
$p_T^{\ell_1} > 100$ GeV;
\item[(iii)] the two-jet invariant masses $m_{j_1 j_2}$ and $m_{j_3 j_4}$,
which reconstruct the $W/Z/H$ bosons decaying hadronically, must be both within 50 and 150 GeV.
\end{itemize}
The kinematical distributions of these variables are presented in Fig.~\ref{fig:2opp-ptmiss}. The missing energy cut requirement greatly reduces the $t \bar t nj$ background,
being also convenient to require an energetic lepton. Unfortunately, these cuts also remove most of the signal in scenario T2, although in T1 it is practically unaffected.

\begin{figure}[htb]
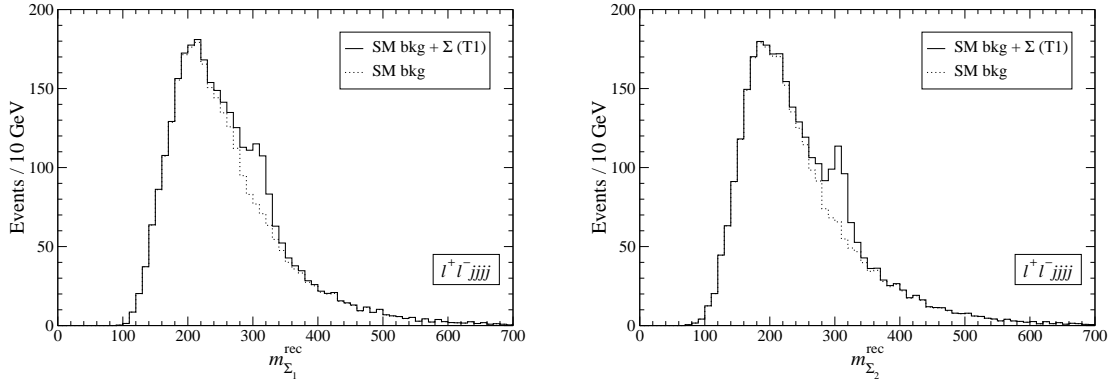

\begin{center}
\begin{tabular}{ccc}
\epsfig{file=Figs/mS1rec-2opp-bkgT1.eps,height=5cm,clip=} & \quad &
\epsfig{file=Figs/mS2rec-2opp-bkgT1.eps,height=5cm,clip=}
\end{tabular}
\caption{$m_{\Sigma_1}^\text{rec}$ (left) and $m_{\Sigma_2}^\text{rec}$ (right) distributions for the SM and the SM plus the fermion triplet signal in scenario T1 at the selection level. The luminosity is 30 fb$^{-1}$.}
\label{fig:2opp-m3rec2}
\end{center}
\end{figure}

Due to the large background, the peaks in the reconstructed mass distributions produced by the signal are small. They are shown in Fig.~\ref{fig:2opp-m3rec2}, for the two states $\Sigma_1$ and $\Sigma_2$ involving the leading and sub-leading charged leptons, respectively.

The significance of the peaks is moderate. We select the peak regions
\begin{align}
260 < m_{\Sigma_1}^\text{rec} < 340~\text{GeV} \,, \notag \\
260 < m_{\Sigma_2}^\text{rec} < 340~\text{GeV} \,,
\end{align}
and give in Table~\ref{tab:sig-2oppT} the number of signal and background events, as well as the luminosity needed to have $5\sigma$ significance, if: (a) the background normalisation is neglected, and (b) if background is determined from off-peak data. The first case is unrealistic, because typical normalisation uncertainties are of the order of 20\%, and important in this final state where the background is large. In case that the background has to be normalised from data, the off-peak signal contributions will be taken as part of the background and so the peak significance decreases.

\begin{table}[ht]
\begin{center}
\begin{tabular}{ccccccc}
& \multicolumn{3}{c}{Case (a)} & \multicolumn{3}{c}{Case (b)} \\
   & $S$   & $B$ & $L$  & $S$   & $B$ & $L$    \\[1mm]
T1 (cut on $m_{\Sigma_1}^\text{rec}$) 
  & 171.0  & 681.5 & 17.4 fb$^{-1}$ & 78.1 & 774.4 & 95 fb$^{-1}$ \\
T1 (cut on $m_{\Sigma_2}^\text{rec}$) 
  & 178.7  & 548.3 & 12.9 fb$^{-1}$ & 78.8 & 648.2 & 78 fb$^{-1}$
\end{tabular}
\end{center}
\caption{Number of signal ($S$) and background ($B$) events in the $m_E^\text{rec}$ and $m_N^\text{rec}$ peaks (defined in the text) for 30 fb$^{-1}$, and luminosity $L$ required to have a $5\sigma$ discovery in the $\ell^+ \ell^- jjjj$ final state.}
\label{tab:sig-2oppT}
\end{table}

Alternatively, one can perform simultaneous cuts on $m_{\Sigma_1}^\text{rec}$
and $m_{\Sigma_2}^\text{rec}$. The number of events in this case is
$S = 160.1$, $B = 265.0$. Assuming a 20\% background uncertainty, the signal
significance is $2.88\sigma$ for 30 fb$^{-1}$, and the excess of events does not reach discovery significance even for much larger integrated luminosities because of the background systematic uncertainty. Of course, kinematical cuts can be optimised for given $E$, $N$ masses, additional variables can be used and a likelihood method can improve the analysis, but the discovery potential is expected to remain below the other channels.


\subsection{Final state $\ell^\pm jjjj$}
\label{sec:6.9}

We finally turn our attention to signals with only one charged lepton, which can be produced in the decays
\begin{align}
& E^+ N \to \bar \nu W^+ \, \ell^\mp W^\pm \,,
  && \quad W \to  q \bar q' 
  & (\mathrm{Br} = 1.28 \times 10^{-1}) \,, \nonumber \\
& E^+ N \to \ell^+ Z/H \, \nu Z/H \,,
  && \quad Z \to  q \bar q'/\nu \bar \nu,H \to q \bar q 
  & (\mathrm{Br} = 1.82 \times 10^{-1}) \,, \nonumber \\
& E^+ E^- \to \ell^+ Z/H \, \nu W^- \,,
  && \quad Z \to q \bar q/\nu \bar \nu,H \to q \bar q,W \to q \bar q' 
  & (\mathrm{Br} = 1.52 \times 10^{-1}) \,, \nonumber \\
& E^+ E^- \to \bar \nu W^+ \, \ell^- Z/H  \,,
  && \quad Z \to q \bar q/\nu \bar \nu,H \to q \bar q,W \to q \bar q' 
  & (\mathrm{Br} = 1.52 \times 10^{-1}) \,,
\label{ec:ch1l}
\end{align}
and thus benefit from a large branching ratio. Nevertheless, backgrounds are very large as well. We concentrate on final states with four jets, asking as pre-selection criteria:
\begin{itemize}
\item[(i)] one charged lepton with $p_T^\ell > 30$ GeV and missing energy $\ptmiss > 50$ GeV, which must have a transverse mass
\begin{equation}
M_T^2 \equiv 2 p_T^\ell \ptmiss (1-\cos \phi_T) > (200~\text{GeV})^2 \,,
\end{equation}
where $\phi_T$ is the angle in the transverse plane between the charged lepton and the missing energy;
\item[(ii)] four jets with $p_T > 20$ GeV, and no $b$-tagged jets.
\end{itemize}
The number of signal and background events fulfilling these criteria are gathered in Table~\ref{tab:nsnb-1l4j}.
\begin{table}[htb]
\begin{center}
\begin{tabular}{cccccccc}
            & Pre-selection & Selection & \quad & & Pre-selection & Selection \\[1mm]
$E^+ E^-$ (T1) & 188.5  & 109.8 & & $tW$              & 440.7 & 171.6 \\
$E^\pm N$ (T1) & 443.7  & 262.9 & & $W nj$            & 15146 & 5538 \\
$E^+ E^-$ (T2) & 26.9   & 11.2 & & $Z^*/\gamma^* nj$ & 455.9 & 161.4 \\
$E^\pm N$ (T2) & 64.4   & 28.1 & & $WWnj$            & 661.7 & 200.8 \\
$t \bar t nj$  & 3678.7 & 1204.2
\end{tabular}
\end{center}
\caption{Number of events for the $\ell^\pm jjjj$ signals and main backgrounds for a luminosity of 30 fb$^{-1}$.}
\label{tab:nsnb-1l4j}
\end{table}
The cut on transverse mass is performed to reduce backgrounds in which the charged lepton and neutrino result from a $W$ boson decay, such as $Wnj$, for which $M_T < M_W$. However, due to energy mismeasurements the $Wnj$ background cannot be completely removed, and dominates due to its huge cross section. We also point out that $4/5$ of the $Wnj$ background comes from parton-level multiplicities $n=0,1,2$, which contribute to four jet final states due to pile-up. The second largest background is $t \bar t nj$ in the dilepton decay channel, with one $W$ boson decaying into a $\tau$ which subsequently decays hadronically, or decaying into an electron or muon which is missed by the detector. This background is also reduced by the requirement of no $b$ tags.

\begin{figure}[htb]
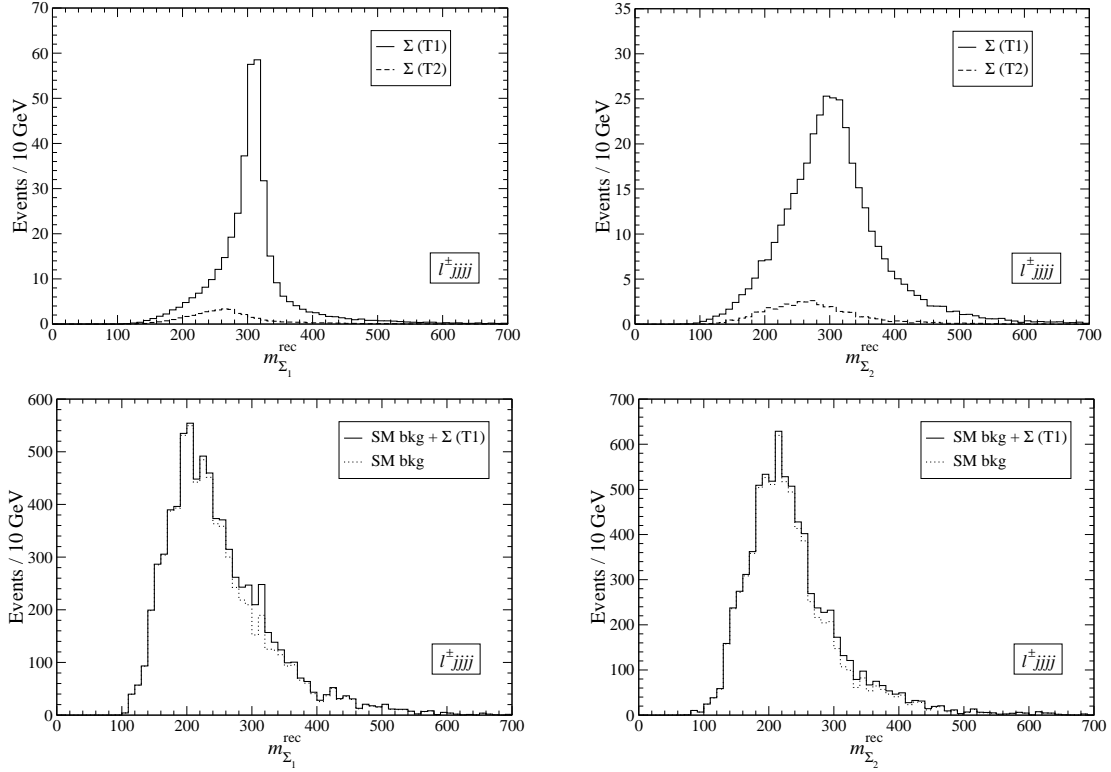

\begin{center}
\begin{tabular}{ccc}
\epsfig{file=Figs/mS1rec-1l4j-T12.eps,height=5cm,clip=} & \quad &
\epsfig{file=Figs/mS2rec-1l4j-T12.eps,height=5cm,clip=} \\
\epsfig{file=Figs/mS1rec-1l4j-bkgT1.eps,height=5cm,clip=} & \quad &
\epsfig{file=Figs/mS2rec-1l4j-bkgT1.eps,height=5cm,clip=}
\end{tabular}
\caption{Up: Reconstructed triplet masses in scenarios T1 and T2, without background, in the $\ell^\pm jjjj$ final state at the selection level.
Down: reconstructed masses with background, in scenario T1. The luminosity in all cases is 30 fb$^{-1}$.}
\label{fig:1l4j-m3rec}
\end{center}
\end{figure}

Events can be reconstructed as in the dilepton final states but replacing the second lepton by the missing momentum vector (taking the third component equal to zero). The dijet invariant mass distributions are very similar.
As selection criteria we impose that:
\begin{itemize}
\item[(i)] The charged lepton must have transverse momentum
$p_T^{\ell} > 100$ GeV.
\item[(ii)] The two-jet invariant masses $m_{j_1 j_2}$ and $m_{j_3 j_4}$
which reconstruct the $W/Z/H$ bosons decaying hadronically must be both within 50 and 150 GeV.
\end{itemize}
The number of events after selection are given in Table~\ref{tab:nsnb-1l4j}. For completeness, the heavy lepton reconstructed masses $m_{\Sigma_1}$ (corresponding to the charged lepton plus two jets) and $m_{\Sigma_2}$ (formed with the missing energy and the other two jets) are presented in Fig.~\ref{fig:1l4j-m3rec} for the two scenarios, without background.
While the reconstruction of $\Sigma_1$ is still very good, the reconstruction of the other state from missing energy and two jets gives a distribution much wider.
The small bumps caused by the presence of the signal would be very difficult to spot over the background if this is not theoretically predicted with a good accuracy, as it is apparent from Fig.~\ref{fig:1l4j-m3rec} (down), and the significance is difficult to estimate without a detailed calculation of the background uncertainty, which is beyond the scope of this work. 
The number of events in the peak regions
\begin{align}
260 < m_{\Sigma_1}^\text{rec} < 340~\text{GeV}  \,, \nonumber \\
200 < m_{\Sigma_1}^\text{rec} < 400~\text{GeV} \,,
\end{align}
is $S = 232.4$, $B = 1328.0$ for 30 fb$^{-1}$. This amounts to $6.3\sigma$ if the background uncertainty is neglected, but less than $1\sigma$ if a 20\% uncertainty is assumed. Further improvements can be made optimising the analysis with a likelihood method, and it is expected that $3\sigma$ could be reached for 30 fb$^{-1}$.


\subsection{Outlook}
\label{sec:6.10}

Fermion triplet production $pp \to E^+ E^- ,\, E^\pm N$ leads to a plethora of possible final state signatures after the heavy leptons decay
\begin{align}
& E^\pm \to \nu W^\pm ~/~ E^\pm \to \ell^\pm Z ~/~  E^\pm \to \ell^\pm H \,, \nonumber \\
& N \to \ell^- W^+ ~/~  N \to \ell^+ W^- ~/~ N \to \nu Z ~/~ N \to \nu H \,,
\end{align}
and the $W$, $Z$ and $H$ bosons decay to hadrons or leptons.
In this section we have studied the fermion triplet signals classifying the experimental signatures by their charged lepton multiplicity. Some final states are characteristic of $E^+ E^-$, $E^\pm N$ production and are not present in other seesaw scenarios, namely
\begin{itemize}
\item Six lepton final states
\item Five lepton final states
\item Four leptons $\ell^\pm \ell^\pm \ell^\pm \ell^\mp$ with total charge $Q=\pm 2$
\item Three like-sign leptons $\ell^\pm \ell^\pm \ell^\pm$
\end{itemize}
The six lepton signal has a too small branching ratio to be useful. 
The signals with five leptons and three like-sign leptons are small but could be observed with a luminosity of 30 fb$^{-1}$ in the case of a triplet mass
$m_{E,N} = 300$ GeV, coupling to either the electron or muon.
The $Q=\pm 2$ four lepton signal is more interesting, and it could be observed with half the luminosity. Additionally, in this channel the $E$ mass can be reconstructed as a three-lepton invariant mass, providing evidence for $E$ production. For much larger luminosities, the opening angle distribution can be tested as well.

Final states common to seesaw II signals are
\begin{itemize}
\item Four leptons $\ell^+ \ell^+ \ell^- \ell^-$ with total charge $Q=0$
\item Three leptons $\ell^\pm \ell^\pm \ell^\mp$
\item Two like-sign leptons
\item Two opposite sign leptons
\item One lepton
\end{itemize}
Among these, the first two final states are very interesting. The
$\ell^+ \ell^+ \ell^- \ell^-$ signal can provide $5\sigma$ evidence already with 6 fb$^{-1}$ for a triplet with $m_{E,N} = 300$ GeV, coupling to the electron or muon. Moreover, the $E$ mass can be reconstructed as a sharp peak in a three-lepton invariant mass, providing clear evidence for $E$ production, and the $E$ opening angle distribution can be clearly reconstructed.
The $\ell^\pm \ell^\pm \ell^\mp$ signal has even better discovery potential: for the same triplet parameters it can be discovered with less than 3 fb$^{-1}$. Additionally, it provides evidence for $N$ production, whose mass can be reconstructed (in final states with two additional jets) from two opposite sign leptons plus the missing energy, the resulting distribution displaying a sharp peak at $m_N$. These two signals can be cleanly distinguished from seesaw II signals by (i) the analysis of the like-sign dilepton invariant mass distribution, and (ii) the $E$, $N$ reconstructed mass distributions. We also point out that if heavy neutrino triplets are not Majorana but (quasi-)Dirac particles, the trilepton final state is  the most interesting one as well \cite{corto}.

The like-sign dilepton final state has been considered in previous literature
\cite{Bajc:2006ia,Franceschini:2008pz} as the most important triplet signal. We have found that its discovery potential equals the one of the $\ell^\pm \ell^\pm \ell^\mp$. The invariant mass of the heavy states can be reconstructed in final states with four additional jets, giving evidence for the production of two resonances with equal mass. Their nature, however, cannot be established because the jet charge is very difficult to measure. Apart from the mass reconstruction, the fermion triplet signal
in this final state can be distinguished from scalar triplet production by the dilepton invariant mass and also by the presence of additional jets. Heavy neutrino singlet production also leads to like-sign dilepton signals, but with smaller jet multiplicity and smaller dilepton invariant mass in general.

The $\ell^+ \ell^-$ final state is very demanding, and in order to make the backgrounds manageable four hard jets are required in addition to the charged leptons.
Still, backgrounds are very large, in particular from $Z^*/\gamma^* \, nj$ and
$t \bar t nj$ production. The mass reconstruction can be performed as in the like-sign dilepton channel to reduce backgrounds, but still the signal is very difficult to observe unless the background uncertainty (theoretical, as well as from the luminosity and detector effects) is under very good control.
In contrast with recent claims~\cite{Franceschini:2008pz} (and in agreement with previous literature~\cite{delAguila:2007em}), we have found that looking for explicitly LFV $e^\pm \mu^\mp$ final states represents little improvement.
In a maximally LFV scenario with $V_{eN} = V_{\mu N}$, the improvement in the signal significance $S/\sqrt B$ is of only 30\% over a LFC scenario with $V_{eN} = 0$ or
$V_{\mu N} = 0$. This is due to the large $t \bar t nj$ background, which gives $e^\pm \mu^\mp$ final states in the dilepton decay channel.

Finally, signals with only one charged lepton suffer large backgrounds
from all sort of SM processes, even if one requires four hard jets and large missing energy in the events. The background with the largest cross section is $Wnj$, and it cannot be fully removed as suggested in Ref.~\cite{Franceschini:2008pz} by the use of the transverse mass between the charged lepton and the missing momentum. $t \bar t nj$ in the dilepton decay channel is another large source of background, when one of the charged leptons is a tau or it is missed by the detector. This background cannot be removed with transverse mass cuts because of the presence of at least two neutrinos from different $W$ decays.
After event reconstruction and invariant mass cuts the significance of the signals is below $1\sigma$, but can be certainly improved if a more sophisticated analysis is used, which is beyond the scope of this study.

\section{Conclusions}
\label{sec:7}

In this paper we have classified and studied possible multi-lepton signals of seesaw messengers at LHC,
namely heavy neutrino singlets (seesaw I), scalar triplets (seesaw II) and fermion triplets (seesaw III). For seesaw II scenarios we have assumed that the triplet vev $\vt$ is small, so that dilepton channels dominate the decay of the new scalars. Although this may not necessarily be  the case it is a plausible assumption, provided that the triplet Yukawa couplings to the charged leptons are not much smaller than the smallest of the SM Yukawa couplings to the Higgs boson. For seesaw III scenarios we have assumed that the mixing between the light and heavy sectors is not unnaturally suppressed, so that the triplet decays to leptons plus gauge or Higgs bosons dominate.
We have set a relatively low scale for the new states, 100 GeV for heavy neutrino singlets and 300 GeV for scalar and fermion triplets. For the latter two cases, such masses would allow a discovery with the first few fb$^{-1}$ of LHC data.

Seesaw signals involve multi-lepton production, thus an adequate classification is in terms of charged lepton multiplicity. This is also very convenient in terms of the experimental searches. Although some of the signals involve the production of extra quarks, further discrimination by using jet multiplicity is difficult, because the correspondence between the number of quarks at the partonic level and the number of jets measured in the detector is not univocal, due to radiation and pile-up.

Nine different final states have been studied, involving up to six charged leptons.
An important feature of our analysis is that for each final state {\em all} signal contributions have been included, within a given scenario. For heavy neutrino production we have used the {\tt Alpgen} extension in Ref.~\cite{delAguila:2007em}, while for scalar and fermion triplet production we have developed a Monte Carlo generator \tria. For scalar triplet production a realistic assessment of the discovery potential must include scalar decays to tau leptons, especiallly for NH where they are dominant, and the subsequent tau decay.
In the case of fermion triplet production, generating all signal contributions is involved.
For $E^+ E^-$ production there are 289 different final states with 128 different matrix elements,
and for $E^\pm N$ there are 748 final states with 72 matrix elements. Backgrounds have been generated with {\tt Alpgen}. Signals and backgrounds have been passed through the parton-shower Monte Carlo {\tt Pythia} to add initial and final state radiation and pile-up, and perform the hadronisation of final state quarks and gluons. Finally, a fast simulation of the detector has been used.

We have found that generating the complete signals and not just particular production or decay channels is essential, because for scalar and fermion triplet production most final states have contributions from several competing decay chains. For instance, in scalar triplet production the most interesting final state is the one with three charged leptons $\ell^\pm \ell^\pm \ell^\mp$, which receives important contributions from both the $\Dpp \Dmm$ and $\Dppmm \Dmp$ production processes. For fermion triplet production this final state is the most interesting one as well, and it receives contributions from 8 decay channels in $E^\pm N$ production. Moreover, in general decay channels with larger lepton multiplicities contribute to signals with a smaller number of leptons when one or more of them are missed by the detector.

The most relevant features of the signals studied have been outlined in sections~\ref{sec:4.3}, \ref{sec:5.6} and \ref{sec:6.10}, and we refer the reader to those sections
for a brief outlook. Table~\ref{tab:summ} summarises the
possible signals which appear in each of the seesaw (I, II and III) scenarios studied, with
the luminosity required to discover them. In seesaw I we only give the numbers for a heavy neutrino coupling to the muon. For seesaw II we give values for both NH and IH, while for seesaw III we include results for a triplet coupling to the electron (for a muon the sensitivity is expected to be very similar). We also indicate whether the mass of the new particles can be reconstructed, which gives an additional evidence for the signal.
For seesaw I the discovery luminosities indicated are rather conservative because the analysis is not optimised, to keep consistency with seesaw II and III, and, since heavy neutrino signals are much harder to see, the improvements are expected to be more significant in this case.

\begin{table}[htb]
\begin{center}
\begin{tabular}{|c|c|c|c|}
\hline
& \begin{tabular}{c}Seesaw I \\ $m_N = 100$ GeV \end{tabular}
& \begin{tabular}{c}Seesaw II \\ $m_\Delta = 300$ GeV \end{tabular}
& \begin{tabular}{c}Seesaw III \\ $m_\Sigma = 300$ GeV \end{tabular} \\
\hline
Six leptons                           & --       & --        & $\times$ \\
\hline
Five leptons                          & --       & --        & 28 fb$^{-1}$ \\
\hline
$\ell^\pm \ell^\pm \ell^\pm \ell^\mp$ & --       & --
  & \begin{tabular}{c}15 fb$^{-1}$ \\ $m_E$ rec\end{tabular} \\
\hline
$\ell^+ \ell^+ \ell^- \ell^-$         & --       
  & \begin{tabular}{c}19 / 2.8 fb$^{-1}$ \\ $m_{\Dpp}$ rec\end{tabular}
  & \begin{tabular}{c}7 fb$^{-1}$ \\ $m_E$ rec\end{tabular} \\
\hline
$\ell^\pm \ell^\pm \ell^\pm$          & --       & --        & 30 fb$^{-1}$ \\
\hline
$\ell^\pm \ell^\pm \ell^\mp$
  & $< 180~\text{fb}^{-1}$
  & \begin{tabular}{c}3.6 / 0.9 fb$^{-1}$ \\ $m_{\Dpp}$ rec\end{tabular}
  & \begin{tabular}{c}2.5 fb$^{-1}$ \\ $m_N$ rec\end{tabular} \\
\hline
$\ell^\pm \ell^\pm$
  & \begin{tabular}{c}$< 180~\text{fb}^{-1}$ \\ $m_N$ rec\end{tabular}
  & \begin{tabular}{c}17.4 / 4.4 fb$^{-1}$ \\ $m_{\Dpp}$ rec\end{tabular}
  & \begin{tabular}{c}1.7 fb$^{-1}$ \\ $m_\Sigma$ rec\end{tabular} \\
\hline
$\ell^+ \ell^-$
  & $\times$
  & \begin{tabular}{c}15 / 27 fb$^{-1}$ \\ $m_{\Delta}$ rec\end{tabular}
  & \begin{tabular}{c}80 fb$^{-1}$ \\ $m_\Sigma$ rec\end{tabular}        \\
\hline
$\ell^\pm$   & $\times$   & $\times$ & $\times$ \\
\hline
\end{tabular}
\caption{Summary of the final states studied for each seesaw scenario. A dash indicates that a given final state is not present. A cross indicates that the signal is produced but it cannot be seen, either because it is too small or because the background is too large.
In each case we include the luminosity required for discovery for the parameters assumed, also indicating when the heavy mass(es) can be reconstructed. For seesaw II we give the results for NH and IH.}
\label{tab:summ}
\end{center}
\end{table}

In this work we have made special emphasis on the discrimination of the different seesaw models if a positive signal is observed.
Table~\ref{tab:summ} shows that, if a positive signal is found, it should be possible to identify whether it corresponds to heavy neutrinos, scalar/fermion triplets, or other new physics, by analysing the different channels and performing the mass reconstruction.
For larger luminosities the study of production angular distributions is possible, as it has been shown in several final states for scalar and fermion triplet signals.
A rough estimate of the LHC discovery reach can be made by applying a simple rescaling of the results presented, obtaining that scalar triplets can be discovered up to 600 GeV (800 GeV) for NH (IH) with a luminosity of 30 fb$^{-1}$. A fermion triplet coupling to the electron or muon can be discovered up to 750 GeV with the same luminosity. For neutrino singlets the more detailed results of Ref.~\cite{delAguila:2007em} can be rescaled with the new upper bounds
$|V_{eN}|^2 \leq 0.0030$, $|V_{\mu N}|^2 \leq 0.0032$, to obtain that a heavy neutrino coupling to the electron can be discovered up to 120 GeV, and if it couples to the muon up to 155 GeV. These results are consistent with the ones in Table~\ref{tab:summ} because for larger $m_N$ the heavy neutrino signals can be separated from the background more easily, and also because the analysis in Ref.~\cite{delAguila:2007em} is much more sophisticated.

New interactions, which could lead to new processes in addition to the ones discussed here,
have not been considered in this work.
It is interesting to discuss briefly, without a detailed simulation of those signals, how one might in principle distinguish some of these scenarios from the ones
studied here.  In models with left-right symmetry heavy neutrino singlets can be produced through $s$-channel $W'$ ($W_R$) exchange \cite{Keung:1983uu,Datta:1992qw,Ferrari:2000sp,Gninenko:2006br}
\begin{equation}
q \bar q' \to W' \to \ell N \,.
\end{equation}
This process gives the same dilepton and trilepton final states as $W$ exchange but with large transverse momenta, which are not possible without new interactions. Four and five lepton signals are not produced. The $W'$ transverse mass could also be kinematically reconstructed \cite{Ferrari:2000sp,Gninenko:2006br}.
A relatively light leptophobic $Z'_\lambda$ boson can produce neutrino pairs in the process \cite{delAguila:2007ua}
\begin{equation}
q \bar q \to Z'_\lambda \to N N \,.
\end{equation}
The $\ell^\pm \ell^\pm$ and $\ell^\pm \ell^\pm \ell^\mp$ signals would be similar to the ones of a fermion triplet, but the $E$ reconstruction in four lepton final states (from three charged leptons) would be a clear indication for fermion triplet production, and in $Z'_\lambda \to NN$ the $Z'$ mass could be reconstructed. New gauge bosons might also contribute to the production of scalar and fermion triplets, and their presence would be seen by examining the invariant mass distribution of all the produced particles. Besides, it is likely that new interactions would be observed in other production processes not involving seesaw messengers.

Finally, it is worth mentioning that, since this study has been done at the level of a fast detector simulation, we have not addressed charge misidentification. Its effect may be important in like-sign dilepton final states, because the cross section for $Z^*/\gamma^*$ production at LHC is very large. However, for final states with more leptons, {\em e.g.} $\ell^\pm \ell^\pm \ell^\mp$, which are the most interesting ones
in terms of discovery potential, its impact in the final results is expected to be small. Hence, our results are not expected to change dramatically with a full detector simulation, which must anyway be performed to confirm them and to compare with real data.

\vspace{1cm}
\noindent
{\Large \bf Acknowledgements}
\vspace{0.3cm}

\noindent
This work has been supported by MEC project FPA2006-05294 and
Junta de Andaluc{\'\i}a projects FQM 101, FQM 437 and FQM03048. The work of J.A.A.S.
has been supported by a MEC Ram\'on y Cajal contract.

\appendix
\section{Feynman rules}
\label{sec:a}

We give here the Feynman rules used in our matrix element calculations.
For LNV processes we use the method in Ref.~\cite{Denner:1992vza}, which allows to use standard propagators avoiding to introduce explicitly the charge conjugation matrix in the Feynman rules. With these prescriptions, the resulting Feynman rules are especially useful for their implementation in Monte Carlo generators. The rules for propagators are the usual ones, and external legs are taken following the flow defined in the fermion lines. More details can be found in Ref.~\cite{Denner:1992vza}.

\begin{table}[p]
\begin{center}
\begin{footnotesize}
\begin{tabular}{clccl}
\raisebox{-11mm}{\epsfig{file=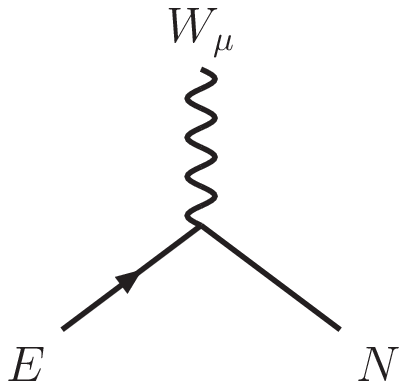,height=22mm,clip=}}
  & $\displaystyle -ig \gamma^\mu$ & \quad &
\raisebox{-11mm}{\epsfig{file=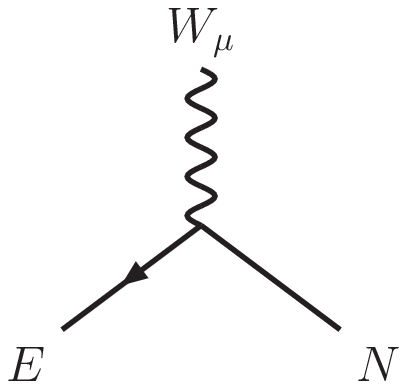,height=22mm,clip=}}
  & $\displaystyle -ig \gamma^\mu$ \\ \\
\raisebox{-11mm}{\epsfig{file=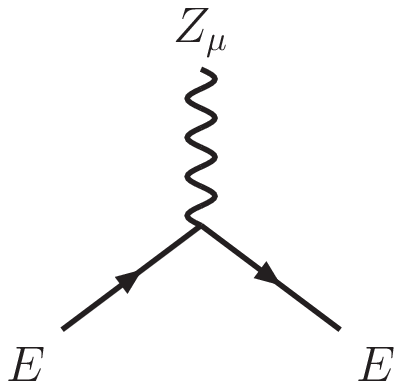,height=22mm,clip=}}
  & $\displaystyle igc_W \gamma^\mu$ & \quad &
\raisebox{-11mm}{\epsfig{file=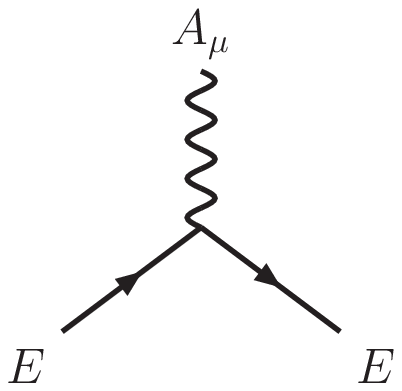,height=22mm,clip=}}
  & $\displaystyle ie \gamma^\mu$ \\ \\
\end{tabular}
\end{footnotesize}
\caption{Feynman rules for heavy fermion triplet gauge interactions.}
\end{center}
\end{table}

\begin{table}[p]
\begin{center}
\begin{footnotesize}
\begin{tabular}{clccl}
\raisebox{-11mm}{\epsfig{file=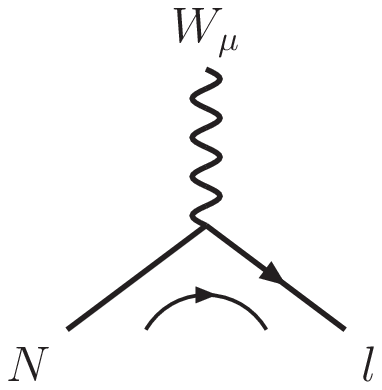,height=22mm,clip=}}
  & $\displaystyle -\frac{ig}{\sqrt 2} V_{lN} \gamma^\mu P_L$ & \quad &
\raisebox{-11mm}{\epsfig{file=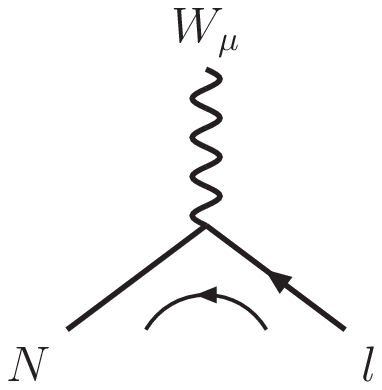,height=22mm,clip=}}
  & $\displaystyle -\frac{ig}{\sqrt 2} V_{lN}^* \gamma^\mu P_L$ \\ \\
\raisebox{-11mm}{\epsfig{file=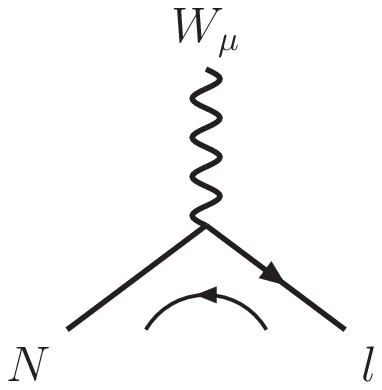,height=22mm,clip=}}
  & $\displaystyle \frac{ig}{\sqrt 2} V_{lN} \gamma^\mu P_R$ & \quad &
\raisebox{-11mm}{\epsfig{file=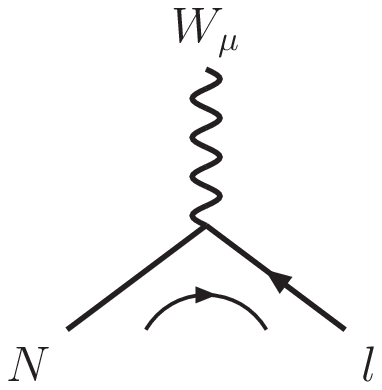,height=22mm,clip=}}
  & $\displaystyle \frac{ig}{\sqrt 2} V_{lN}^* \gamma^\mu P_R$ \\ \\
\raisebox{-11mm}{\epsfig{file=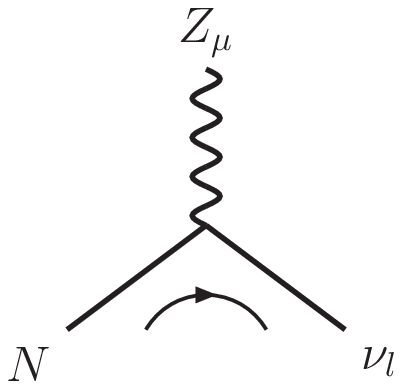,height=22mm,clip=}}
  & $\displaystyle - \eta \frac{ig}{2c_W} \gamma^\mu \left( 
  V_{lN} P_L - V_{lN}^* P_R \right) $ & \quad &
\raisebox{-11mm}{\epsfig{file=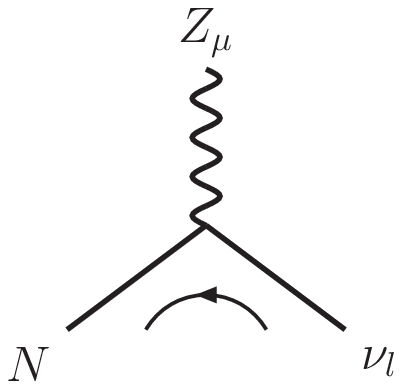,height=22mm,clip=}}
  & $\displaystyle - \eta \frac{ig}{2c_W} \gamma^\mu \left( 
  V_{lN}^* P_L - V_{lN} P_R \right) $ \\ \\
\raisebox{-11mm}{\epsfig{file=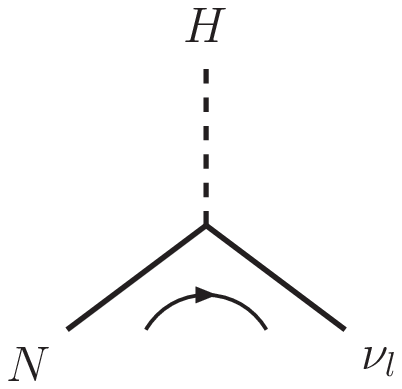,height=22mm,clip=}}
  & $\displaystyle - \eta \frac{ig m_N}{2 M_W} \left( 
   V_{lN}^* P_L + V_{lN} P_R \right)$ & \quad &
\raisebox{-11mm}{\epsfig{file=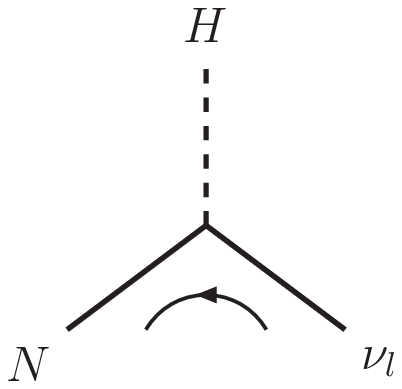,height=22mm,clip=}}
  & $\displaystyle - \eta \frac{ig m_N}{2 M_W} \left( 
   V_{lN}^* P_L + V_{lN} P_R \right)$ \\ \\
\end{tabular}
\end{footnotesize}
\caption{Feynman rules for heavy neutrino singlet ($\eta=1$) and triplet
($\eta=-1$) interactions with SM fermions. The fermion flow is indicated with an arrow.}
\end{center}
\end{table}

\clearpage

\begin{table}[p]
\begin{center}
\begin{footnotesize}
\begin{tabular}{clccl}
\raisebox{-11mm}{\epsfig{file=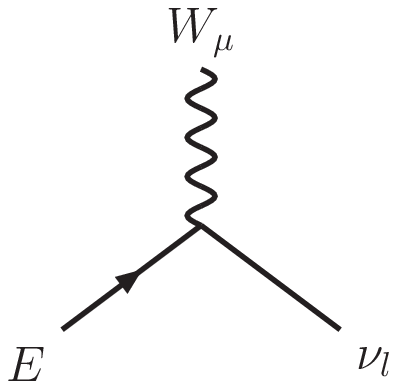,height=22mm,clip=}}
  & $\displaystyle -ig V_{lN}^* \gamma^\mu P_R$ & \quad &
\raisebox{-11mm}{\epsfig{file=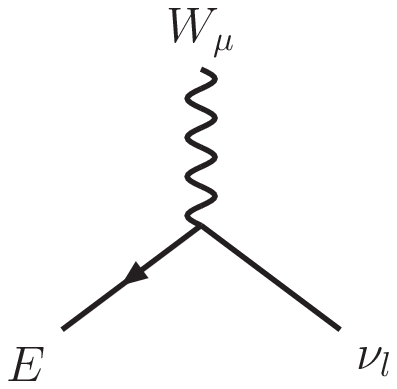,height=22mm,clip=}}
& $\displaystyle -ig V_{lN} \gamma^\mu P_R$ \\ \\
\raisebox{-11mm}{\epsfig{file=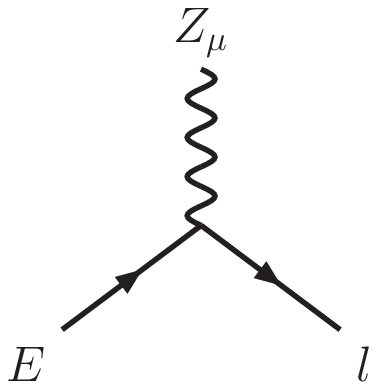,height=22mm,clip=}}
  & $\displaystyle \frac{ig}{\sqrt 2c_W} V_{lN} \gamma^\mu  P_L  $ & \quad &
\raisebox{-11mm}{\epsfig{file=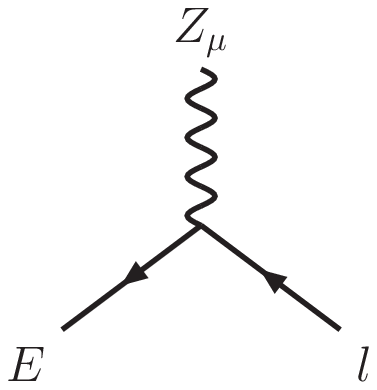,height=22mm,clip=}}
  & $\displaystyle \frac{ig}{\sqrt 2c_W} V_{lN}^* \gamma^\mu P_L  $ \\ \\
\raisebox{-11mm}{\epsfig{file=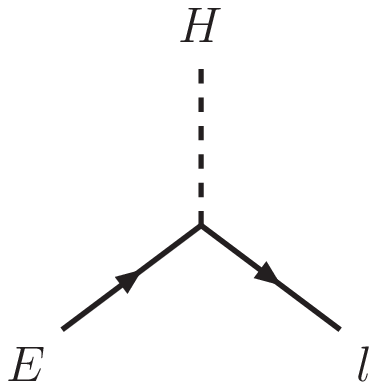,height=22mm,clip=}}
  & $\displaystyle \frac{ig m_E}{\sqrt 2 M_W} V_{lN} P_R$ & \quad &
\raisebox{-11mm}{\epsfig{file=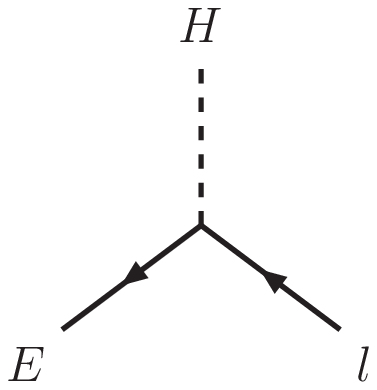,height=22mm,clip=}}
  & $\displaystyle \frac{ig m_E}{\sqrt 2 M_W} V_{lN}^* P_L$ \\ \\
\end{tabular}
\end{footnotesize}
\caption{Feynman rules for heavy charged lepton interactions with SM fermions.}
\end{center}
\end{table}

\begin{table}[p]
\begin{center}
\begin{footnotesize}
\begin{tabular}{clccl}
\raisebox{-11mm}{\epsfig{file=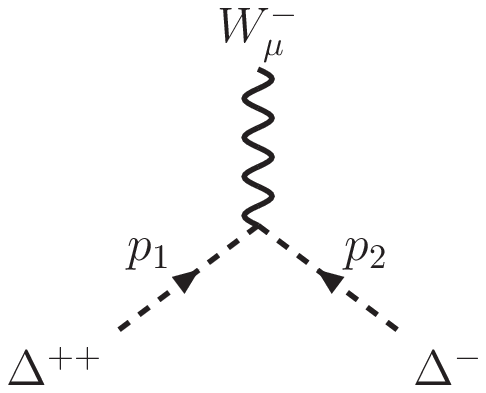,height=22mm,clip=}}
  & $-ig(p_2-p_1)^\mu$ & \quad &
\raisebox{-11mm}{\epsfig{file=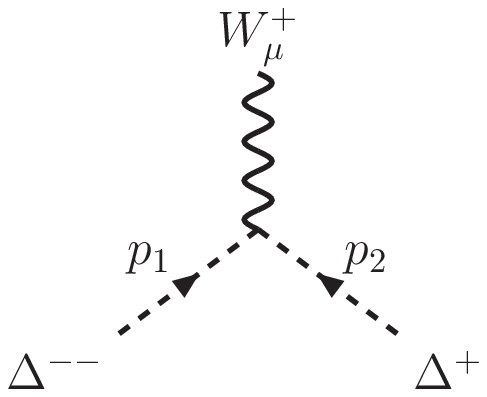,height=22mm,clip=}}
  & $-ig(p_1-p_2)^\mu$ \\ \\
\raisebox{-11mm}{\epsfig{file=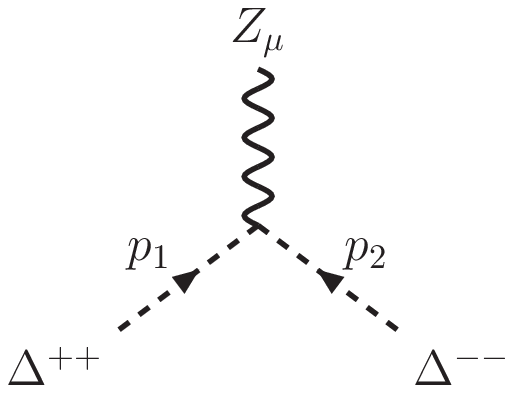,height=22mm,clip=}}
  & $\displaystyle \frac{ig}{c_W} (1-2s_W^2) (p_2-p_1)^\mu$ & \quad &
\raisebox{-11mm}{\epsfig{file=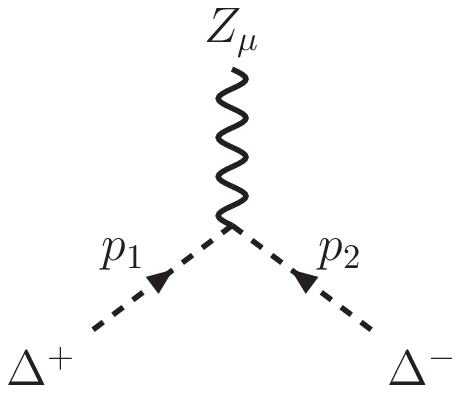,height=22mm,clip=}}
  & $\displaystyle -\frac{ig}{c_W} s_W^2 (p_2-p_1)^\mu$ \\ \\
\raisebox{-11mm}{\epsfig{file=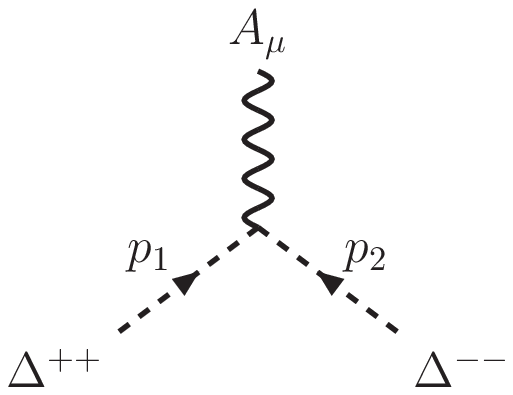,height=22mm,clip=}}
  & $i2e (p_2-p_1)^\mu$ & \quad &
\raisebox{-11mm}{\epsfig{file=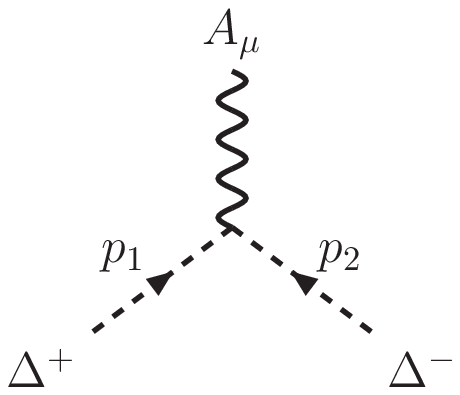,height=22mm,clip=}}
  & $ie (p_2-p_1)^\mu$ \\ \\
\end{tabular}
\end{footnotesize}
\caption{Feynman rules for heavy scalar triplet gauge interactions.}
\end{center}
\end{table}

\clearpage

\begin{table}[t]
\begin{center}
\begin{footnotesize}
\begin{tabular}{clccl}
\raisebox{-11mm}{\epsfig{file=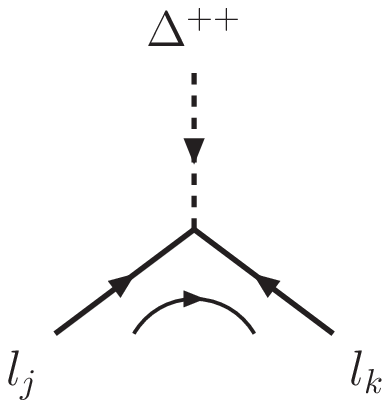,height=22mm,clip=}}
  & $i 2Y_{jk} P_L$ & \quad &
\raisebox{-11mm}{\epsfig{file=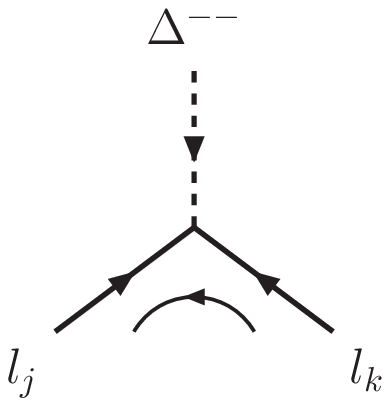,height=22mm,clip=}}
  & $i 2Y_{jk}^* P_R$ \\ \\
\raisebox{-11mm}{\epsfig{file=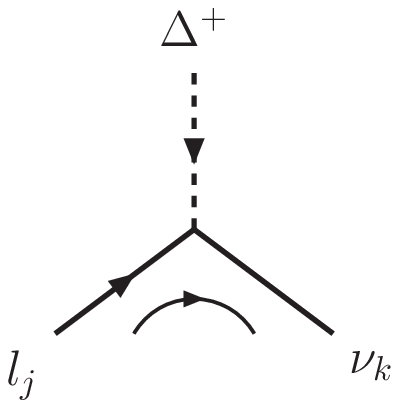,height=22mm,clip=}}
  & $i \sqrt 2 Y_{jk} P_L$ & \quad &
\raisebox{-11mm}{\epsfig{file=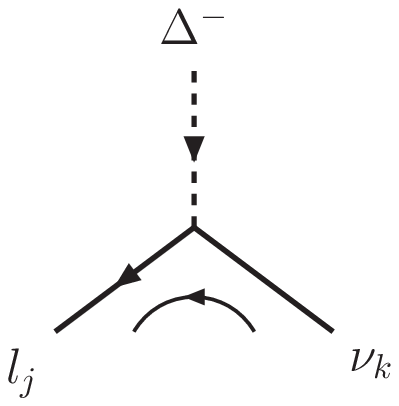,height=22mm,clip=}}
  & $i \sqrt 2 Y_{jk}^* P_R$ \\ \\
\end{tabular}
\end{footnotesize}
\caption{Feynman rules for heavy scalar triplet Yukawa interactions with charged leptons. The fermion flow is indicated with an arrow.}
\end{center}
\end{table}

\end{document}